\documentclass{article}

\usepackage{amssymb,amsfonts,amsmath}
\usepackage{enumerate,float}
\usepackage{color}
\usepackage{tikz}
\usepackage{subfigure}
\usetikzlibrary{arrows,snakes,backgrounds}

\def\be{\begin{eqnarray}}
\def\ee{\end{eqnarray}}
\def\nn{\nonumber}

\date{}

\def\Tr{{\rm Tr}\,}

\def\PPhi{\phi}
\def\PPPhi{\Phi}
\def\horr{{{\smallsmile}\atop{\smallfrown}}}

\def\ld{lock diagram}
\def\lR{lock Reidemester}

\definecolor{red}{rgb}{1,0,0}
\definecolor{orange}{rgb}{1,0.5,0}
\definecolor{violet}{rgb}{0.7,0,1}

\usepackage{setspace}
\usepackage[utf8]{inputenc}
\usepackage[english]{babel}
\usepackage{amsfonts,amsmath,amssymb,mathtools}
\usepackage[left=2cm,right=2cm,top=2cm,bottom=2cm,bindingoffset=0cm]{geometry}
\usepackage{graphicx}
\usepackage{authblk}
\usepackage{xcolor}
\usepackage{hyperref}
\usepackage{dsfont}
\usepackage{authblk}
\usepackage{slashed}
\usepackage{braket}
\usepackage{comment}
\usepackage{pb-diagram}
\usepackage{wasysym}
\usepackage{amsthm}
\usepackage{cancel}
\usepackage{tikz}\usetikzlibrary{tikzmark}
\usetikzlibrary{fadings}
\usetikzlibrary{positioning}
\usetikzlibrary{backgrounds} 
\usetikzlibrary{decorations.pathreplacing}
\usetikzlibrary{arrows}
\definecolor{airforceblue}{rgb}{0.36, 0.54, 0.66}	
\definecolor{beige}{rgb}{0.96, 0.96, 0.86}
\definecolor{bittersweet}{rgb}{1.0, 0.44, 0.37}
\definecolor{melon}{rgb}{0.99, 0.74, 0.71}
\definecolor{mustard}{rgb}{1.0, 0.86, 0.35}
\definecolor{lava}{rgb}{0.81, 0.06, 0.13}
\definecolor{magnolia}{rgb}{0.97, 0.96, 1.0}
\definecolor{lavendermist}{rgb}{0.9, 0.9, 0.98}
\definecolor{lavendergray}{rgb}{0.77, 0.76, 0.82}
\definecolor{palepink}{rgb}{0.98, 0.85, 0.87}
\definecolor{palesilver}{rgb}{0.79, 0.75, 0.73}
\definecolor{cadetgrey}{rgb}{0.57, 0.64, 0.69}
\definecolor{anti-flashwhite}{rgb}{0.95, 0.95, 0.96}
\colorlet{Light0anti-flashwhite}{anti-flashwhite!70!white}
\colorlet{Lightanti-flashwhite}{anti-flashwhite!50!white}
\colorlet{Light2anti-flashwhite}{anti-flashwhite!30!white}
\definecolor{linkcolor}{rgb}{0,0,1}
\definecolor{urlcolor}{rgb}{0,0,1}

\usepackage{tikz-cd}
\usetikzlibrary{decorations.markings}
\usetikzlibrary{positioning}
\usepackage{braids}
\usepackage{array}
\usepackage{ytableau}
\usepackage{longtable}
\usepackage{empheq}
\usepackage{multicol}
\usepackage{mathrsfs}
\usepackage{diagbox}
\usepackage[vcentermath]{youngtab}
\usepackage[natbib=true, backend = bibtex, style=numeric-comp, sorting=none,maxbibnames=99]{biblatex}
\hypersetup{pdfstartview=FitH,  linkcolor=linkcolor,urlcolor=urlcolor,citecolor=urlcolor, colorlinks=true}

\newtheorem{theorem}{Theorem}

\def\theequation{\arabic{section}.\arabic{equation}}
\begin{document}

\title{\bf Planar decomposition of the HOMFLY polynomial \\ for bipartite knots and links
}

\author[2,3]{{\bf A. Anokhina}\thanks{\href{mailto:anokhina@itep.ru}{anokhina@itep.ru}}}
\author[1,2,3]{{\bf E. Lanina}\thanks{\href{mailto:lanina.en@phystech.edu}{lanina.en@phystech.edu}}}
\author[1,2,3]{{\bf A. Morozov}\thanks{\href{mailto:morozov@itep.ru}{ morozov@itep.ru}}}

\vspace{5cm}

\affil[1]{Moscow Institute of Physics and Technology, 141700, Dolgoprudny, Russia}
\affil[2]{Institute for Information Transmission Problems, 127051, Moscow, Russia}
\affil[3]{NRC "Kurchatov Institute", 123182, Moscow, Russia}
\affil[4]{Institute for Theoretical and Experimental Physics, 117218, Moscow, Russia}
\renewcommand\Affilfont{\itshape\small}

\maketitle

\vspace{-7.5cm}

\begin{center}
	\hfill MIPT/TH-17/24\\
	\hfill ITEP/TH-22/24\\
	\hfill IITP/TH-18/24
\end{center}

\vspace{4.5cm}

\begin{abstract}

{
The theory of the Kauffman bracket, which
describes the Jones polynomial as a sum over closed circles
formed by the planar resolution of vertices in a knot diagram,
can be straightforwardly lifted from $\mathfrak{sl}(2)$ to $\mathfrak{sl}(N)$
at arbitrary $N$ -- but for a special class of {\it bipartite}  diagrams
made entirely from the antiparallel lock tangle.
Many amusing and important knots and links can be described in this way,
from twist and double braid knots
to the celebrated Kanenobu knots for even parameters --
and for all of them the entire HOMFLY polynomials
possess planar decomposition.
This provides an approach to evaluation of HOMFLY polynomials,
which is complementary to the arborescent calculus,
and this opens a new direction to homological techniques, parallel to Khovanov-Rozansky
generalizations of the Kauffman calculus.
Moreover, this planar calculus is also applicable to other symmetric representations
beyond the fundamental one,
and to links which are not fully bipartite what is illustrated by examples of Kanenobu-like links.
}
\end{abstract}

















\tableofcontents

\section{Introduction}

Knot calculus in the three-dimensional Chern--Simons theory~\cite{CS,Witten,MoSmi} provides an example of exact non-perturbative calculation in quantum field theory
and emphasizes the role of hidden symmetries and integrability in this kind of problems.
Of special importance there is the possibility to substitute the complicated algebraic machinery by a much simpler
geometrical and homological one -- which is still under-investigated and therefore somewhat restricted.
This paper enlarges applicability of this kind of approaches and shows a possible way out of the long-standing constraints,
by releasing a parameter $N$ (related to the rank of a gauge group in this case) at the expense of restricting the moduli space (from all knots and links to bipartite ones).

From the point of view of quantum field theory,
we consider the three-dimensional topological Chern--Simons theory with $SU(N)$ gauge group
defined by the following action:
\begin{equation}
         S_{\text{CS}}[{\cal A}] = \frac{\kappa}{4 \pi} \int_{S^3} \text{tr} \left( {\cal A} \wedge d{\cal A} +  \frac{2}{3} {\cal A} \wedge {\cal A} \wedge {\cal A} \right).
     \end{equation}
The belief is that the gauge invariant Wilson loops in this theory are quantum knot invariants called the colored HOMFLY polynomials:
\begin{equation}
     \label{WilsonLoopExpValue}
         H_{R}^{\mathcal{K}}(q, A) = \left\langle \text{tr}_{R} \ P \exp \left( \oint_{\mathcal{K}} {\cal A} \right) \right\rangle_{\text{CS}},
     \end{equation}
where the gauge fields are taken in an arbitrary $\mathfrak{sl}(N)$ representation ${\cal A}={\cal A}_\mu^a T_a^R dx^\mu$ with $T_a$ being generators of $\mathfrak{sl}(N)$ Lie algebra and the integration contour can be tied in an arbitrary knot. A generalization to the link case is straightforward. Wilson loops are of great importance in the theoretical physics.
For example, these operators are long known to govern one of the most intriguing physical phenomena such as the
quark confinement in QCD~\cite{polyakov1977quark,polyakov1987gauge}.

A crucial property of the Chern--Simons theory is that correlators, in particular~\eqref{WilsonLoopExpValue}, can be computed exactly, and the answers are
strongly non-perturbative -- polynomials in the variables $q = \exp\left(\frac{2\pi i}{\kappa +N}\right)$ and $A=q^N$. There are several methods for non-perturbative calculations. One of them is the so-called Reshetikhin--Turaev approach~\cite{Reshetikhin,guadagnini1990chern2,reshetikhin1990ribbon,turaev1990yang,reshetikhin1991invariants} coming from hidden quantum group $\mathfrak{sl}_q(N)$ symmetry, and another one is related to an intriguing duality between the 3d Chern--Simons theory and the 2d Wess--Zumono--Novikov--Witten theory~\cite{moore1989classical,Gu_2015,gerasimov1990wess}. However, these methods imply calculations of concrete correlators in concrete representations
and become drastically complicated even to computer calculations
with the growth of complexity of the representation and the knot/link.
Moreover, they are not well suited for examination of properties common for different representations
and particular families of knots/links.
Thus, other methods to calculate Wilson loops in the Chern--Simons theory are also needed.

So far, these alternative methods are applicable only to special kinds of links -- but within these families they are remarkably efficient
and provide a lot of interesting information about the hidden properties of the Wilson observables.
One of such techniques is
the so-called tangle calculus~\cite{kaul1998chern, MMMtangles, morozov2018knot, anokhina2021khovanov,
anokhina2024towards} which allows to calculate the HOMFLY polynomials for links that can be split into some
simple tangles.
Another one is applicable only to the family of the arborescent knots~\cite{mironov2015colored}. 

In this paper, we introduce a very different method -- of planar calculus for the HOMFLY polynomial
generalizing to arbitrary $N$ the celebrated Kauffman-bracket approach to the Jones polynomials at $N=2$.
It works for a specific but rather large family
of bipartite knots, and moreover, can be applied to simplify calculations of the HOMFLY polynomials for some other
knot families. Some examples of calculation of the HOMFLY polynomials via the planar technique are provided in Sections~\ref{sec:examples},~\ref{sec:Kanenobu}.
Here we mostly consider the fundamental representation\footnote{Throughout the paper, we utilize standard enumeration of $\mathfrak{sl}(N)$ representations by Young diagrams, and in particular, we use the box $\Box$ to denote the fundamental representation.} and give just a brief announcement for other
symmetric representations (see Section~\ref{sec:symm}) where additional ideas and techniques are needed.
They will be considered in more details in a companion paper. We also speculate on a possible Khovanov--Rozansky calculus for bipartite links in Section~\ref{sec:Kh-bip}.

A relative simplicity of the Jones polynomials, i.e. of representation theory of $\mathfrak{sl}_q(2)$,
is explained by the Kauffman rule for the quantum $\mathcal{R}$-matrix:
\be
{\cal R}^{ij}_{kl} \ \stackrel{N=2}{=} \  \epsilon^{ij}\epsilon_{kl} -q\cdot \delta^i_k \delta^j_l\,.
\label{Krule}
\ee
In application to knot/link calculus this implies 
the planar decomposition of the Jones polynomials in a sum of powers of $ D_2 =\Tr_{\mathfrak{sl}_q(2)}\, \mathds{1}  =[2]=q+q^{-1}$,
associated with cycles\footnote{We denote by $\mathds{1}$ the identity matrix.}, which are formed by various resolutions of vertices of the knot/link diagram. This provides the Jones polynomial as a sum over vertices of the {\it hypercube} of resolutions,
which afterwards can be $T$-deformed to give the Khovanov polynomial.
See \cite{DM12,DM12-2} for comprehensive reviews.
This works so easily for $N=2$ (i.e. for two colors), because in this case the contravariant invariant tensor $\epsilon^{ij}$
has just two variables -- and becomes far more sophisticated for the HOMFLY polynomials at an arbitrary $N$
\cite{DM3,AnoM}.

\begin{figure}[h]
\begin{picture}(100,100)(-200,-50)

\put(-60,-20){\line(1,1){40}}
\put(-20,-20){\line(-1,1){18}}
\put(-60,20){\line(1,-1){18}}

\put(10,-2){\mbox{$=$}}

\qbezier(30,20)(50,0)(70,20)
\qbezier(30,-20)(50,0)(70,-20)

\put(85,-2){\mbox{$- \ \ \ q $}}

\qbezier(125,20)(145,0)(125,-20)
\qbezier(150,20)(130,0)(150,-20)

\put(-68,22){\mbox{$i$}}  \put(-15,22){\mbox{$j$}}  \put(-68,-28){\mbox{$k$}}  \put(-15,-28){\mbox{$l$}}

\end{picture}
\caption{\footnotesize The  celebrated "Kauffman bracket" -- the planar decomposition (\ref{Krule})
of the ${\cal R}$-matrix vertex for the fundamental representation of $\mathfrak{sl}_q(2)$.
In this case ($N=2$) the conjugate of the fundamental representation is isomorphic to it,
thus, tangles in the picture has no orientation.
Indices can be raised with the help of the invariant tensor $\epsilon^{ij}$, which is exactly of rank $N=2$.
}
\label{fig:Kauff}
\end{figure}
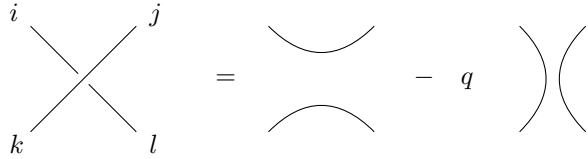

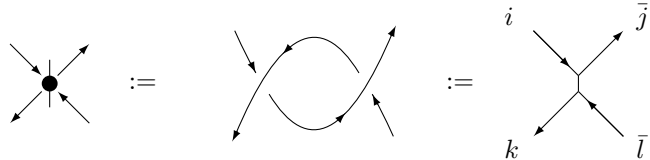
\begin{figure}[h]
\begin{picture}(100,60)(-140,-30)

\put(70,0){
\put(-105,15){\vector(1,-1){12}} \put(-87,3){\vector(1,1){12}}
\put(-93,-3){\vector(-1,-1){12}} \put(-75,-15){\vector(-1,1){12}}
\put(-90,0){\circle*{6}}  \put(-90,-9){\line(0,1){18}}

\put(-60,-2){\mbox{$:=$}}
}

\qbezier(50,20)(55,9)(58,4) \qbezier(63,-4)(85,-40)(110,20)
\put(56,8){\vector(1,-2){2}} \put(90,-13){\vector(1,1){2}} \put(109,18){\vector(1,2){2}}
\qbezier(50,-20)(75,40)(97,4)  \qbezier(102,-4)(105,-9)(110,-20)
\put(104,-8){\vector(-1,2){2}} \put(70,13){\vector(-1,-1){2}} \put(51,-18){\vector(-1,-2){2}}

\put(180,0){
\put(-50,-2){\mbox{$:=$}}

\put(0,0){\put(-17,20){\line(1,-1){17}}\put(-17,20){\vector(1,-1){14}}   \put(0,3){\vector(1,1){17}}
 \put(0,-3){\vector(-1,-1){17}}   \put(17,-20){\line(-1,1){17}} \put(17,-20){\vector(-1,1){14}}
 \put(0,-3){\line(0,1){6}}}

\put(-28,22){\mbox{$i$}}  \put(22,22){\mbox{$\bar j$}}  \put(-28,-28){\mbox{$k$}}  \put(22,-28){\mbox{$\bar l$}}
}

\end{picture}
\caption{\footnotesize
The  antiparallel (AP) lock -- the main hero of this paper.
It is in the picture in the middle, while to the left and to the right sides are its two abbreviated notation,
which will be used in the text. We mark the orientation explicitly -- and sometimes we call the lock in the picture "vertical".
The name and the small segments in these picture are {\it vertical} -- point in the direction of {\it locking}.
This is the standard notation in the theory of bipartite knots \cite{BipKnots,Lewark}.
}
\label{fig:lock}
\end{figure}

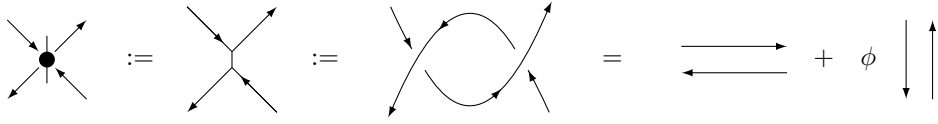
\begin{figure}[h!]
\begin{picture}(100,60)(-150,-30)

\put(20,0){
\put(-105,15){\vector(1,-1){12}} \put(-87,3){\vector(1,1){12}}
\put(-93,-3){\vector(-1,-1){12}} \put(-75,-15){\vector(-1,1){12}}
\put(-90,0){\circle*{6}}  \put(-90,-9){\line(0,1){18}}

\put(-60,-2){\mbox{$:=$}}
}

\put(0,0){\put(-17,20){\line(1,-1){17}}\put(-17,20){\vector(1,-1){14}}   \put(0,3){\vector(1,1){17}}
 \put(0,-3){\vector(-1,-1){17}}   \put(17,-20){\line(-1,1){17}} \put(17,-20){\vector(-1,1){14}}
 \put(0,-3){\line(0,1){6}}}

\put(10,0){

\put(20,-2){\mbox{$:=$}}

\qbezier(50,20)(55,9)(58,4) \qbezier(63,-4)(85,-40)(110,20)
\put(56,8){\vector(1,-2){2}} \put(90,-13){\vector(1,1){2}} \put(109,18){\vector(1,2){2}}
\qbezier(50,-20)(75,40)(97,4)  \qbezier(102,-4)(105,-9)(110,-20)
\put(104,-8){\vector(-1,2){2}} \put(70,13){\vector(-1,-1){2}} \put(51,-18){\vector(-1,-2){2}}

}

\put(0,0){
\put(140,-2){\mbox{$=$}}

\put(100,65){
\put(70,-60){\vector(1,0){40}}
\put(110,-70){\vector(-1,0){40}}
\put(120,-67){\mbox{$+\ \ \ \phi$}}
\put(155,-50){\vector(0,-1){30}}
\put(165,-80){\vector(0,1){30}}
}
}

\end{picture}
\caption{\footnotesize
The planar decomposition of the lock vertex -- the main statement of the present text.
}\label{fig:pladeco}
\end{figure}

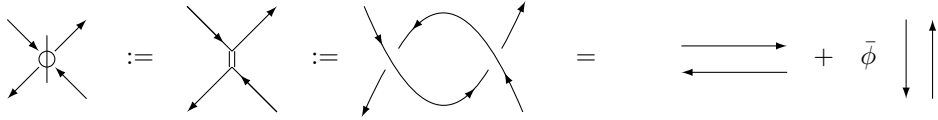
\begin{figure}[h!]
\begin{picture}(100,50)(-150,-20)

\put(20,0){
\put(-105,15){\vector(1,-1){12}} \put(-87,3){\vector(1,1){12}}
\put(-93,-3){\vector(-1,-1){12}} \put(-75,-15){\vector(-1,1){12}}
\put(-90,0){\circle{6}}  \put(-90,-9){\line(0,1){18}}

\put(-60,-2){\mbox{$:=$}}
}

\put(0,0){

\put(0,0){\put(-17,20){\line(1,-1){17}}\put(-17,20){\vector(1,-1){14}}   \put(0,3){\vector(1,1){17}}
 \put(0,-3){\vector(-1,-1){17}}   \put(17,-20){\line(-1,1){17}} \put(17,-20){\vector(-1,1){14}}
\put(-1,-3){\line(0,1){6}} \put(1,-3){\line(0,1){6}}
}

\put(30,-2){\mbox{$:=$}}

\qbezier(50,20)(75,-40)(97,-4)  \qbezier(102,4)(105,9)(110,20)
\qbezier(50,-20)(55,-9)(58,-4) \qbezier(63,4)(85,40)(110,-20)
\put(55,9){\vector(1,-2){2}} \put(90,-13){\vector(1,1){2}} \put(109,18){\vector(1,2){2}}
\put(105,-9){\vector(-1,2){2}} \put(70,13){\vector(-1,-1){2}} \put(51,-18){\vector(-1,-2){2}}

\put(130,-2){\mbox{$=$}}

\put(100,65){
\put(70,-60){\vector(1,0){40}}
\put(110,-70){\vector(-1,0){40}}
\put(120,-67){\mbox{$+\ \ \ \bar\PPhi$}} \put(-25,0){
\put(180,-50){\vector(0,-1){30}}
\put(190,-80){\vector(0,1){30}}
}
}}

\end{picture}
\caption{\footnotesize The "opposite" of the vertical lock denoted by a double segment.
It is made from inverse vertices, thus $\bar{\phi}=\phi\big|_{A\rightarrow A^{-1},\,q\rightarrow q^{-1}}$.
} \label{fig:oppvert}
\end{figure}

In this paper, we note that the same trick can be applied for an arbitrary $N$ -- but for a very special
kind of knot diagrams, which are made entirely from the antiparallel (AP) lock tangles,
depicted in Figs.\ref{fig:lock},\ref{fig:pladeco},\ref{fig:oppvert} (and also ones having inverse orientation) and studied in some detail in \cite{evo,MMMtangles}.
Knots and links possessing such realization are sometimes called {\it bipartite} \cite{BipKnots,Lewark},
and we discuss examples and the intriguing question of their abundancy in the special Section \ref{sec:bipknots}.

\noindent
The AP lock tangle  is the convolution of two ${\cal R}$-matrices $\tau^{i\bar j}_{k\bar l} =
({\cal R}*{\cal R})^{i\bar j}_{k \bar l} := {\cal R}^{in}_{km}{\cal R}^{m\bar j}_{n\bar l}$
and admits the following planar decomposition\footnote{
Throughout the paper we use the standard notation $\{x\} :=x-x^{-1}$, and $A=q^N$. Then the quantum number is $[n]=\frac{\{q^n\}}{\{q\}}$, and the dimension of the fundamental representation of $\mathfrak{sl}_q(N)$,
which will be associated with each closed cycle, is $D_N:=[N]=\frac{\{A\}}{\{q\}}$. Sometimes we omit the subscript $N$ of $D_N$. 
We denote incoming indices either by a superscript or by a conjugate subscript,
and outcoming indices are the opposite: either a subscript or a conjugate superscript.
The invariant tensor is either $\delta^i_k$ and $\delta^{\bar j}_{\bar l}$  with upper and lower indices of the same type
or $\delta^{i\bar j}$ and $\delta_{k\bar l}$ with both upper or lower indices of different types.}:
\be
\tau^{i\bar j}_{k\bar l}  = \delta^{i\bar j}\delta_{k\bar l} + \PPhi\, \delta^i_k \delta^{\bar j}_{\bar l}\quad \text{with}
\quad
\PPhi =A(q-q^{-1}):= A\{q\}\,.
\label{tau}
\ee
Pictorially it is shown in Fig.\ref{fig:pladeco} and derived in Section~\ref{sec:planar-deriv}. The "opposite" element in Fig.\ref{fig:oppvert} is
\be\label{bar-tau}
{({\bar\tau})}^{i\bar j}_{k\bar l}
= \delta^{i\bar j}\delta_{k\bar l} +  \bar{\PPhi}\, \delta^i_k \delta^{\bar j}_{\bar l}\quad \text{with}\quad \bar{\PPhi} = -A^{-1}\{q\}\,.
\ee

Emphasize that the decomposition in Fig.\ref{fig:pladeco} turns into the Kauffman decomposition in Fig.\ref{fig:Kauff} when transforming $\phi\rightarrow -q$ (for inverse crossings $\bar{\phi}\rightarrow -q^{-1}$). This fact allows to provide general computations valid both for the bipartite HOMFLY polynomial and for the Jones polynomial for a link obtained from a bipartite link by shrinking every AP lock to a single crossing. We call such a diagram a {\it precursor} diagram. Thus, the computation of the HOMFLY polynomial of a bipartite link can be made as easy as the computation of the Jones polynomial for the corresponding precursor diagram. One just needs to substitute $\phi$ and $\bar{\phi}$ instead of $-q$ and $-q^{-1}$ correspondingly in the Kauffman-bracket calculus for the Jones polynomial.
\begin{figure}[h!]
\centering
\includegraphics[width=12cm]{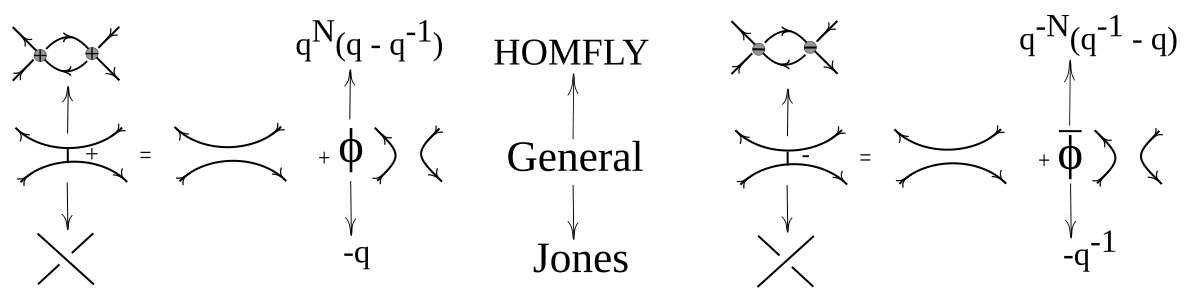}
\caption{\footnotesize
The Kauffman and lock decompositions can be generalized to describe both the bipartite HOMFLY polynomial and the precursor Jones polynomial.
}\label{fig:}
\end{figure}

Substituting the decompositions in Figs.\ref{fig:pladeco},\ref{fig:oppvert} for all vertices in a link we get a sum over various non-intersecting (planar) cycles,
each contributing $D$
with coefficients, which are made from products of
$\PPhi$ and $\bar{\PPhi}$. From this planar resolution we write down the polynomial 
\be
\boxed{
P^{\,\rm bipartite}_\Box = \sum^{2^v} D^a \PPhi^b \bar\PPhi^c\,,
}
\label{abc}
\ee
where $v$ is the number of vertices in the diagram. Note that all the  items here come with unit coefficients (!), thus the only needed information is pure
combinatorial/geometrical, reflecting the selection of powers $a,b,c$ for each planar resolution of the AP lock diagram.
In fact, the powers $b$ and $c$ are distributed in a very simple way, see (\ref{bc}) below -- thus, the only issue
is the distribution of $a$.

Due to the above discussion, the polynomial $P^{\,\rm bipartite}_\Box$ from~\eqref{abc} corresponds to both the bipartite HOMFLY polynomial and to the Jones polynomial for the corresponding precursor diagram. The final answers for these polynomials are obtained by the substitutions from Table~\ref{tab:subs}, or explicitly:
\begin{equation}\label{H-J-gen}
\begin{aligned}
    H^{\,\rm bipartite}_\Box &= A^{-w}\sum^{2^v} D_N^a \PPhi^b \bar\PPhi^c\,, \quad \phi=A\{q\}\,,\; \bar{\phi}=-A^{-1}\{q\}\,, \\
    J^{\,\rm precursor}_\Box &= (-1)^{\frac{1}{2}(w+n_+-n_-)}\cdot q^{-\frac{3}{2}w-\frac{1}{2}(n_+-n_-)}\sum^{2^v} D_2^a (-q)^b (-q)^{-c}\,.
\end{aligned}
\end{equation}

\begin{table}[h!]
    \centering
    \begin{tabular}{|c||c|c|c|c|c|}
    \hline
         & \raisebox{-0.1cm}{knot diagram} & \raisebox{-0.1cm}{$\phi$} & \raisebox{-0.1cm}{$\bar{\phi}$} & \raisebox{-0.1cm}{$D$} & \raisebox{-0.1cm}{framing factor} \\ [1.5ex]
    \hline
    \hline
        \raisebox{-0.1cm}{HOMFLY} & \raisebox{-0.1cm}{bipartite} & \raisebox{-0.1cm}{$A\{q\}$} & \raisebox{-0.1cm}{$-A^{-1}\{q\}$} & \raisebox{-0.1cm}{$D_N = \frac{\{A\}}{\{q\}}$} & \raisebox{-0.1cm}{$A^{-w}$} \\ [1.5ex]
    \hline   
        \raisebox{-0.1cm}{Jones} & \raisebox{-0.1cm}{precursor} & \raisebox{-0.1cm}{$-q$} & \raisebox{-0.1cm}{$-q^{-1}$} & \raisebox{-0.1cm}{$D_2=q+q^{-1}$} & \raisebox{-0.1cm}{$(-1)^{\frac{1}{2}(w+n_+-n_-)}\cdot q^{-\frac{3}{2}w-\frac{1}{2}(n_+-n_-)}$} \\ [1.5ex]
    \hline
    \end{tabular}
    \caption{\footnotesize These substitutions turn~\eqref{abc} into the HOMFLY and Jones polynomials. Here $w$ is the write number of the corresponding diagram, $n_+$ is the number of ``$+$''-crossings shown in Fig.\ref{fig:Kauff} and $n_-$ is the number of the opposite ``$-$''-crossings in a link diagram.}
    \label{tab:subs}
\end{table}

A possible formulation of our statement is that a single knot/link diagram
can describe the Jones polynomial by a direct application of the Kauffman rule (\ref{Krule}) at all vertices
and simultaneously a whole bunch of the HOMFLY polynomials for a variety of knots/links,
obtained by insertion of differently oriented AP lock tangles at every vertex, see Section~\ref{sec:hypercube} for more details.
Expressions for these HOMFLY polynomials are just the same as for the Jones polynomials, with powers of $q$ in a cycle decomposition changed
for $\PPhi$ and $\bar{\PPhi}$, and $D_2$ changed for $D_N$.

Note that one can consider one intersection in the lock tangle to be virtual what leads to the construction of virtual links~ \cite{1406.7331,1407.6319}. The planar technique for such virtual links is an interesting question for future research. 

\setcounter{equation}{0}
\section{Derivation of planar decomposition}\label{sec:planar-deriv}

\begin{figure}[h]
\begin{picture}(100,245)(-170,-210)


\put(0,0){\put(-17,20){\line(1,-1){17}}\put(-17,20){\vector(1,-1){14}}   \put(0,3){\vector(1,1){17}}
 \put(0,-3){\vector(-1,-1){17}}   \put(17,-20){\line(-1,1){17}} \put(17,-20){\vector(-1,1){14}}
 \put(0,-3){\line(0,1){6}}}

\put(-28,22){\mbox{$i$}}  \put(22,22){\mbox{$\bar j$}}  \put(-28,-28){\mbox{$k$}}  \put(22,-28){\mbox{$\bar l$}}

\put(30,-2){\mbox{$:=$}}

\qbezier(50,20)(55,9)(58,4) \qbezier(63,-4)(85,-40)(110,20)
\put(56,8){\vector(1,-2){2}} \put(90,-13){\vector(1,1){2}} \put(109,18){\vector(1,2){2}}
\qbezier(50,-20)(75,40)(97,4)  \qbezier(102,-4)(105,-9)(110,-20)
\put(104,-8){\vector(-1,2){2}} \put(70,13){\vector(-1,-1){2}} \put(51,-18){\vector(-1,-2){2}}

\put(80,-40){\mbox{$||$}}

\put(-40,-80){
\qbezier(50,20)(55,9)(58,4) \qbezier(63,-4)(80,-40)(98,-5)    \qbezier(102,3)(105,9)(110,20)
\put(56,8){\vector(1,-2){2}} \put(93,-13){\vector(1,1){2}} \put(109,18){\vector(1,2){2}}
\qbezier(50,-20)(80,40)(110,-20)
\put(105,-11){\vector(-1,2){2}} \put(71,7){\vector(-1,-1){2}} \put(51,-18){\vector(-1,-2){2}}

\put(120,-2){\mbox{$+\ \ \{q\}$}}

\put(155,20){\vector(1,-2){9}}
\qbezier(169,-10)(190,-40)(190,0)
\qbezier(155,-20)(190,40)(190,0)
\put(156,-18){\vector(-1,-2){2}}
\put(200,-22){\vector(0,1){43}}

\put(120,-40){\mbox{$||$}}

\put(0,-20){
\put(70,-60){\vector(1,0){40}}
\put(110,-70){\vector(-1,0){40}}
\put(120,-67){\mbox{$+\ \ \ A\{q\}$}}
\put(170,-50){\vector(0,-1){30}}
\put(180,-80){\vector(0,1){30}}
}
}

\put(-35,-165){
\put(0,0){\put(-17,20){\line(1,-1){17}}\put(-17,20){\vector(1,-1){14}}   \put(0,3){\vector(1,1){17}}
 \put(0,-3){\vector(-1,-1){17}}   \put(17,-20){\line(-1,1){17}} \put(17,-20){\vector(-1,1){14}}
 \put(0,-3){\line(0,1){6}}}

\put(30,-2){\mbox{$=$}}
}

\put(-70,-135){\line(1,0){230}}
\put(-70,-195){\line(1,0){230}}
\put(-70,-135){\line(0,-1){60}}
\put(160,-135){\line(0,-1){60}}

\end{picture}
\caption{\footnotesize
Substituting the skein relation from Fig.\ref{skein} for the right vertex, we get the two diagrams in the second line.
As shown in the third line, the first is trivial and the second is trivialized by the first Reidemeister relation.
Altogether we get the decomposition of the AP lock into two planar trivialities -- exactly like it was in the
Kauffman case.
The three differences are: applicability at arbitrary $N$, the presence of orientation, a more sophisticated
$N$-dependent coefficient
$A\{q\}= q^N(q-q^{-1})$.
Note that of crucial importance is the presence of two vertices in the lock and the antiparallel choice --
this allows the arrows to be directed in a peculiar way, consistent with {\it planar} decomposition.
Direct generalization of the Kauffman rule for a single vertex gives rise to non-planar decomposition \cite{DM3}.
{\bf The planar decomposition rule in the box is the main statement of this paper.}
All the rest results are illustrative examples and possible applications.
} \label{planar}
\end{figure}
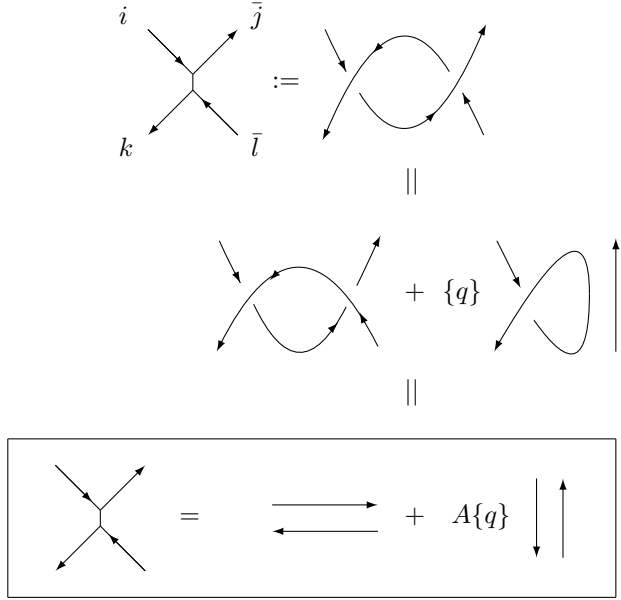

\begin{figure}[h]
\begin{picture}(100,60)(-200,-30)

\put(-60,-20){\vector(1,1){40}}
\put(-20,-20){\line(-1,1){18}}
\put(-42,2){\vector(-1,1){18}}

\put(10,-2){\mbox{$=$}}

\put(45,-20){\line(1,1){18}}
\put(67,2){\vector(1,1){18}}
\put(85,-20){\vector(-1,1){40}}

\put(95,-2){\mbox{$+ \ \ \ \{q\}$}}

\put(10,0){
\qbezier(125,20)(145,0)(125,-20)
\qbezier(150,20)(130,0)(150,-20)
\put(127,18){\vector(-1,1){2}}
\put(148,18){\vector(1,1){2}}
}

\end{picture}
\caption{\footnotesize The skein relation following from the knowledge of ${\cal R}$-matrix eigenvalues
in the fundamental representation, see (\ref{eigenR}).
This relation is used in the derivation of the planar decomposition of AP lock in Fig.\ref{planar}.
} \label{skein}
\end{figure}
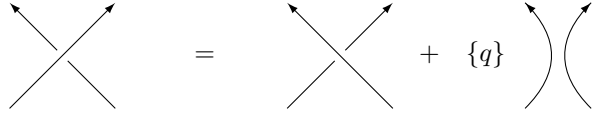

In this section we derive the planar decomposition for the AP lock in Fig.\ref{fig:pladeco}. As shown in  Fig.\ref{planar}, equation (\ref{tau}) is a direct corollary of the skein relation, Fig.\ref{skein}, saying that the fundamental ${\cal R}$-matrix has two eigenvalues $q$ and $-q^{-1}$:
\be
({\cal R}-q)({\cal R}+q^{-1}) = 0 \ \ \Longrightarrow \ \ {\cal R} = {\cal R}^{-1} + \{q\} \cdot \mathds{1}
\ \ \Longrightarrow \ \ \tau = {\cal R}*{\cal R} = {\cal R}*\Big({\cal R}^{-1} + \{q\}\cdot \mathds{1}\Big) = || + A\{q\}\cdot \horr\,.
\label{eigenR}
\ee

\newpage

\noindent
At this point, we have fixed the {\bf framing}, and preferable choice is {\it vertical} \footnote{
This is the standard term for the representation-theory framing --
not to be mixed with the word "vertical", which we use to characterize our way of drawing the locks.
}
rather than topological --
it allows to keep  the coefficients in front of $||$ unit.
The price to pay is the overall framing factor in the HOMFLY polynomial, see Table~\ref{tab:subs}, which should be restored to provide a topological invariant quantity.

In Sections~\ref{sec:examples},~\ref{sec:Kanenobu} we provide various examples of the fundamental HOMFLY polynomials, calculated in by this planar decomposition method. After that in Section~\ref{sec:symm}, we discuss generalizations to other symmetric representations and to the Khovanov calculus 
in Section~\ref{sec:Kh-bip} (which can provide not the usual Khovanov--Rozansky polynomials for bipartite knots but still looks potentially interesting).

\setcounter{equation}{0}
\section{Which knots are bipartite?
\label{sec:bipknots}}

In this section, we present some known results about biparticity of links. We call links {\it bipartite} if they can be represented by a diagram obtained by gluing together antiparallel lock tangles only. Such diagrams are often called in literature as {\it matched diagrams}~\cite{BipKnots,Lewark}, we mostly call them bipartite or lock diagrams in this paper. Matched diagrams were first introduced in 1987 in paper~\cite{PP87}. Until the work~\cite{BipKnots} (for 24 years), nobody succeeded in finding or even proving that non-bipartite knots do exist. 

To find the obstacles for existence of bipartite realization of a given knot/link 
one needs to find a link invariant which has some special property in bipartite case
but not in general. In \cite{BipKnots}, it was suggested to look at the Alexander ideals.
Since the HOMFLY polynomial has a symmetry $H_{R^\vee}(q,A) = H_R(q^{-1},A)$ 
and transposition (denoted by $\vee$) does not change the fundamental representation,
$\Box^\vee=\Box$, the Alexander polynomial at $A=q^N=1$ satisfies 
${\rm Al}_\Box(q) = {\rm Al}_\Box(q^{-1})$
and depends on $q$ only through the Conway variable  $z^2 = \{q\}^2$. 
However, the Alexander polynomial has an alternative realization in terms of the determinant
of the Alexander matrix $\mathbb{A}=q^2 S-S^\vee$ obtained from the Seifert matrix $S$.
It turns out that the $m-$th ideal being defined as the ideal in $\mathbb{Z}[q^2,q^{-2}]$ generated by all minors of an arbitrary Alexander matrix of size $n - m + 1$, where $n$ is the smallest among the number of columns and rows in $\mathbb{A}$, are also invariants -- not directly expressible
through the HOMFLY polynomials.
Thus, these {\it Alexander ideals} can depend on $q^2$ and $q^{-2}$ separately.
The claim of \cite{BipKnots} is that for bipartite knots this does not happen --
the entire Alexander matrix, and thus, all its minors can be chosen to depend polynomially on $z^2$.
At the same time there are knots with Alexander ideals generated by $1+q^2$ 
not expressible through $z^2$ -- which therefore cannot be bipartite. For computations of the second Alexander ideals see Appendix~\ref{sec:App}.
\begin{theorem}[Duzhin, Shkolnikov, 2011,~\cite{BipKnots}]
    If a higher Alexander ideal of a knot contains the polynomial $1+q^2$, then this knot is not bipartite. 
\end{theorem}
Due to this observation one immediately gets a series of non-bipartite knots by the tables and the computer program of Knot Atlas~\cite{KA}. The first examples in the Rolfsen table are
$9_{35}$, $9_{37}$, $9_{41}$, $9_{46}$, $9_{47}$, $9_{48}$, $9_{49}$,
$10_{74}$, $10_{75}$, $10_{103}$, $10_{155}$, $10_{157}$, 
$11a_{123}$, $11a_{135}$, $11a_{155}$, $11a_{173}$, $11a_{181}$, $11a_{196}$, $11a_{249}$, $11a_{277}$, 
$11a_{291}$, $11a_{293}$,  $11a_{314}$, $11a_{317}$, $11a_{321}$, $11a_{366}$, 
$11n_{126}$, $11n_{133}$, $11n_{167}$. The second Alexander ideals of these knots are $\langle b,\,1+q^2 \rangle$ with some integer number $b$, and due to the discussed criteria these knots are non-bipartite. Among these non-bipartite knots, there is the pretzel knot ${\rm pretzel}(3,3,-3)=9_{46}$. In fact, this generalizes to ${\rm pretzel}(p,\, q,\, r)$ knots with $p,\, q,\, r$ odd and the greatest common divisor of these numbers $\lambda = {\rm gcd}(p,\, q,\, r) > 1$, because their second Alexander ideal is generated by $\lambda$ and $1 + q^2$,~\cite{Lewark}. 

It seems that many links are bipartite. For all knots in the Rolfsen table with up to 8 crossings, the authors of \cite{BipKnots} and \cite{8_18}
claim that explicit bipartite realizations can be provided. Note that sometimes the $z^2$-dependence of an ideal can be a bit obscure,
e.g. for $8_{18}$ from Knot Atlas, the second ideal is $\langle q^4-q^2+1 \rangle$. However, these ideals are in $\mathbb{Z}[q^2,q^{-2}]$ what in particular means that their generators can by multiplied by $q^{2n}$, $n\in \mathbb{Z}$. Thus, $\langle q^4-q^2+1 \rangle=\langle z^2+1 \rangle$, and the knot $8_{18}$ is bipartite. 

Additional claims is that all rational knots \cite{RatKnots} and Montesinos links\footnote{A {\it Montesinos link} $M(p_1/q_1,...,p_n/q_n)$ is a generalization of pretzel links, where the 2-strand braids are replaced by rational tangles. In particular, a knot $M(p/q)$ is a rational knot, and thus, bipartite (see Theorem 2). Montesinos realizations of knots with up to 12 crossings are available in~\cite{knotinfo}.} of special kind \cite{Lewark} are bipartite. The explicit statements are provided below.
\begin{theorem}[Duzhin, Shkolnikov, 2010,~\cite{RatKnots}]
    All rational knots are bipartite.
\end{theorem} 
\begin{theorem}[Lewark, Lobb, 2015,~\cite{Lewark}]
    Consider the unoriented Montesinos link $L = M(p_1/q_1,...,p_n/q_n)$. If $L$ has more than one component, then it is bipartite. If $L$ is a knot and one of the denominators $q_i$ is even, then $L$ is bipartite.
\end{theorem}
It is worth noting that non-Montesinos knots can also be bipartite. For example, the knots $8_{16}$, $8_{17}$, $8_{18}$ are not Montesinos, but was proved to be bipartite, as we have already discussed. The actual abundance of bipartite knots 
-- even if they are of measure zero or one in the space of all knots --
remains unknown (at least to us).

\begin{figure}[h!]
$$\begin{array}{c}
\includegraphics[width=7.5cm]{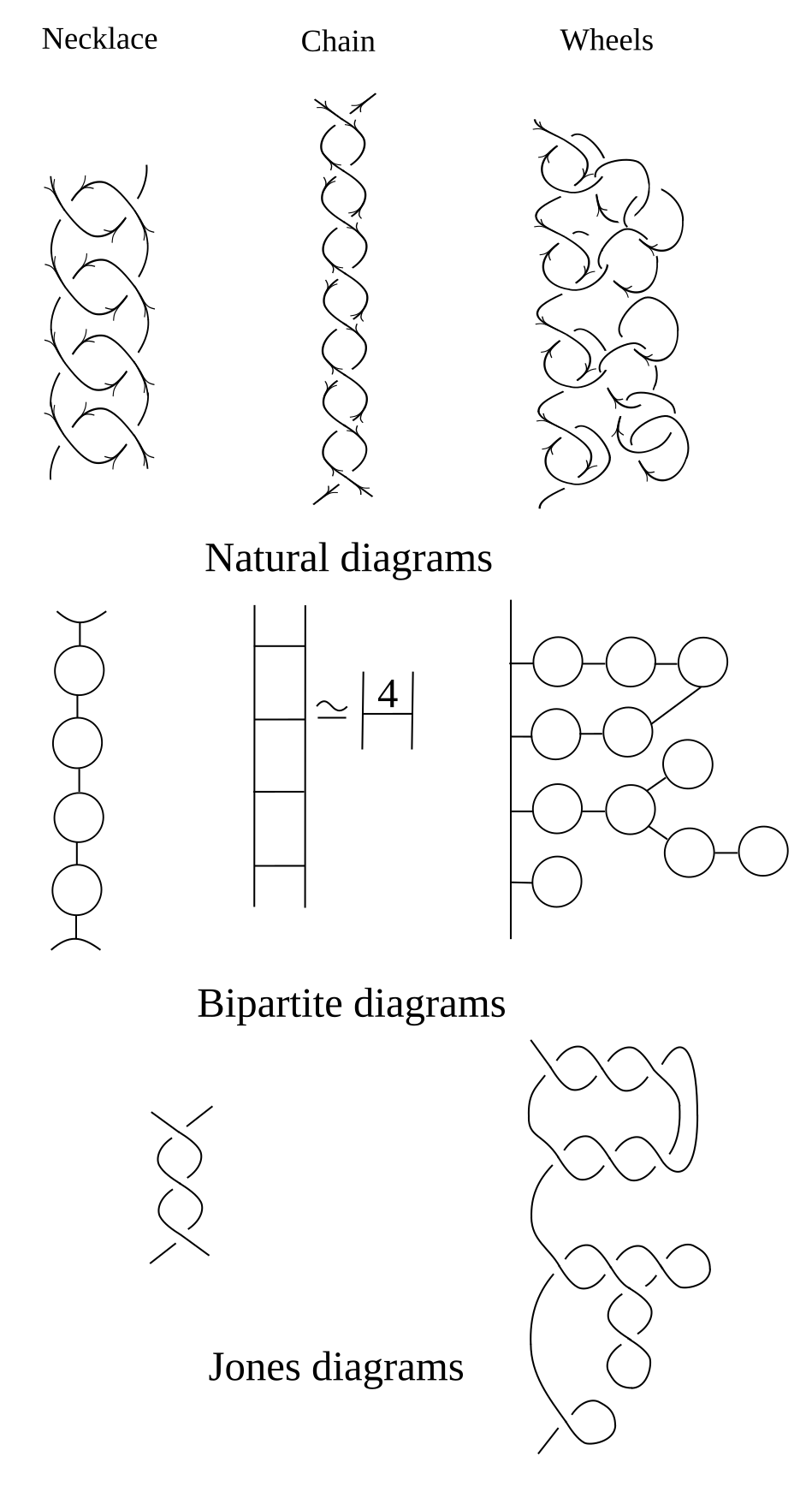}
\end{array}$$
\caption{\footnotesize
The simplest examples of bipartite diagrams for knots and links under our consideration.
}\label{fig:Examplefigs}
\end{figure}

Some examples of bipartite links and knots  are shown in Fig.\ref{fig:Examplefigs}.
As usual in the Kauffman calculus, separation between knots and links is artificial --
in this formalism they form closely related families, as we will also see in sample calculations
of Sections~\ref{sec:examples},~\ref{sec:Kanenobu}.

\setcounter{equation}{0}
\section{Lock diagrams and lock hypercube(s)}\label{sec:hypercube}

A natural step in the study of bipartite diagrams is to try to substitute them by the simpler ones,
obtained by shrinking of AP locks to points.
We call them {\it precursor diagrams}.
The idea is that the expression for the HOMFLY polynomial of a bipartite diagram can be obtained from that
for the (Kauffman-expanded) Jones polynomial of the corresponding precursor diagram by a simple substitution of $\phi$ and $\bar{\phi}$ dictated by a lock diagram
in the expansion coefficients.
In particular, it should be possible to read this expression from the {\it lock hypercube},
i.e. the hypercube of resolutions of a lock diagram.

{\footnotesize
One immediate word of caution is that we need not just the final expression for the Jones polynomial as a function of $q$,
but rather its cycle expansion $J^{\cal K}_\Box(q,D_2)$, where we are going to substitute $D_2$ by $D_N$
and the $q$-dependent coefficients by $\phi$ and $\bar{\phi}$.
But this double expansion is exactly what is provided by hypercube calculus, see \cite{DM12}. Note that the mentioned substitution depends on a bipartite diagram.
}

\begin{figure}[!ht]
    \centering
    \includegraphics[scale=0.5]{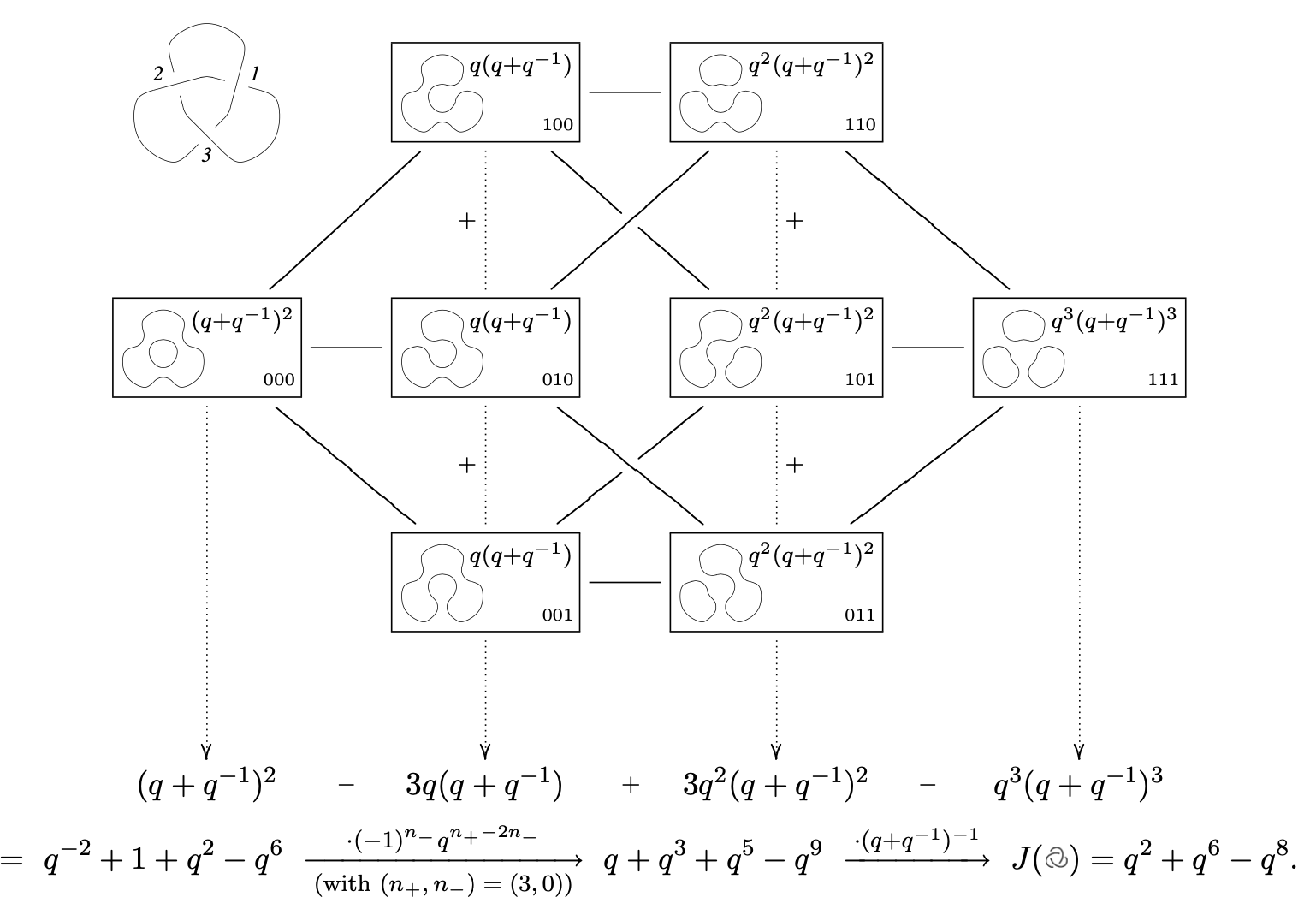}
\caption{\footnotesize Calculation of the Jones polynomial by the hypercube of resolutions for the trefoil knot from~\cite{BarN}. A calculation of the bipartite HOMFLY polynomial is the same and provided by the change $D_2=q+q^{-1}\rightarrow D_N$ and $q$-powers to $\phi$- and $\bar\phi$-powers depending on orientation of AP locks to be inserted instead of single vertices.}\label{fig:trefoil}
\end{figure}

The idea of a hypercube of resolutions is that there are two choices at every vertex of the diagram,
thus a total of $2^n$ choices for $n$ vertices.
With each choice one associates a {\it vertex} of the hypercube,
and its {\it edges} correspond to the flips of choices at all possible vertices --
thus there are $n$ {\it edges} at each hypercube {\it vertex}, and the total number of
{\it edges} is $2^{n-1}n$. In Fig.\ref{fig:trefoil}, one can see an example of the Jones hypercube of resolutions from~\cite{BarN}.
We use italic to denote {\it vertices} and {\it edges} of the hypercube, to distinguish them with
vertices of the diagram.
For the Jones and the bipartite HOMFLY counting we need just a hypercube with a system of closed planar cycles
hanged over each {\it vertex} -- formula (\ref{abc}) is then a sum over {\it vertices}.
For categorification leading to Khovanov-like calculus,
one introduces a vector space at each {\it vertex} and associate nilpotent maps (differentials)  between them
with all {\it edges}.
This is a well known construction for the Jones-Khovanov polynomials, but what would it give for the bipartite HOMFLY polynomials remains to be studied.

The next issue is what is the "choice", mentioned in the previous paragraph.
In the HOMFLY case, we have a big variety of "choices" leading potentially to different constructions.
The standard one for the Jones and the Khovanov polynomials \cite{DM12}
implies that the "choice" is between two resolutions at the r.h.s. of Fig.\ref{fig:Kauff}.
But in general, we have many more options.

\begin{figure}[h!]
\center{\includegraphics[width=0.55\linewidth]{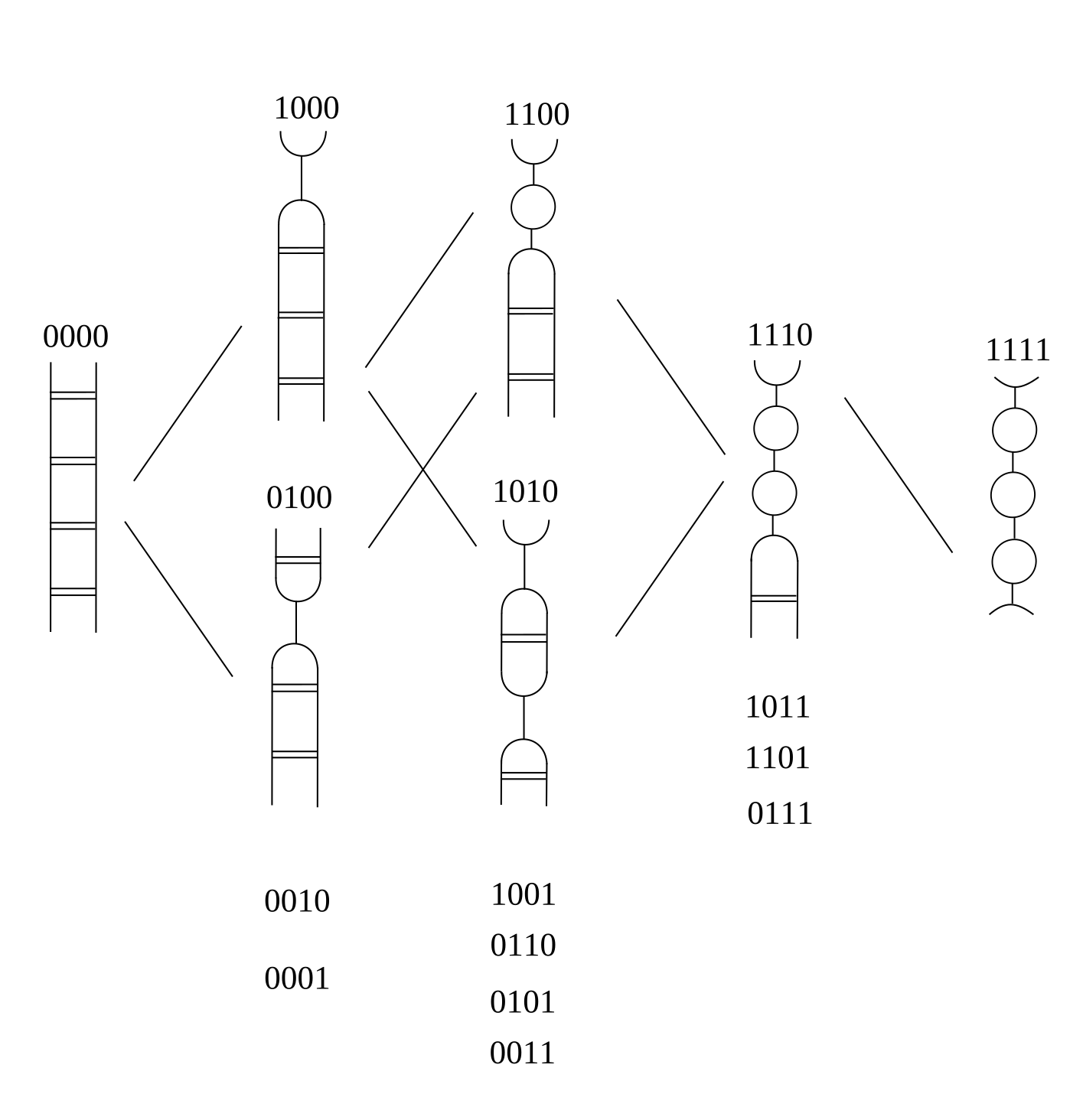}}
    \caption{\footnotesize Hypercube of bipartite diagrams for a chain with 4 AP locks; only some diagrams and edges are shown explicitly.}\label{fig:SameJones} 
\end{figure}

\begin{itemize}
    \item First, our diagram can be interpreted as the ordinary one or the lock diagram --
we build the bipartite HOMFLY polynomial in the latter case or the precursor Jones polynomial in the former case.
    \item Second, if this is a lock diagram, its resolutions depends on the orientation of locks. The locks can be vertical or opposite horizontal, as in Fig.\ref{fig:SameJones}), but after shrinking leading to the same single intersection in Fig.\ref{fig:Kauff}. Thus, there are many different hypercubes, associated with different orientations of locks inside a bipartite diagram but still reduced to the same precursor diagram.
    \item Third, since $\bar\PPhi$ is now essentially different from $\PPhi$, the change
    to the opposite vertices is no longer a part of the same hypercube --
    hypercube depends on the diagram stronger than usually.
    \item Fourth, one can choose the flips to switch not between the resolutions, but between orientations. In this case, we distinguish this hypercube by calling it a {\it hypercube of bipartite diagrams} (not to be confused with a hypercube of resolutions!).
    This looks like an exotic option, see Fig.\ref{fig:SameJones},
    and it is not immediately clear what kind if counting one can ascribe to this kind of lock hypercubes, --
    still it can be interesting.
\end{itemize}

All this diversity of {\it lock hypercubes} opens roads to various categorifications.
Probably, no one of them will reproduce the conventional Khovanov and Khovanov--Rozansky homologies invariant under all the Reidemeister moves.
Instead, there can be a whole family of Khovanov-like polynomials for bipartite diagrams,
which are not fully topological invariant, still can be interesting, see also the discussion in Section~\ref{sec:Kh-bip}.

\setcounter{equation}{0}
\section{Differential expansion}\label{sec:diffexp}

In the abelian case, when $N=1$ and $A=q$, all knot and link polynomials become trivial.
This means that the fundamental HOMFLY polynomials for knots in the topological framing
(which we actually deal with in this paper)
have the form
\be\label{DEdef}
H^{\cal K}_\Box(q,A) = D\cdot\Big(1 + F_{[1]}^{\cal K}(q,A)\{Aq\}\{A/q\}\Big)\,
\ee
where the only dependence on a knot is contained in a polynomial $F_{[1]}^{\cal K}(q,A)$ called the {\it cyclotomic function}. This {\it differential expansion} \cite{DE} (also known as the cyclotomic expansion \cite{cyclotomic,garoufalidis2011asymptotics,arxiv.1512.07906,Kameyama_2020,berest2021cyclotomic,beliakova2021cyclotomic,chen2021cyclotomic})
has a less trivial generalization to higher representations, at least symmetric \cite{DEgen,Kononov_2016,Morozov_2018,Morozov_2019,morozov2020kntz,BM1,morozov2022differential},
which we do not touch in this paper, but will appear in our forthcoming paper.
As to the fundamental representation, in this case the differential expansion is a direct consequence of (\ref{H-J-gen}) --
what we need is to substitute $D=D_1=1$, and note that the number of terms with powers $b$ and $c$
is defined just by the number of black and white vertices defined in Figs.\ref{fig:pladeco},\ref{fig:oppvert}\,:
\be
H^{\rm bipartite}_\Box \ \stackrel{D=1}{=} \  A^{-2(v\bullet - v_\circ)}\cdot \sum^{2^v}  \PPhi^b \bar\PPhi^c \ \stackrel{A=q}{=} \
q^{-2(v\bullet - v_\circ)}\cdot (1+\PPhi)^{v_\bullet}(1+\bar\PPhi)^{v_\circ} = 1\,.
\label{bc}
\ee
For higher representations the interference between the two structures -- planar and differential --
seem an interesting subject for further investigations.

The planar calculus for the bipartite HOMFLY polynomials also allows to further investigate the property of factorization of the cyclotomic function. It was first discovered for the family of double-braid knots, see~\cite{DEgen}. Namely, it turned out that
\begin{equation}\label{F-DB}
    F_{[1]}^{DB(2m,2n)}=-\frac{\left(1-A^{-2m}\right)\left(1-A^{-2n}\right)}{\{A\}^2}
\end{equation}
what is also obtained in Section~\ref{sec:DB}. Here the dependences on two parameters split. It is an interesting topological phenomena whose nature is still to be understood. Moreover, this factorization property helped to calculate some previously unknown Racah matrices~\cite{DEgen}. We guess that the same property can hold if vertical and horizontal braids are connected. If these braids additionally are antiparallel and have even numbers of intersections, then the planar technique is applicable, and such knots/links can be easily calculated. Exactly this happens with our generalization of Kanenobu knots, and for them the cyclotomic function hopefully factorizes~\eqref{factor-Kan-cyc}, see Section~\ref{sec:Kan-factor} for details.


\setcounter{equation}{0}
\section{Simple examples}\label{sec:examples}

In this section we consider several simple examples of the planar technique for calculation of the HOMFLY polynomials for lock diagrams.

\subsection{Necklace\label{sec:necklace}}

\begin{figure}[h!]

\begin{picture}(100,270)(-310,-240)

\put(-100,0){

\put(-130,-2){\mbox{$\tau_|:=  $}}

\put(20,0){
\put(-105,15){\vector(1,-1){12}} \put(-87,3){\vector(1,1){12}}
\put(-93,-3){\vector(-1,-1){12}} \put(-75,-15){\vector(-1,1){12}}
\put(-90,0){\circle*{6}} \put(-90,-9){\line(0,1){18}}

\put(-60,-2){\mbox{$:=$}}
}

\put(0,0){\put(-17,20){\line(1,-1){17}}\put(-17,20){\vector(1,-1){14}}   \put(0,3){\vector(1,1){17}}
 \put(0,-3){\vector(-1,-1){17}}   \put(17,-20){\line(-1,1){17}} \put(17,-20){\vector(-1,1){14}}
 \put(0,-3){\line(0,1){6}}}

\put(-100,0){
\put(140,-2){\mbox{$=$}}

\put(100,65){
\put(70,-60){\vector(1,0){40}}
\put(110,-70){\vector(-1,0){40}}
\put(120,-67){\mbox{$+\ \ \ \phi$}}
\put(155,-50){\vector(0,-1){30}}
\put(165,-80){\vector(0,1){30}}
}
}

}


\put(-250,-210){

\put(0,110){
\qbezier(0,3)(-20,15)(0,27) \put(-4,6){\vector(1,-1){2}}
\qbezier(0,3)(20,15)(0,27) \put(8,20){\vector(-1,1){2}}
\put(0,-3){\line(0,1){6}}

\put(0,27){\line(0,1){6}}
\put(0,33){\vector(1,1){10}}
\put(-10,43){\vector(1,-1){10}}

\qbezier(0,-3)(-10,-10)(-10,-18) \put(-10,-18){\vector(0,-1){2}}
\qbezier(0,-3)(10,-10)(10,-18) \put(8,-11){\vector(-1,1){2}}

\put(0,30){
\put(30,-69){\mbox{$=$}}
\put(50,-67){\mbox{$ \widetilde \PPPhi_n$}}
\put(70,-60){\vector(1,0){40}}
\put(110,-70){\vector(-1,0){40}}
\put(120,-67){\mbox{$+\ \ \ \phi^n$}}
\put(155,-50){\vector(0,-1){30}}
\put(165,-80){\vector(0,1){30}}
}
}

\put(-6,84){\mbox{$\ldots$}}

\put(0,60){
\qbezier(0,3)(-10,7)(-10,15) \put(-4.5,5){\vector(1,-1){2}}
\qbezier(0,3)(10,7)(10,15) \put(10,14){\vector(0,1){2}}
\put(0,-3){\line(0,1){6}}
}

\put(0,30){
\qbezier(0,3)(-20,15)(0,27) \put(-4,6){\vector(1,-1){2}}
\qbezier(0,3)(20,15)(0,27) \put(8,20){\vector(-1,1){2}}
\put(0,-3){\line(0,1){6}}
}

\qbezier(0,3)(-20,15)(0,27) \put(-4,6){\vector(1,-1){2}}
\qbezier(0,3)(20,15)(0,27) \put(8,20){\vector(-1,1){2}}
\put(0,-3){\line(0,1){6}}

\put(0,-3){\vector(-1,-1){10}}
\put(10,-13){\vector(-1,1){10}}
}

\put( 0,-210){

\put(0,110){

\qbezier(0,3)(-20,15)(0,27) \put(-4,6){\vector(1,-1){2}}
\qbezier(0,3)(20,15)(0,27) \put(8,20){\vector(-1,1){2}}
\put(0,-3){\line(0,1){6}}

\put(0,27){\line(0,1){6}}
\qbezier(0,33)(-25,50)(0,50)
\qbezier(0,33)(25,50)(0,50)
\put(-9,40){\vector(1,-1){2}}

\qbezier(0,-3)(-10,-10)(-10,-18) \put(-10,-18){\vector(0,-1){2}}
\qbezier(0,-3)(10,-10)(10,-18) \put(8,-11){\vector(-1,1){2}}

}

\put(-6,84){\mbox{$\ldots$}}

\put(0,60){
\qbezier(0,3)(-10,7)(-10,15) \put(-4.5,5){\vector(1,-1){2}}
\qbezier(0,3)(10,7)(10,15) \put(10,14){\vector(0,1){2}}
\put(0,-3){\line(0,1){6}}
}

\put(0,30){
\qbezier(0,3)(-20,15)(0,27) \put(-4,6){\vector(1,-1){2}}
\qbezier(0,3)(20,15)(0,27) \put(8,20){\vector(-1,1){2}}
\put(0,-3){\line(0,1){6}}
}

\qbezier(0,3)(-20,15)(0,27) \put(-4,6){\vector(1,-1){2}}
\qbezier(0,3)(20,15)(0,27) \put(8,20){\vector(-1,1){2}}
\put(0,-3){\line(0,1){6}}

\qbezier(0,-3)(-25,-20)(0,-20)
\qbezier(0,-3)(25,-20)(0,-20)
\put(9,-10){\vector(-1,1){2}}
}

\put(100,-210){

\put(0,110){
\qbezier(0,3)(-20,15)(0,27) \put(-4,6){\vector(1,-1){2}}
\qbezier(0,3)(20,15)(0,27) \put(8,20){\vector(-1,1){2}}
\put(0,-3){\line(0,1){6}}

\put(0,27){\line(0,1){6}}
\put(0,33){\vector(1,1){10}}
\put(-10,43){\vector(1,-1){10}}

\qbezier(0,-3)(-10,-10)(-10,-18) \put(-10,-18){\vector(0,-1){2}}
\qbezier(0,-3)(10,-10)(10,-18) \put(8,-11){\vector(-1,1){2}}
}

\put(-6,84){\mbox{$\ldots$}}

\put(0,60){
\qbezier(0,3)(-10,7)(-10,15) \put(-4.5,5){\vector(1,-1){2}}
\qbezier(0,3)(10,7)(10,15) \put(10,14){\vector(0,1){2}}
\put(0,-3){\line(0,1){6}}
}

\put(0,30){
\qbezier(0,3)(-20,15)(0,27) \put(-4,6){\vector(1,-1){2}}
\qbezier(0,3)(20,15)(0,27) \put(8,20){\vector(-1,1){2}}
\put(0,-3){\line(0,1){6}}
}

\qbezier(0,3)(-20,15)(0,27) \put(-4,6){\vector(1,-1){2}}
\qbezier(0,3)(20,15)(0,27) \put(8,20){\vector(-1,1){2}}
\put(0,-3){\line(0,1){6}}

\put(0,-3){\vector(-1,-1){10}}
\put(10,-13){\vector(-1,1){10}}

\qbezier(-10,153)(-30,160)(-30,70)
\qbezier(-10,-13)(-30,-20)(-30,70)
\qbezier(10,153)(30,160)(30,70)
\qbezier(10,-13)(30,-20)(30,70)
}



\end{picture}
\caption{\footnotesize
The vertical AP lock and the necklace operator obtained by its vertical iterations.
It can be closed in two different ways giving rise to open and closed necklace links.
If a necklace operator contains $n$ locks, then open chain is a link with $n+1$ components,
while the closed one has $n$ components.
A better way (not shown) to draw the closed necklace is by a circle.
}\label{fig:necklace}
\end{figure}
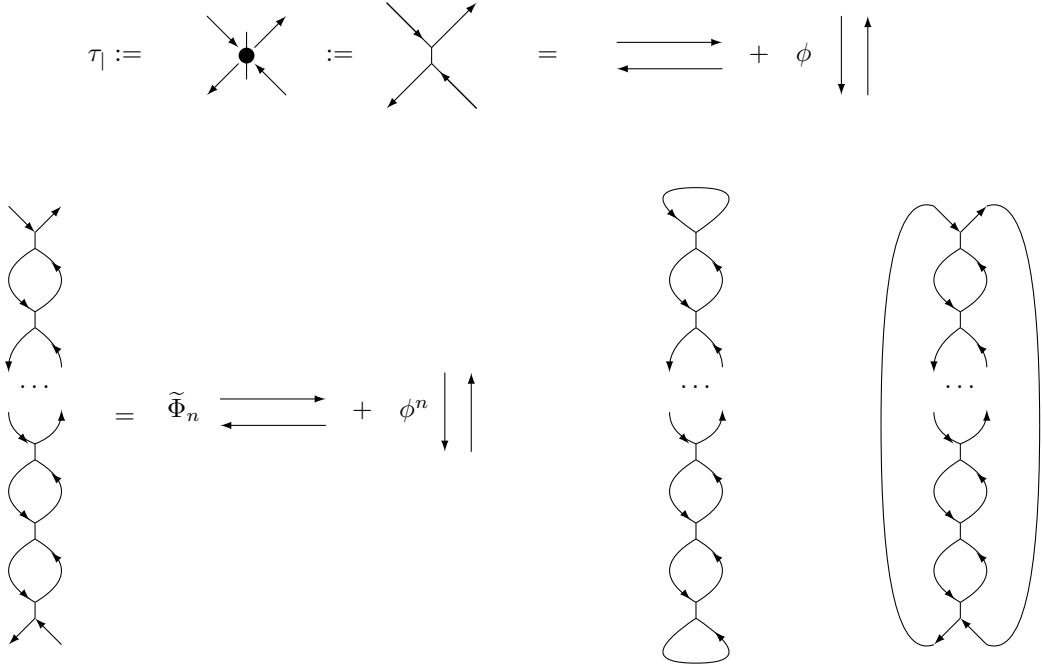

As we have discussed, the vertical lock $\tau_|=\ \horr \ +\, \PPhi \ \cdot \ || \ $ with $\PPhi=A\{q\}$ in the HOMFLY case.
Its vertical iteration is a necklace tangle with an explicit expression
\be
{\rm ON} = \ \widetilde \PPPhi_n\cdot \horr \ +\  \PPhi^n\cdot ||
\label{ON}
\ee 
where
\be
\widetilde \PPPhi_{n} = \frac{1}{D}\sum\limits_{k=1}^n C^k_n\cdot D^k\cdot \phi^{n-k}=
\frac{1}{D}\Big( (D+\phi)^n-\phi^n\Big)\,.
\label{tildePhin}
\ee
Pictorially the vertical AP lock, a necklace tangle and its open and closed closures are shown in Fig.\ref{fig:necklace}. Considering the necklace with $n+m$ lock elements, one can easily prove that
\begin{equation}
    \boxed{\widetilde{\Phi}_{n+m}= \widetilde{\Phi}_n \widetilde{\Phi}_m D + \phi^m\cdot \widetilde{\Phi}_n  + \phi^n\cdot \widetilde{\Phi}_m\,. }
\end{equation}
Gluing the remaining legs provides closed or open necklace
 -- the $n$ and $n+1$ component links respectively, and we immediately write down
\begin{equation}
\begin{aligned}
    P_\Box^{{\rm Neck}^{\rm cl}_n} &= \widetilde \PPPhi_n D + \PPhi^n D^2\,, \\
    P_\Box^{{\rm Neck}^{\rm op}_{n+1}} &= \widetilde \PPPhi_n D^2 + \PPhi^n D\,. 
\end{aligned}
\end{equation}
Recall that these polynomials provide both the bipartite HOMFLY and the precursor Jones polynomials by substitutions from Table~\ref{tab:subs}. In particular, for the HOMFLY polynomial we get:
\be\label{H-neck}
\begin{aligned}
 H_\Box^{{\rm Neck}^{\rm cl}_n} &= A^{-2n}\Big((D+\phi)^n+\phi^n(D^2-1)\Big)=D^{n}\left(1+\frac{\{Aq\}\{A/q\}}{\{A\}^{n-1}}\cdot F_{[1]}^{{\rm Neck}^{\rm cl}_n}\right)\,, \\
H_\Box^{{\rm Neck}^{\rm op}_{n+1}} &= A^{-2n}\cdot D\cdot (D+\phi)^n=D^{n+1}\left(1+\frac{\{Aq\}\{A/q\}}{\{A\}^{n}}\cdot F_{[1]}^{{\rm Neck}^{\rm op}_{n+1}}\right)
\end{aligned}
\ee
where $A^{-2n}$ is a framing factor, $\phi=A\{q\}$ and $F_{[1]}$ being some polynomial cyclotomic functions. Note that to provide a polynomiality of the cyclotomic function, one must factorize $\{A\}^{-1}$ to the power of number of link components minus one and also keep an overall factor being the quantum dimension to the power of link components~\cite{bai2018differential} (to be compared with the knot case~\eqref{DEdef}). 

In the particular case of $n+1=2$ we get the Hopf link:
\be
H^{\rm Hopf}_\Box = H_\Box^{{\rm Neck}^{\rm op}_{2}} = A^{-1}\frac{\{A\}}{\{q\}^2}\left(-A^{-2}+q^2+q^{-2}-1\right)=D^2\left(1-A^{-1}\cdot\frac{\{Aq\}\{A/q\}}{\{A\}}\right)\,.
\ee
All formulae for $\bar{\PPhi}$-necklaces are obtained from the above ones by the substitution $q\rightarrow q^{-1}$, $A\rightarrow A^{-1}$. So that, for example, inverse necklace tangle is $\ \widetilde \PPPhi_{-n}\cdot \ \horr = \ + \  \bar{\PPhi}^{n}\cdot ||$ with
\be
\widetilde \PPPhi_{-n} = \frac{1}{D}\sum\limits_{k=1}^n C^k_n\cdot D^k\cdot {\bar{\phi}}^{n-k}=\frac{1}{D}\left(\left[D+\bar{\phi}\,\right]^n-\bar{\phi}^{n}\right)\,.
\ee
All these expressions for the HOMFLY polynomial, beginning from (\ref{ON}) and (\ref{tildePhin}), 
are just the same   for their Jones precursors -- these are obtained by substitutions from Table~\ref{tab:subs}.
As the result, we get for the precursor diagrams:
\begin{equation}\label{J-neck}
{
\begin{aligned}
    J^{\overline{T[2,n]}}_\Box&=(-1)^{\frac{1}{2}(w+n)}\cdot q^{-\frac{3}{2}w-\frac{1}{2}n}\Big((D-q)^n+(-q)^n(D^2-1)\Big)
    =q^{-n}\left((-q^2)^{-n}+q^2+q^{-2}+1\right)\,, \\
    J^{\rm unknot}_\Box&=q^n\cdot D \cdot(D-q)^n=D\,.
\end{aligned}
}
\end{equation}

\subsection{Chain\label{sec:chain}}

\begin{figure}[h]

\begin{picture}(100,220)(-320,-190)

\put(-100,0){
\put(-130,-2){\mbox{$\tau_-:=  $}}

\put(20,0){
\put(-105,15){\vector(1,-1){12}} \put(-87,3){\vector(1,1){12}}
\put(-93,-3){\vector(-1,-1){12}} \put(-75,-15){\vector(-1,1){12}}
\put(-90,0){\circle*{6}} \put(-99,0){\line(1,0){18}}

\put(-60,-2){\mbox{$:=$}}
}

\put(0,0){
 \put(-20,17){\line(1,-1){17}}\put(-20,17){\vector(1,-1){14}}   \put(3,0){\vector(1,1){17}}
 \put(-3,0){\vector(-1,-1){17}}   \put(20,-17){\line(-1,1){17}} \put(20,-17){\vector(-1,1){14}}
 \put(-3,0){\line(1,0){6}}
 }

\put(0,0){
\put(30,-2){\mbox{$=$}}

\put(0,65){
\put(70,-50){\vector(0,-1){30}}
\put(80,-80){\vector(0,1){30}}

\put(90,-67){\mbox{$+\ \ \ \phi$}}

\put(125,-60){\vector(1,0){40}}
\put(165,-70){\vector(-1,0){40}}
}
}
}


\put(-210,-170){

\put(-40,38){\mbox{$:=$}}

\put(-70,40){
\put(-13,10){\vector(1,-1){10}}
\put(7,0){\vector(1,1){10}}
\put(17,-10){\vector(-1,1){10}}
\put(-3,0){\vector(-1,-1){10}}

\put(-3,0){\line(1,0){10}}
\put(0,5){\mbox{$n$}}
}

\put(0,100){
\qbezier(-3,0)(-13,-7)(-12,-10) \put(-12,-10){\vector(0,-1){2}}
\qbezier(7,0)(17,-7)(15,-10) \put(9,-2){\vector(-1,1){2}}
\put(-3,0){\line(1,0){10}}

\put(-13,10){\vector(1,-1){10}}
\put(7,0){\vector(1,1){10}}
}

\put(-4,80){\mbox{$\ldots$}}

\put(0,60){
\qbezier(-3,0)(-13,7)(-12,10) \put(-6,2){\vector(1,-1){2}}
\qbezier(7,0)(17,7)(15,10) \put(15,9){\vector(0,1){2}}
\put(-3,0){\line(1,0){10}}
}

\put(0,40){
\qbezier(-3,0)(-20,10)(-3,20) \put(-6,2){\vector(1,-1){2}}
\qbezier(7,0)(20,10)(7,20) \put(9,18){\vector(-1,1){2}}
\put(-3,0){\line(1,0){10}}
}

\put(0,20){
\qbezier(-3,0)(-20,10)(-3,20) \put(-6,2){\vector(1,-1){2}}
\qbezier(7,0)(20,10)(7,20) \put(9,18){\vector(-1,1){2}}
\put(-3,0){\line(1,0){10}}
}

\qbezier(-3,0)(-20,10)(-3,20) \put(-6,2){\vector(1,-1){2}}
\qbezier(7,0)(20,10)(7,20) \put(9,18){\vector(-1,1){2}}
\put(-3,0){\line(1,0){10}}

\put(-3,0){\vector(-1,-1){10}}
\put(17,-10){\vector(-1,1){10}}

\put(-20,105){
\put(50,-67){\mbox{$=$}}

\put(70,-50){\vector(0,-1){30}}
\put(80,-80){\vector(0,1){30}}

\put(90,-67){\mbox{$+\ \ \ \PPPhi_n$}}

\put(125,-60){\vector(1,0){40}}
\put(165,-70){\vector(-1,0){40}}
}

}


\put( 0,-170){

\put(0,100){
\qbezier(-3,0)(-13,-7)(-12,-10) \put(-12,-10){\vector(0,-1){2}}
\qbezier(7,0)(17,-7)(15,-10) \put(9,-2){\vector(-1,1){2}}
\put(-3,0){\line(1,0){10}}
 \put(-6,2){\vector(1,-1){2}}

\qbezier(-3,0)(-27,18)(2,18)
\qbezier(7,0)(31,18)(2,18)
 \put(-6,2){\vector(1,-1){2}}
}

\put(-4,80){\mbox{$\ldots$}}

\put(0,60){
\qbezier(-3,0)(-13,7)(-12,10) \put(-6,2){\vector(1,-1){2}}
\qbezier(7,0)(17,7)(15,10) \put(15,9){\vector(0,1){2}}
\put(-3,0){\line(1,0){10}}
}

\put(0,40){
\qbezier(-3,0)(-20,10)(-3,20) \put(-6,2){\vector(1,-1){2}}
\qbezier(7,0)(20,10)(7,20) \put(9,18){\vector(-1,1){2}}
\put(-3,0){\line(1,0){10}}
}

\put(0,20){
\qbezier(-3,0)(-20,10)(-3,20) \put(-6,2){\vector(1,-1){2}}
\qbezier(7,0)(20,10)(7,20) \put(9,18){\vector(-1,1){2}}
\put(-3,0){\line(1,0){10}}
}

\qbezier(-3,0)(-20,10)(-3,20) \put(-6,2){\vector(1,-1){2}}
\qbezier(7,0)(20,10)(7,20) \put(9,18){\vector(-1,1){2}}
\put(-3,0){\line(1,0){10}}

\qbezier(-3,0)(-27,-18)(2,-18)
\qbezier(7,0)(31,-18)(2,-18)
\put(9,-2){\vector(-1,1){2}}
}


\put(100,-170){

\put(0,100){
\qbezier(-3,0)(-13,-7)(-12,-10) \put(-12,-10){\vector(0,-1){2}}
\qbezier(7,0)(17,-7)(15,-10) \put(9,-2){\vector(-1,1){2}}
\put(-3,0){\line(1,0){10}}

\put(-13,10){\vector(1,-1){10}}
\put(7,0){\vector(1,1){10}}
}

\put(-4,80){\mbox{$\ldots$}}

\put(0,60){
\qbezier(-3,0)(-13,7)(-12,10) \put(-6,2){\vector(1,-1){2}}
\qbezier(7,0)(17,7)(15,10) \put(15,9){\vector(0,1){2}}
\put(-3,0){\line(1,0){10}}
}

\put(0,40){
\qbezier(-3,0)(-20,10)(-3,20) \put(-6,2){\vector(1,-1){2}}
\qbezier(7,0)(20,10)(7,20) \put(9,18){\vector(-1,1){2}}
\put(-3,0){\line(1,0){10}}
}

\put(0,20){
\qbezier(-3,0)(-20,10)(-3,20) \put(-6,2){\vector(1,-1){2}}
\qbezier(7,0)(20,10)(7,20) \put(9,18){\vector(-1,1){2}}
\put(-3,0){\line(1,0){10}}
}

\qbezier(-3,0)(-20,10)(-3,20) \put(-6,2){\vector(1,-1){2}}
\qbezier(7,0)(20,10)(7,20) \put(9,18){\vector(-1,1){2}}
\put(-3,0){\line(1,0){10}}

\put(-3,0){\vector(-1,-1){10}}
\put(17,-10){\vector(-1,1){10}}

\qbezier(-13,110)(-30,120)(-30,50)
\qbezier(-13,-10)(-30,-20)(-30,50)
\qbezier(17,110)(34,120)(34,50)
\qbezier(17,-10)(34,-20)(34,50)
}

\end{picture}
\caption{\footnotesize
The horizontal AP lock tangle and the vertical chain operator obtained by vertical iteration of this lock.
Since it will appear in many further examples, we also introduce a special abbreviated notation.
A chain tangle can be closed in two different ways giving rise to the unknot and to the 2-component 2-strand AP torus links
$APT[2,2n]$.
With the exception of Hopf at $n=1$, they are different from the more familiar parallel torus links,
which are more similar to torus 2-strand {\it knots} with odd number of intersections.
Two-strand torus {\it knots} are also bipartite (as being rational, see Theorem 2 in Section~\ref{sec:bipknots}) -- but depicted by more sophisticated diagrams.
Since the decomposition of the chain is just the same as for the single AP lock
(only $\PPhi$ is changed for $\PPPhi_n$),
it is especially simple to make further compositions beginning from the double braids.
}\label{fig:vertchain}
\end{figure}
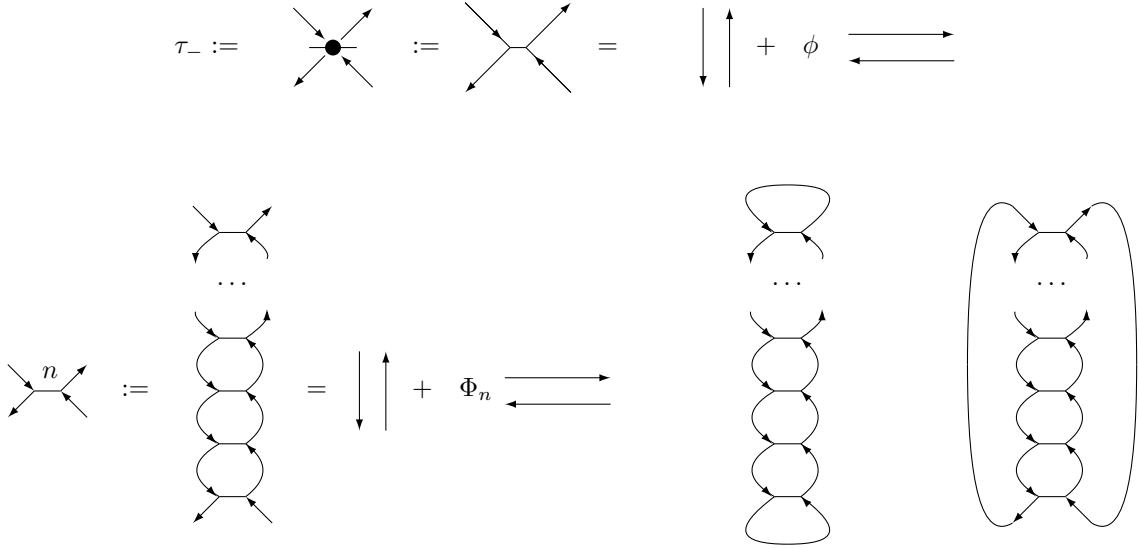

The horizontal lock $\tau_-=\ || \ + \PPhi \ \cdot  \horr \ $ with the same $\PPhi=A\{q\}$ in the HOMFLY case. Its vertical iteration which we call an antiparallel chain is  $\ || \ + \ \PPPhi_n\cdot \horr $, see Fig.\ref{fig:vertchain}. Considering the chain with $n+m$ locks, it is easy to derive the boxed rule below. Knowing the initial conditions, one gets the general formula for $\Phi_n$:
\be
\PPPhi_1 = \PPhi, \ \ \  \PPPhi_0= 0, \ \ \ \PPPhi_{-1}= \bar{\PPhi}, \ \ \ \
\boxed{\PPPhi_{n+m} = \PPPhi_n+\PPPhi_m + \PPPhi_n\PPPhi_m D}
\ \ \Longrightarrow \ \
\PPPhi_n = \frac{(1+\PPhi D)^n-1}{D}\,.
\label{compolocks}
\ee

For the opposite $\bar \phi$-chain we get:
\begin{equation}
    \overline{\PPPhi}_n =\PPPhi_{-n}= \frac{(1+\bar\PPhi D)^n-1}{D}\,.
\end{equation}
Cancellation of two inverse vertices gives the following relation:
\be
\PPhi + \bar{\PPhi} +\PPhi\bar{\PPhi}\cdot D = 0\,. 
\ee
It can be straightforwardly checked by substitutions from Table~\ref{tab:subs}. Again, we have two different closures of a chain tangle. The open closure of a chain gives the unknot, while another closure gives the 2-component antiparallel link $APT[2,2n]$:
\begin{equation}\label{chain-pol}
\begin{aligned}
    P_\Box^{APT[2,2n]}&=D(D+\Phi_n)\,, \\
    P_\Box^{\,\rm unknot}&=D(1+D\Phi_n)\,.
\end{aligned}
\end{equation}
Multiplying by the framing factor from Table~\ref{tab:subs} and putting $\phi=A\{q\}$, one gets for the HOMFLY polynomial
\begin{equation}\label{H-chain}
\begin{aligned}
H_\Box^{APT[2,2n]}&=A^{-2n}\left(D^2-1+(1+\phi D)^n\right)=D^2\left(1+\frac{\{Aq\}\{A/q\}}{\{A\}}\cdot F_{[1]}^{APT[2,2n]}\right)\,, \\
    H_\Box^{\rm unknot}&=A^{-2n}\cdot D(1+\phi D)^n=D\,.
\end{aligned}
\end{equation}
The cyclotomic function gets a simple form:
\begin{equation}\label{F-APT}
   F_{[1]}^{APT[2,2n]}=\frac{A^{-2n}-1}{\{A\}}\,.
\end{equation}
Again, one can obtain the Jones polynomials for precursor diagrams from polynomials~\eqref{chain-pol} by the appropriate substitutions from Table~\ref{tab:subs}\,: 
\begin{equation}\label{J-chain}
\begin{aligned}
    J_\Box^{T[2,n]}&=(-1)^{\frac{1}{2}(w+n)}\cdot q^{-\frac{3}{2}w-\frac{1}{2}n}\left(\left[1-q D\right]^n+D^2-1\right)=q^{n}\left((-q^2)^{n}+q^2+q^{-2}+1\right)\,, \\
    J_\Box^{\rm unknot}&=(-q^2)^{-n}\cdot D \cdot\left[1-qD\right]^n=D\,.
\end{aligned}
\end{equation}
From the discussion in Section~\ref{sec:hypercube}, it follows that the $\phi$-necklace and the $\bar\phi$-chain with $n$ lock tangles reduce to the same precursor diagrams. This fact can be also seen by the correspondence between~\eqref{J-chain} and~\eqref{J-neck}.

\subsection{Composite chains and double braids including twist knots}\label{sec:DB}
\begin{figure}[h!]
\center{\includegraphics[width=0.4\linewidth]{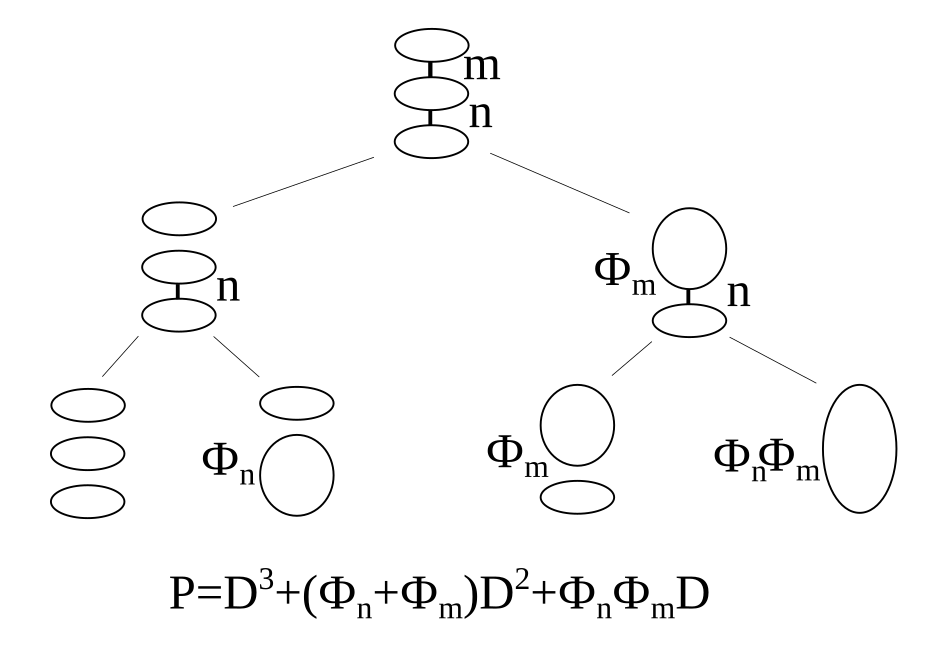}\hspace{2cm}
\includegraphics[width=0.4\linewidth]{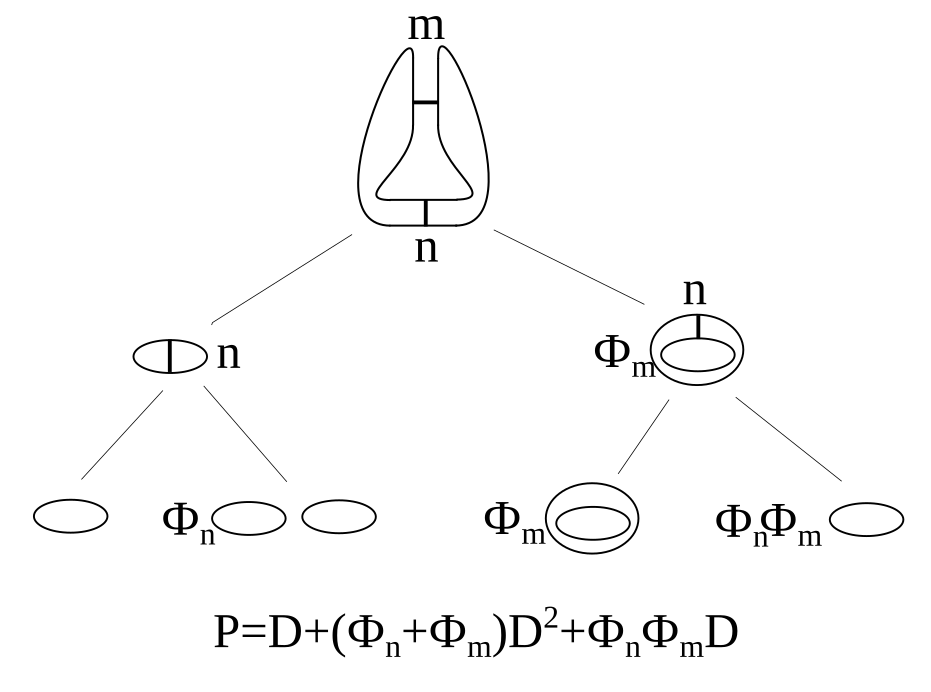}}
    \caption{\footnotesize The computation for the joint sum of two antiparallel braids (on the left) and of the double braids composed of the same two braids (on the right) by subsequent resolution of multiple edges. By $P$ we denote a polynomial in the vertical framing read directly from the diagram. It turns either to the HOMFLY polynomial of a bipartite knot, or to the Jones polynomial of a knot with shrinked locks by a proper fixation of $D$ and $\Phi_n$.}\label{fig:DBcompAP}
\end{figure}

Relation (\ref{compolocks}) allows us to treat a chain with $n$ parallel edges as a single multiple edge, see denotations in Fig.\ref{fig:vertchain}. Here we demonstrate how it works in the two simplest generalizations of a chain. They are the joint sum of two chains and the double braid composed of two orthogonal chains, Fig.\ref{fig:DBcompAP}. In each case, we draw a tree that shows consequent resolution of all crossings and the corresponding coefficients. The answer in the vertical framing is read just from the lower level of a tree if we add obtained factors on each leaf with the coefficient $D^c$ where $c$ is the number of cycles. The answers for the HOMFLY polynomials and the Jones polynomials in the topological framing are obtained by multiplication on the corresponding framing factors from Table~\ref{tab:subs}.

In the \textbf{joint sum} case, we read from Fig.\ref{fig:DBcompAP} on the left:
\be
P_\Box^{\scriptscriptstyle APT[2,2m]\#APT[2,2n]}=D^3+D^2(\Phi_n+\Phi_m)+D\Phi_m\Phi_n=
D(D+\Phi_n)(D+\Phi_m)\,,
\ee
i.e. factorization of the polynomials of the connected sum $P^{\scriptscriptstyle APT[2,2m]\#APT[2,2n]}$ takes place already with general $\Phi_n$ and $D$. In particular, it leads to the standard identites $H_\Box^{\scriptscriptstyle APT[2,2m]\#APT[2,2n]}=D^{-1}\cdot H_\Box^{APT[2,2m]}\cdot H_\Box^{APT[2,2n]}$ in the HOMFLY case, and $J_\Box^{\scriptscriptstyle T[2,m]\# T[2,n]}=D^{-1}\cdot J_\Box^{T[2,m]}\cdot J_\Box^{T[2,n]}$ in the Jones case.

\

In the \textbf{double braid} case, we read from Fig.\ref{fig:DBcompAP} on the right:
\be\label{mncyc}
P_\Box^{DB(2m,2n)}=D+(\Phi_n+\Phi_m)D^2+\Phi_n\Phi_m D\,.
\ee
Substituting (\ref{tau}), (\ref{compolocks}) and multiplying by the framing factor $=A^{-2n-2m}$, we obtain the HOMFLY polynomial in the topological framing:
\begin{equation}
\begin{aligned}
    H_\Box^{DB(2m,2n)}&=A^{-2n-2m}\cdot D\left((1+\phi D)^n+(1+\phi D)^m+D^{-2}\left[(1+\phi D)^n-1\right]\cdot\left[(1+\phi D)^m-1\right]-1\right)=\\
    &=D\left(1-\left(1-A^{-2m}\right)\left(1-A^{-2n}\right)\frac{\{Aq\}\{A/q\}}{\{A\}^2}\right)=D\Big(1 + \{Aq\}\{A/q\}\cdot F^{DB(2m,2n)}\Big)
\end{aligned}
\end{equation}
with the cyclotomic function
\begin{equation}\label{DB-F}
    F_{[1]}^{DB(2m,2n)}=\{Aq\}^{-1}\{A/q\}^{-1}\left(A^{-2n-2m}\left(1+\Phi_n\Phi_m+(\Phi_n+\Phi_m)D\right)-1\right)=-\frac{\left(1-A^{-2m}\right)\left(1-A^{-2n}\right)}{\{A\}^2}
\end{equation}
which factorizes into a product of the cyclotomic functions of antiparallel torus links~\eqref{F-APT}, as we have already discussed in Section~\ref{sec:diffexp}. In particular, we obtain the standard formulae  for twist knots with two orientations of the lock element:
\be
\begin{aligned}
H_\Box^{{\rm Tw}_{2n}}&=H_\Box^{DB(2,2n)}=D\left(A^{-2}+A^{-1}\frac{\{q\}^2}{\{A\}}+\frac{\{A/q\}\{Aq\}}{\{A\}}A^{-2n-1}\right)\,, \\
H_\Box^{\overline{{\rm Tw}}_{-2n}}&=H_\Box^{DB(-2,2n)}=D\left(A^2-A\frac{\{q\}^2}{\{A\}}-\frac{\{A/q\}\{Aq\}}{\{A\}}A^{-2n+1}\right)\,.
\end{aligned}
\ee
Similarly, one obtains the Jones polynomials from (\ref{mncyc}) by substituting (\ref{compolocks}), $\phi\rightarrow -q$ and by multiplying by the framing factor $=(-1)^{\frac{1}{2}(w+n+m)}\cdot q^{-\frac{3}{2}w-\frac{1}{2}(n+m)}$:
\begin{equation}
\begin{aligned}
    J_\Box^{DB(m,n)}&=(-1)^{\frac{1}{2}(w+n+m)}\cdot q^{-\frac{3}{2}w-\frac{1}{2}(n+m)}\cdot D\left((1-q D)^n+(1-q D)^m+D^{-2}\left[(1-q D)^n-1\right]\cdot\left[(1-q D)^m-1\right]-1\right)= \\
    &=(-1)^{\frac{1}{2}(w+n+m)}\cdot q^{-\frac{3}{2}w-\frac{1}{2}(n+m)}\cdot D \left((-q^2)^n+(-q^2)^m+D^{-2}\left[(-q^2)^n-1\right]\cdot \left[(-q^2)^m-1\right]-1\right)\,.
\end{aligned}
\end{equation}
In particular, for knots obtained from twist knots by the change of locks to single intersections:
\begin{equation}
\begin{aligned}
    J_\Box^{DB(1,n)}&=J_\Box^{T[2,n-1]}=-q^{n-3} \left((-q^{2})^n-q^4-q^2-1\right), \\
    J_\Box^{DB(-1,n)}&=J_\Box^{T[2,n+1]}=-q^{n-1} \left(\left(-q^2\right)^{n+2}-q^4-q^2-1\right).
\end{aligned}
\end{equation}
These are the same formulae as \eqref{J-chain} for the Jones polynomials corresponding to the chain HOMFLY polynomials. Indeed, the precursor diagrams are the same for a twist knot and for a chain with the corresponding number of crossings (see also Section~\ref{sec:hypercube}). We discuss this phenomenon in more details in the next section.

\subsection{Same Jones, different HOMFLY: general vertex in the hypercube}
In Sections \ref{sec:necklace} and \ref{sec:chain}, we have considered two examples of bipartite diagrams that correspond to the same precursor diagram. In fact, this precursor diagram (which is actually a 2-strand torus knot/link with $n$ crossings) generates a hypercube of $2^n$ bipartite diagrams obtained from each other by flipping between horizontal and vertical edges (see Section \ref{sec:hypercube}). Part of this hypercube of bipartite diagrams for $n=4$ is shown in Fig.\ref{fig:SameJones}.

\begin{figure}[h!]
\center{\includegraphics[width=0.19\linewidth]{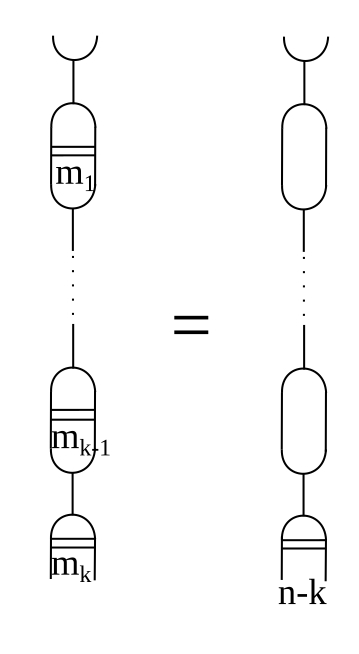}}
    \caption{\footnotesize 
    A general vertex in the hypercube of bipartite diagrams with $k$ vertical edges and $n-k=m_1+\dots+m_k$ horizontal inverse lock elements, so that the total number of locks equals $n$. It is easy to see that all these tangles are topologically equivalent for the fixed sum $m_1+\dots+m_k$.}
    \label{fig:Gen-vert-hyp}
\end{figure}

In this section, we compute a general vertex of the hypercube (see Fig.\ref{fig:Gen-vert-hyp}) giving the same Jones polynomial. Remind that double edge means taking inverse lock elements. In this tangle, all horizontal edges can be localized in a single circle due to the topology of the tangle. The resolution of the horizontal edge of multiplicity $n-k$ in the r.h.s. of the equality in Fig.\ref{fig:Gen-vert-hyp} gives the necklace and the necklace with one torn edge, so that this tangle gives 
\begin{equation}\label{Gen-vert-tan}
    \underbrace{\phi^k\cdot ||+\cfrac{(D+\phi)^k-\phi^k}{D}\,\cdot\horr}_{\rm necklace}\, + \; \overline{\Phi}_{n-k}\cdot \underbrace{(D+\phi)^k\cdot\horr}_{\text{torn necklace}}=\phi^k\cdot ||+\Theta_{n,k}\,\cdot\horr
\end{equation}
with
\begin{equation}
    \Theta_{n,k}=\cfrac{(D+\phi)^k(1+\bar{\phi} D)^{n-k}-\phi^k}{D}\,.
\end{equation}
In Section~\ref{sec:Kanenobu-gen}, we also need opposite to Fig.\ref{fig:Gen-vert-hyp} tangle. Its resolution is obtained by the change $q\rightarrow q^{-1}$ and $A\rightarrow A^{-1}$ so that it takes the following form:
\begin{equation}\label{theta-bar}
    \bar{\phi}^k\cdot ||+\bar{\Theta}_{n,k}\,\cdot\horr\quad \text{with}\quad \bar{\Theta}_{n,k}=\cfrac{(D+\bar{\phi})^k(1+\phi D)^{n-k}-\bar{\phi}^k}{D}\,.
\end{equation}
Considering a general vertex with $n+m$ AP locks and $k+l$ vertical edges, one can easily prove the following relation:
\begin{equation}\label{rel-gen-vert}
    \boxed{\Theta_{n+m,k+l}=\bar{\phi}^k \Theta_{m,l}+\bar{\phi}^l \Theta_{n,k} + \Theta_{m,l}\Theta_{n,k}D\,.}
\end{equation}

Two different closures of the tangle in Fig.\ref{fig:Gen-vert-hyp} give the following polynomials:
\begin{equation}
\begin{aligned}
    P^{(m_1,\dots,m_k)_{\rm op}}_\Box&=P_\Box^{{\rm Neck}^{\rm op}_{k+1}}=\phi^k D+\Theta_{n,k}D^2=D (D+\phi)^k(1+\bar{\phi} D)^{n-k}= D(D+\phi)^k\,,\\
    P^{(m_1,\dots,m_k)_{\rm cl}}_\Box&=\phi^k D^2+\Theta_{n,k}D\,.
\end{aligned}
\end{equation}
It is easily seen that the open closure of the tangle in Fig.\ref{fig:Gen-vert-hyp} is the same as the open closure of the necklace with $k$ vertical edges. The bipartite HOMFLY polynomials are given by the second formula in~\eqref{H-neck} and 
\begin{equation}\label{H-gen-2}
\begin{aligned}
    H^{(m_1,\dots,m_k)_{\rm cl}}_\Box&=A^{2n-4k}\cdot \Big((D+\phi)^k(1+\bar{\phi} D)^{n-k}+\phi^k(D^2-1)\Big)=D^k\left(1+\frac{\{Aq\}\{A/q\}}{\{A\}^{k-1}}\cdot F_{[1]}^{(m_1,\dots,m_k)_{\rm cl}}\right)  
\end{aligned}
\end{equation}
with some polynomial cyclotomic functions. For the precursor Jones polynomial we get by the substitution from Table~\ref{tab:subs}:
\begin{equation}
\begin{aligned}
    J_\Box^{\rm unknot}&=(-1)^{n-k}q^{2n-k}\cdot D (D-q)^k(1-q^{-1} D)^{n-k}=D\,, \\
    J_\Box^{\overline{T[2,n]}}&=(-1)^k q^{-n-k}\cdot \Big((D-q)^k(1-q^{-1} D)^{n-k}+(-q)^k(D^2-1)\Big)=q^{-n}\left((-q^2)^{-n}+q^2+q^{-2}+1\right)\,.
\end{aligned}
\end{equation}
As it has been discussed at the beginning of this section, these are the same answers as in~\eqref{J-neck}.

\subsection{Wheels}

One more natural family of bipartite diagrams is the family of wheel diagrams. 
Example is shown in Fig.\ref{fig:Examplefigs} on the right. All ``wheels'' should rotate in the same direction and be linked only pairwise (not Borromean). Below we compute several examples.

\begin{figure}[h!]
\center{\includegraphics[width=0.2\linewidth]{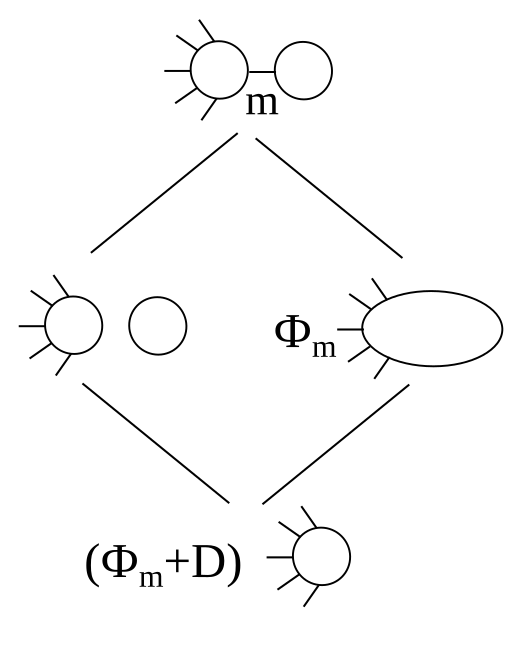}\hspace{2cm}
\includegraphics[width=0.19\linewidth]{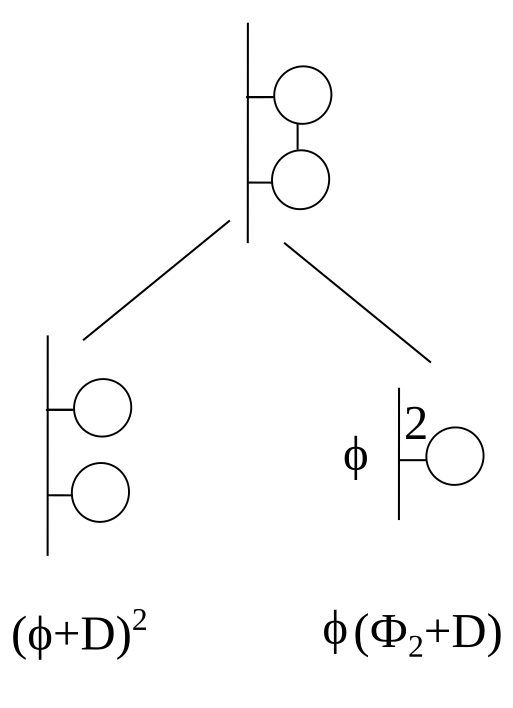}}
    \caption{\footnotesize Examples of simple wheel configurations.
    Left picture: A wheel attached to an arbitrary diagram through a multilock.
    Right picture: Two lock-connected wheels attached at the neighboring positions.
    In both cases, wheels can be fully eliminated and substituted by certain factors.  
    \label{fig:Wheels2}}

\end{figure}

\begin{figure}[h!]
\center{\includegraphics[width=0.3\linewidth]{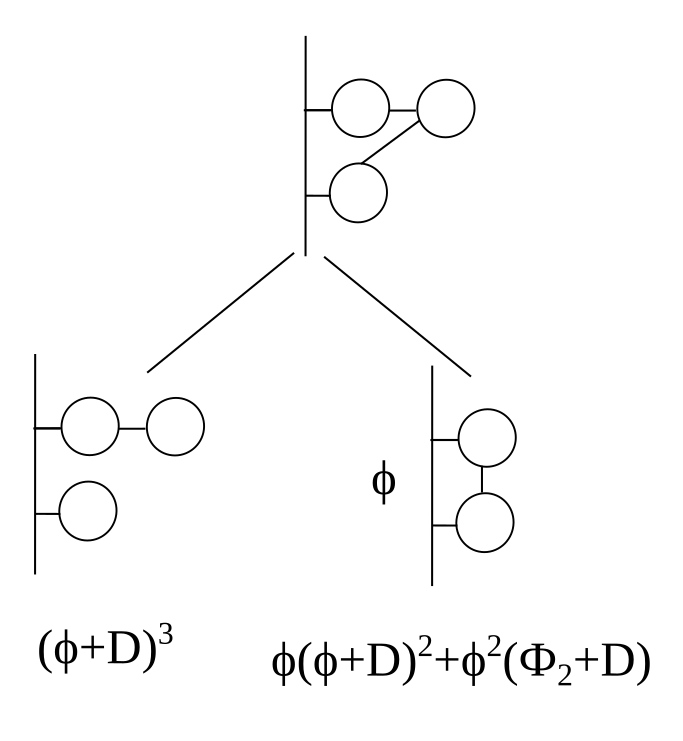}\hspace{2cm}
\includegraphics[width=0.39\linewidth]{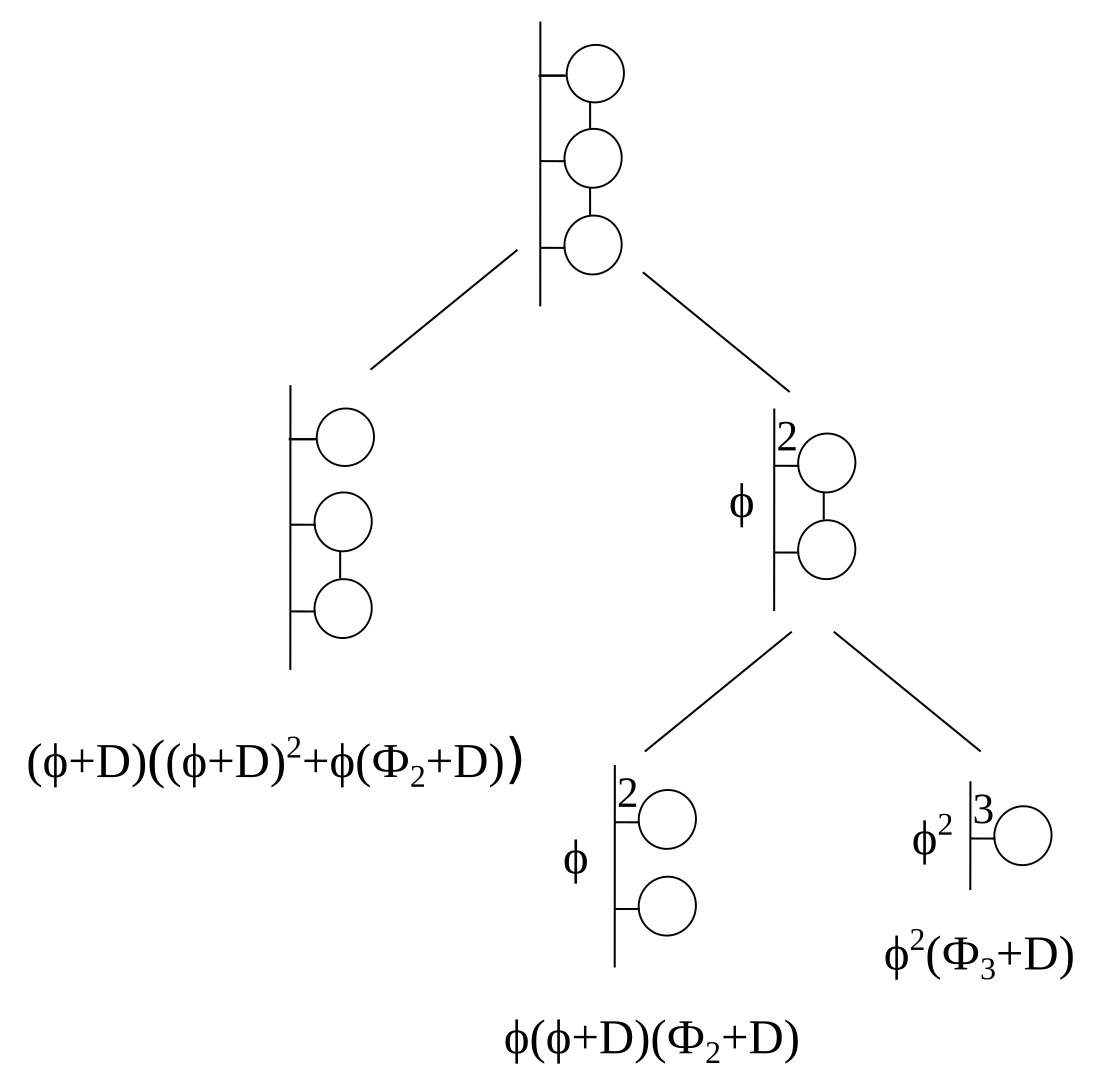}}
    \caption{\footnotesize Further examples of wheel elimination. 
    \label{fig:Wheels3}}
\end{figure}

The diagram technique is very similar to that shown in Fig.\ref{fig:DBcompAP} but now we reduce the original diagram to any already computed diagram, which is not necessarily a collection of cycles. We substitute the corresponding expressions at the bottom level of the trees in Figs.\ref{fig:Wheels2},\ref{fig:Wheels3}
so that it remains to take the sum over all leaves to obtain an answer. As we shall see, {\bf ``wheels'' can be always eliminated}, leaving behind just an overall factor, multiplying the remaining diagram.

The first observation is demonstrated in Fig.\ref{fig:Wheels2} on the left. Namely, if a ``wheel'' is connected with the rest diagram by a single (maybe multiple) edge, this ``wheel'' contributes to the answer by the factor of
\begin{equation}\label{1-Wheel}
    (\Phi_m+D) 
\end{equation}
with $m$ being the multiplicity of an edge. Then such a tree wheel diagram (which contains only ``wheels'' connected by a single edge) gives $\prod_{i}(\Phi_{m_i}+D)$ factor where the product is taken over all edges, and $m_i$ are their multiplicities.

The simplest non-tree diagram contains two ``wheels'' and is shown in Fig.\ref{fig:Wheels2} on the right. It is immediately computed with help of the rule in Fig.\ref{fig:Wheels2} on the left.
Summing over the leaves in the right part of Fig.\ref{fig:Wheels2}, we get the total factor
\be\label{2-Wheels}
(\phi+D)^2+\phi(\Phi_2+D)\,.
\ee
Two simplest generalizations are shown and computed in Fig.\ref{fig:Wheels3}. Both diagrams are reduced by subsequent resolutions to those in Fig.\ref{fig:Wheels2}. At the last stage, one takes the sum over the the leaves.
The answer for the left diagram in Fig.\ref{fig:Wheels3} is
\be
(\phi+D)^3+\phi(\phi+D)^2+\phi^2(\Phi_2+D).
\ee
The answer for the right diagram in Fig.\ref{fig:Wheels3} is
\be\label{3-Wheels}
(\phi+D)^3+2\phi(\phi+D)(\Phi_2+D)+\phi^2(\Phi_3+D)\,.
\ee

\begin{figure}[h!]
\center{\includegraphics[width=0.28\linewidth]{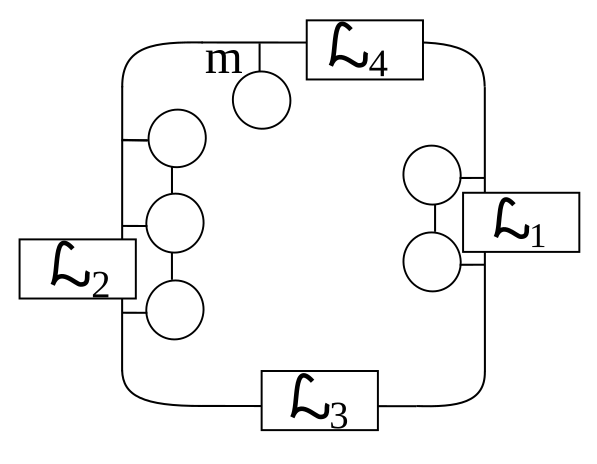}}
    \caption{\footnotesize Example of a link diagram with wheels. Wheels give only a factor in the HOMFLY polynomial.}
    \label{fig:Wheel-diag} 
\end{figure}

Now, it is easy to get sure that wheels always give a factor in the resulting HOMFLY polynomial. This is actually a consequence of the property shown in Fig.\ref{fig:Wheels2} and the HOMFLY factorization for composite knots. For example, for a link shown in Fig.\ref{fig:Wheel-diag} the polynomial is

{\scriptsize \begin{equation}
    P_\Box^{({\cal L}_1,\,{\cal L}_2,\,{\cal L}_3,\,{\cal L}_4,\,m)}=\underbrace{\left((\phi+D)^3+2\phi(\phi+D)(\Phi_2+D)+\phi^2(\Phi_3+D)\right)}_{\eqref{3-Wheels}}\cdot\underbrace{\left((\phi+D)^2+\phi(\Phi_2+D)\right)}_{\eqref{2-Wheels}}\cdot\underbrace{(\Phi_m+D)}_{\eqref{1-Wheel}}\cdot D^{-3}\cdot P_\Box^{{\cal L}_1}\cdot P_\Box^{{\cal L}_2}\cdot P_\Box^{{\cal L}_3}\cdot P_\Box^{{\cal L}_4}
\end{equation}}
with ${\cal L}_1,\,{\cal L}_2,\,{\cal L}_3,\,{\cal L}_4$ being arbitrary links. Thus, we see that wheels contribute as a factor both in the bipartite HOMFLY and in the precursor Jones cases.


\setcounter{equation}{0}
\section{Kanenobu(-like) links}\label{sec:Kanenobu}

\begin{figure}[h!]
\center{\includegraphics[width=0.5\linewidth]{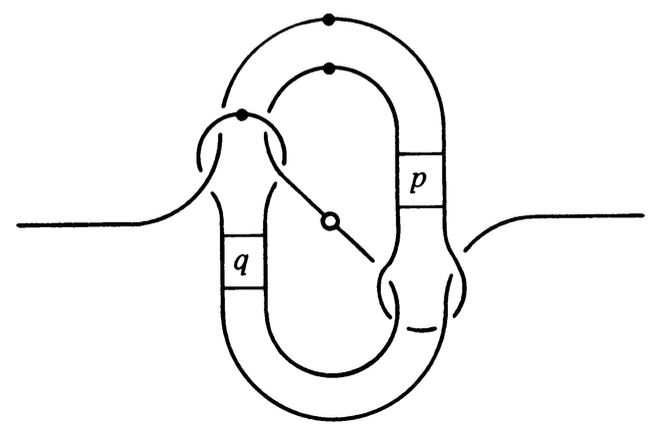}}
    \caption{\footnotesize Closure of this tangle gives a Kanenobu knot Kan$(p,q)$,~\cite{Kanenobu,Kanenobu-gen}.
   We mostly consider even $p$ and $q$ when the diagram is fully bipartite.
   However, this is not a restriction, and the actual arguments do not use this -- 
   sufficient will be planar decomposition of braids shown inside $p,\,q$ rectangles which turn out to be always antiparallel without dependence on parity of $p$ and $q$; 
   see Section~\ref{sec:p+q-dep} for further details. 
   Thus, Kanenobu $p+q$ property holds for arbitrary parities  of $p$ and $q$. 
    }
    \label{fig:Kanenobu}
\end{figure}

In this section we consider a rich family of Kanenobu-like links. One can see a spoiler of the results in Table~\ref{tab:summary}. The closure of a tangle shown in Fig.\ref{fig:Kanenobu} gives a Kanenobu knot Kan$(p,q)$~\cite{Kanenobu-gen}. For even $p,\,q$, this knot provides a famous example of a 2-parametric bipartite family \cite{Kanenobu} 
where the fundamental HOMFLY polynomial depends on the sum of two parameters $p+q$,
though it is not obvious from the picture.  

A Kanenobu knot Kan$(p,q)$ actually provides an exercise about a vertical chain (see Section~\ref{sec:chain}), and the mentioned above property
is a direct corollary of the chain planar decomposition shown in Fig.\ref{fig:vertchain}; for details, see Section~\ref{sec:p+q-dep}. Moreover, this approach allows immediate generalization to families of $3$ and $4$ parameters (see Section~\ref{sec:Kan-factor}) possessing an intriguing factorization property of cyclotomic functions discussed in Section~\ref{sec:diffexp}.

\subsection{Precursor diagram}

\begin{figure}[h!]
\center{\includegraphics[width=0.4\linewidth]{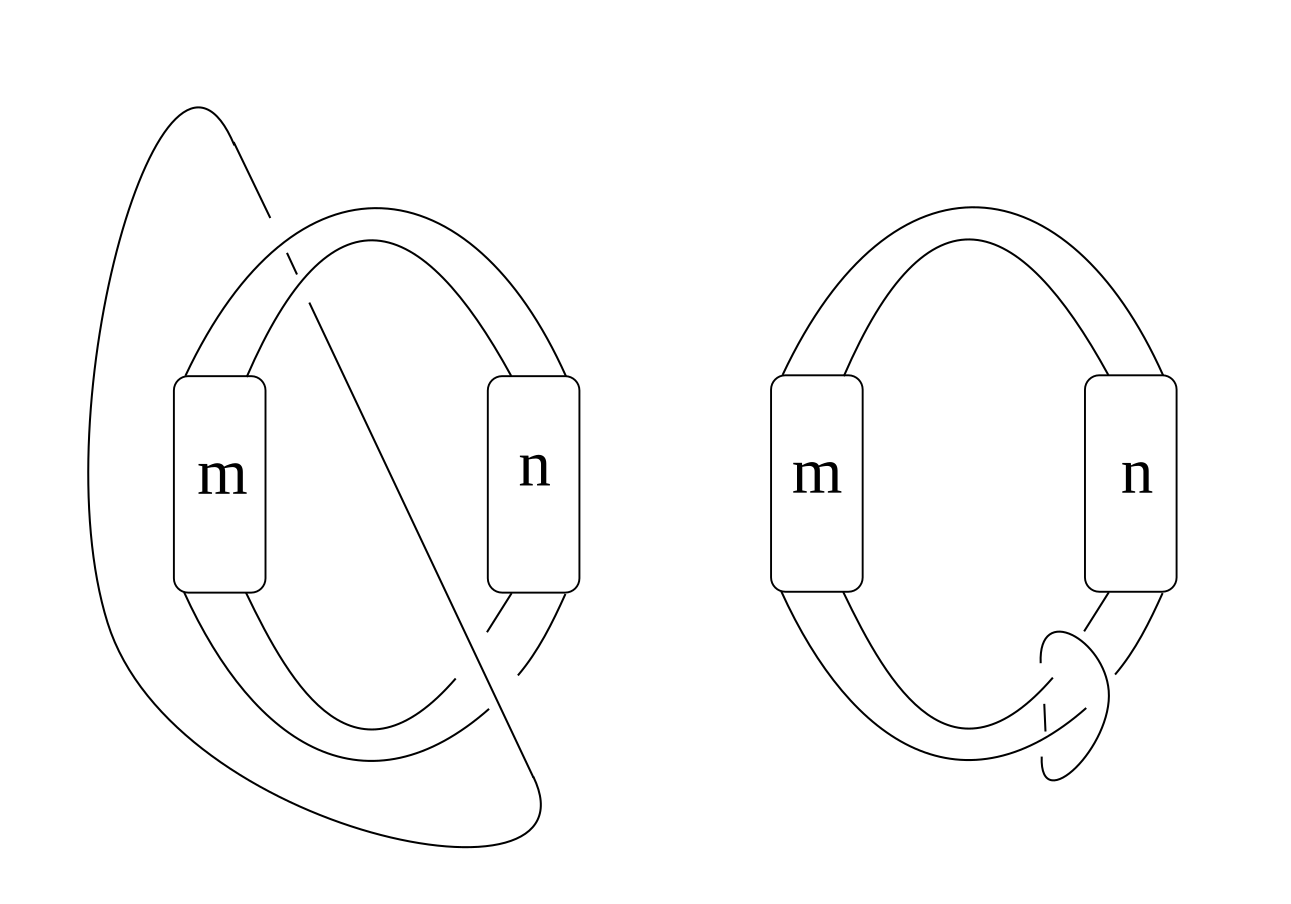}}
    \caption{\footnotesize The precursor of a Kanenobu diagram where each AP lock is substituted by an ordinary vertex.
    Planar decomposition of AP lock from Fig.\ref{fig:pladeco} then turns into the Kauffman rule in Fig.\ref{fig:Kauff}.
    Reidemeister moves now allow to carry the long line through the $m$ box what converts the picture into a linking
    of the unknot with the $(m+n)$-fold (2-strand torus) knot/link.
    Thus, at the precursor level the Kanenobu property just gets trivial.
    \label{fig:KanenobuJones}}
\end{figure}
First of all, a Kanenobu diagram with even numbers of twists Kan$(2m,2n)$ is fully bipartite. 
Then, it has the {\it precursor} diagram 
in Fig.\ref{fig:KanenobuJones} on the left, which is topologically equivalent to the diagram in Fig.\ref{fig:KanenobuJones} on the right. 
The Jones polynomial of the latter one clearly depends on $m+n$, but not on $m$ and $n$ separately, 
and the Jones polynomial of the original diagram is just the same.
Hence, for the precursor diagram, the $(m+n)$-property of the Jones polynomial is trivial.

Yet, there is no immediate analogue of this nice argument for the corresponding bipartite HOMFLY polynomial. 
The first reason is an ambiguity in the bipartite diagram corresponding to the given precursor diagram (see Section \ref{sec:hypercube}), and the fact that not any bipartite diagram lifted from the precursor one possess the $(m+n)$-property, see Section~\ref{Kan-gen-1} for a discussion. 
The second reason is that the III Reidemeister move which we use to pass from the left diagram in Fig.\ref{fig:KanenobuJones} to the right one
becomes non-trivial for {\it lock} diagrams (see Section \ref{sec:RIII}), and we cannot use it without additional analysis. 
The question of extension of the Jones argument to the original bipartite diagram via lock Reidemeister moves should be further explored.

\subsection{Why HOMFLY polynomial for Kan$(p,q)$ depends on $p+q$?}\label{sec:p+q-dep}

\begin{figure}[h!]
\center{\includegraphics[width=0.8\linewidth]{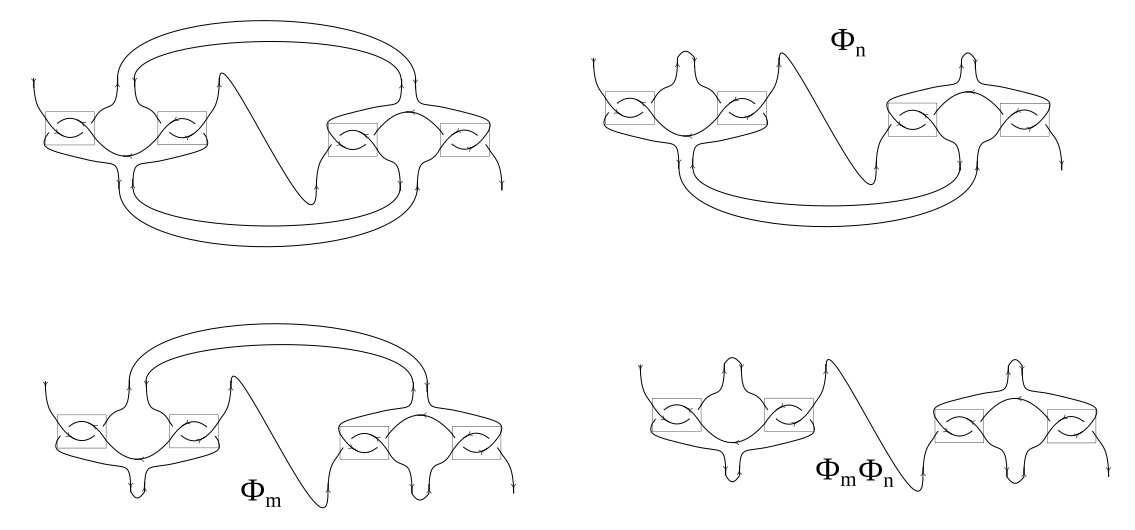}}
    \caption{\footnotesize Four resolutions of the chain parts in the Kanenobu tangle Kan$(2m,2n)$ from Fig.\ref{fig:Kanenobu}. Moreover, substitution of a general vertex of the bipartite hypercube from Fig.\ref{fig:Gen-vert-hyp} (or its mirror) instead of chain parts still leaves these four resolutions the same. Note that resolutions of the chain parts with $m$ and $n$ AP lock elements leave the resolved diagrams independent on the parameters $n$ and $m$. The $(m,n)$-dependences factorize to the multipliers $\Phi_n$, $\Phi_m$ of the corresponding diagrams.
    }
    \label{fig:Kanenobu-resolved-1}
\end{figure}

Now, we proceed to the HOMFLY case. Let us resolve chain parts in a Kanenobu knot Kan$(p,q)$. The resolution results to four contributions, see Fig.\ref{fig:Kanenobu-resolved-1}. The first one is the Kanenobu knot with $\varepsilon_{1,2}=0,\,1$ intersections in the chain parts depending on the parity of the initial knot parameters. This contribution for Kan$(2m,2n)$ is shown in the upper left picture in Fig.\ref{fig:Kanenobu-resolved-1}. The remaining parts are universal -- they are two unknots (upper right and lower left pictures in Fig.\ref{fig:Kanenobu-resolved-1}) with $\Phi_n$ and $\Phi_m$ factors and three unknots (lower right picture in Fig.\ref{fig:Kanenobu-resolved-1}) with $\Phi_n \Phi_m$ factor. In total, we get for the HOMFLY polynomial:
\begin{equation}\label{KanHOMFLY}
\begin{aligned}
H_\Box^{{\rm Kan}(2m+\varepsilon_1,2n+\varepsilon_2)}&=A^{-2m-2n}\left( H_\Box^{{\rm Kan}(\varepsilon_1,\varepsilon_2)}+ D^2(\PPPhi_{n}+\PPPhi_{m}+ \PPPhi_{n}\PPPhi_{m}D)\right)\overset{\eqref{compolocks}}{=}A^{-2m-2n}\left(H_\Box^{{\rm Kan}(\varepsilon_1,\varepsilon_2)}+ D^2\Phi_{n+m}\right)\,. 
\end{aligned}
\end{equation}
This explains why the HOMFLY polynomial for a Kanenobu knot Kan$(p,q)$ depends only on the sum $p+q$ irrespective of  parities of $p$ and $q$, but inside families of fixed parities of these parameters. 
Note that only Kanenobu diagram with even number of intersections is fully bipartite. 
The answers for the smallest numbers of intersections are:

{\scriptsize 
\begin{equation}\label{Kan-small}
\begin{aligned}
    H_\Box^{{\rm Kan}(0,0)}&=D\big(1+(D-D^{-1})(\phi+\bar{\phi})\big)^2=D\cdot \frac{\left(A^4 q^2-A^2 q^4+A^2 q^2-A^2+q^2\right)^2}{A^4 q^4}=D^{-1}\cdot \left(H^{4_1}\right)^2\,, \\
    H_\Box^{{\rm Kan}(1,-1)}&=D\cdot \frac{A^4 q^{10}-A^4 q^8+2 A^4 q^6-A^4 q^4+A^4 q^2-A^2 q^{12}+A^2 q^{10}-3 A^2 q^8+3 A^2 q^6-3 A^2 q^4+A^2 q^2-A^2+q^{10}-q^8+2 q^6-q^4+q^2}{A^2 q^6}\,, \\
    H_\Box^{{\rm Kan}(1,0)}&=H_\Box^{{\rm Kan}(0,1)}=-D\cdot\frac{A^6 q^6-A^6 q^4+A^6 q^2-A^4 q^8+2 A^4 q^6-4 A^4 q^4+2 A^4 q^2-A^4-A^2 q^8+2 A^2 q^6-3 A^2 q^4+2 A^2 q^2-A^2+q^6-q^4+q^2}{A^4 q^4}\,.
\end{aligned}
\end{equation}}
\noindent
so that due to~\eqref{KanHOMFLY} there are three families of Kanenobu knots dependent only on the sum of parameters -- Kan$(2m,2n)$, Kan$(2m+1,2n)$ = Kan$(2m,2n+1)$ and Kan$(2m+1,2n+1)$.

\subsection{Another explanation of Kanenobu $(p+q)$-property}\label{sec:another-Kan}

\begin{figure}[h!]
\center{\includegraphics[width=0.25\linewidth]{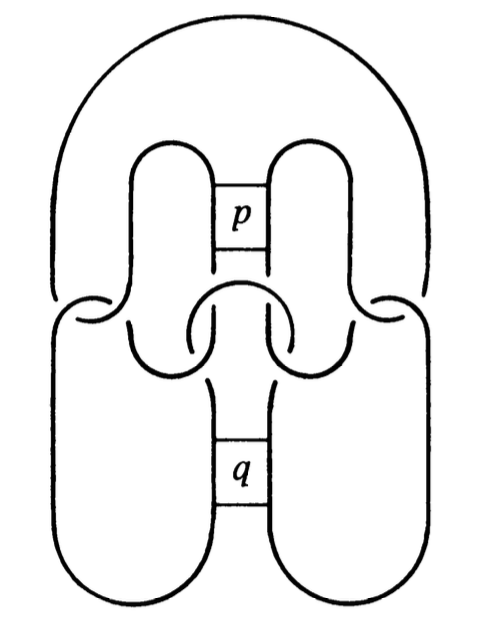}}
    \caption{\footnotesize 
    Another presentation of a Kanenobu knot from~\cite{Kanenobu} used in Section~\ref{sec:another-Kan}.}
    \label{fig:Kanenobu-2}
\end{figure}

\begin{figure}[h!]
$\begin{array}{c}\includegraphics[width=11.5cm]{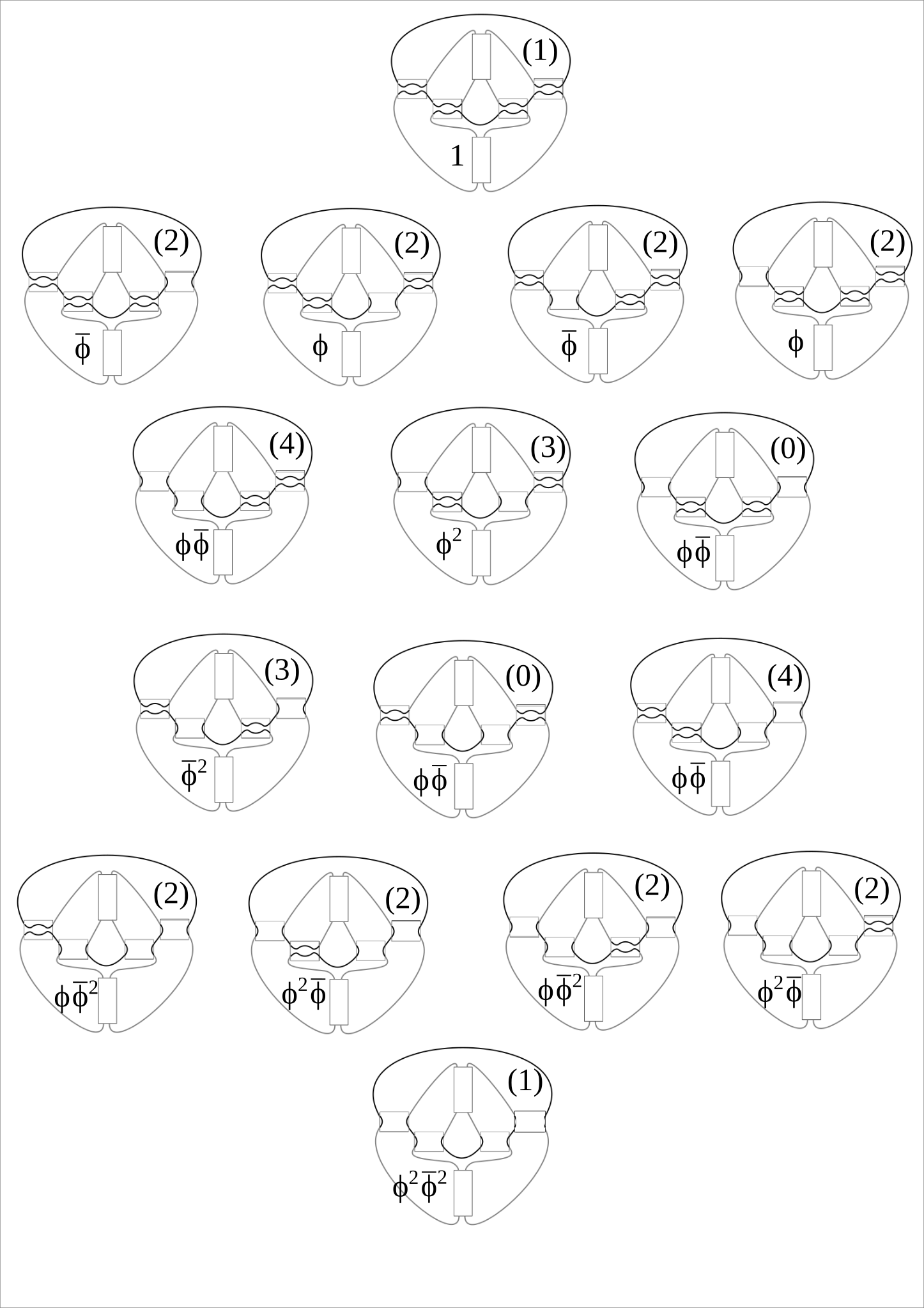}\\A\end{array}$
$\begin{array}{c}\includegraphics[width=5.5cm]{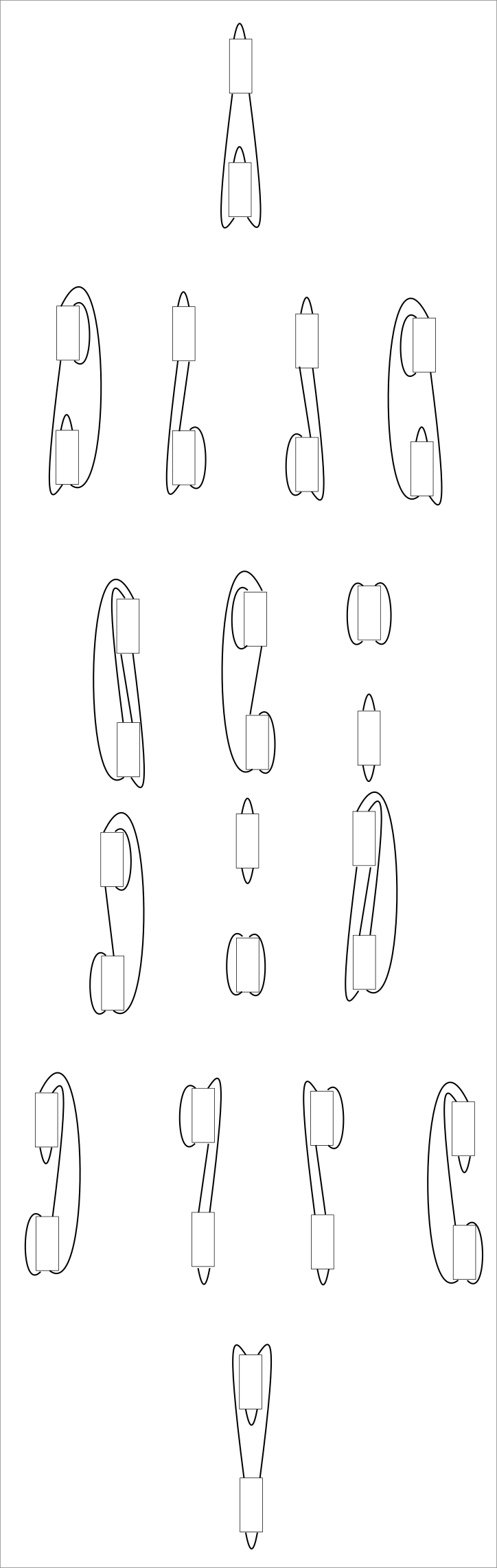}\\B\end{array}$
\caption{\label{Fig:Kan16FD} \footnotesize Resolutions of a Kanenobu knot with $p=2m$, $q=2n$ shown in Fig.\ref{fig:Kanenobu-2}. Each diagram contains $2m$ crossings in the upper box and $2n$ crossings in the lower box. (A) 16 Kanenobu diagrams with the resolved four 2-twists, with coefficients in front of the corresponding resolutions and labels of the diagram type; (B) simplified versions of the same 16 diagrams in the same order, which were used to specify the diagram types.}
\end{figure}

Instead of resolving chain parts in a Kanenobu knot, as it has been done in Section~\ref{sec:p+q-dep}, for $p=2m$, $q=2n$ one can deal with lock elements explicitly shown in Fig.\ref{fig:Kanenobu-2}. Resolution of these four AP locks gives 16 diagrams of 5 types shown in Fig.\ref{Fig:Kan16FD}(A). These diagrams can be topologically transformed to simpler diagrams in Fig.\ref{Fig:Kan16FD}(B). This simplification shows that the type-(1) diagram is the open closure of the chain with $n+m$ locks, and its polynomial is given by the second formula in~\eqref{chain-pol}. The type-(2) diagram gives the composite knots ${\rm Chain}^{\rm cl}_n\, \# \,{\rm Chain}^{\rm op}_m$ or ${\rm Chain}^{\rm cl}_m\, \# \,{\rm Chain}^{\rm op}_n$, and the type-(0) diagram presents the disjoint sums of the same knots, also see Section~\ref{sec:chain}. The type-(3) diagram is the composite knot of two closed chains ${\rm Chain}^{\rm cl}_n\, \# \,{\rm Chain}^{\rm cl}_m$, the polynomial of a closed chain is given by the first formula in~\eqref{chain-pol}. What remains to calculate is the polynomial for the type-(4) diagram. In total, we get:
\be
\small
\begin{array}{c}
P^{(1)}_{n,m}=D(1+D\Phi_{m+n}),\quad P^{(2)}_{m,n}=\cfrac{P^{(0)}_{m,n}}{D}=D(1+D\Phi_n)(D+\Phi_m),
\quad P^{(3)}_{m,n}=P^{(4)}_{m,n}+D(D^2-1)=D(D+\Phi_m)(D+\Phi_n)
\end{array}
\label{PiFD}\ee
where $(i)$-superscript indicates the correspondence to the type-$(i)$ diagram from Fig.\ref{Fig:Kan16FD}. Then we read the answer for the Kanenobu knot in the \textit{vertical} framing from Fig.\ref{Fig:Kan16FD}(A):
\be
P_\Box^{\,{\rm Kan}(2m,2n)} =(1+\phi^2\bar{\phi}^2)P_{m,n}^{(1)} +
\big(\phi+\bar{\phi}+(\phi+\bar{\phi}+D)\phi\bar{\phi}\big)\big(P^{(2)}_{m,n}+P^{(2)}_{n,m}\big)+
\big(\phi^2+\bar{\phi}^2\big)P^{(3)}_{m,n}+2\phi\bar{\phi}\cdot P^{(4)}_{m,n}\,.
\ee
One can now use that $P^{(3)}_{m,n}-P^{(4)}_{m,n}=D(D^2-1)$, substitute (\ref{PiFD}) and simplify this answer by algebraic manipulations:
\be
P_\Box^{\,{\rm Kan}(2m,2n)}=(1+\phi^2\bar{\phi}^2)P_{m,n}^{(1)} +
\Big\{(\phi+\bar{\phi}+D\phi\bar{\phi})\big(1+\textstyle{\frac{1}{D}}(\phi+\bar{\phi})\big)-\textstyle{\frac{1}{D}}(\phi+\bar{\phi})^2\Big\}
\big(P^{(2)}_{m,n}+P^{(2)}_{n,m}\big)+\nn\\+
P^{(3)}_{m,n}\big(\phi+\bar{\phi})^2+2D(1-D^2)\phi\bar{\phi}\stackrel{(\ref{PiFD})}{=}\nn\\
=D(1+\phi^2\bar{\phi}^2)(1+D\Phi_{m+n})+D(\phi+\bar{\phi})^2
\underbrace{\big(D^2-2-\textstyle{\frac{1}{D}}(\Phi_m+\Phi_n)-\Phi_m\Phi_n\big)}_{(D-\frac{1}{D})^2-\frac{1}{D^2}\underline{(1+D\Phi_m)(1+D\Phi_n)}}
+2D(1-D^2)\phi\bar{\phi}+\\
+\,D\underline{\underline{(\phi+\bar{\phi}+D\phi\bar{\phi})}}
\big(\underline{\underline{\underline{1+\textstyle{\frac{1}{D}}(\phi+\bar{\phi})}}}\big)\big((1+D\Phi_n)(D+\Phi_m)+(1+D\Phi_m)(D+\Phi_n)\big)\,.
\nn\ee

We have two identities for the underlined terms valid for both the bipartite HOMFLY and the precursor Jones cases (see Section \ref{sec:chain}):
\be
(1+D\Phi_m)(1+D\Phi_n)&=&1+D\Phi_{m+n}\,,
\nn \\
\phi+\bar{\phi}+D\phi\bar{\phi}&=&0\,.
\label{Dphiphi}\ee
Modulo these identities:
{\small \be\label{Kan-P}
\begin{aligned}
&P_\Box^{\,{\rm Kan}(2m,2n)}=D\big(1+\underline{\underline{\phi^2\bar{\phi}^2-\textstyle{\frac{1}{D^2}}(\phi+\bar{\phi})^2}}\big)(1+D\Phi_{m+n})
+ D(D-\textstyle{\frac{1}{D}})^2(\phi+\bar{\phi})^2+2D(1-D^2)
\big(\underline{\underline{\phi\bar{\phi}+\frac{1}{D}(\phi+\bar{\phi})}}-\frac{1}{D}(\phi+\bar{\phi})\big)\\ 
&\Longrightarrow\quad  \boxed{ P_\Box^{\,{\rm Kan}(2m,2n)}=D^2\Phi_{m+n}+D\big(1+(D-D^{-1})(\phi+\bar{\phi})\big)^2} 
\end{aligned}
\ee}
what obviously depends on $m$ and $n$ through the sum $m+n$. Note that the restoration of the framing factor for the HOMFLY polynomial gives exactly the formula~\eqref{KanHOMFLY} for $\varepsilon_1=\varepsilon_2=0$ together with the first expression in~\eqref{Kan-small}. While the precursor substitution from Table~\ref{tab:subs} 
gives for the link shown in Fig.\ref{fig:KanenobuJones}:

\begin{equation}\label{Jones-Kan-precursor}
    J_\Box^{\,{\rm Kan}(m,n)}=q^{n+m}\left((-q^2)^{n+m}-1+(q^2+q^{-2})^2\right)\,.
\end{equation}
Amusingly, in the Jones-precursor case also
\be
1+\textstyle{\frac{1}{D}}(\phi+\bar{\phi})=0\,.
\ee
what is an algebraic justification
of the Reidemeister trick in Fig.\ref{fig:KanenobuJones}.

\subsection{Generalizing Kanenobu knots}\label{sec:Kanenobu-gen}

In this section, we consider different generalizations of Kanenobu knots in Figs.\ref{fig:Kanenobu},\ref{fig:Kanenobu-2}.

\subsubsection{General vertex of the hypercube instead of $(p,q)$-chain tangles}\label{Kan-gen-1}

\begin{figure}[h!]
\begin{picture}(100,350)(-20,-50)

\put(138,60){\mbox{$k$ locks}}
\put(147,117){\mbox{$q-2k$}}
\put(287,154){\mbox{$p-2l$}}
\put(264,240){\mbox{$\ldots$}}
\put(290,240){\mbox{$l$ locks}}

\qbezier(180,140)(179,149)(176,151)
\qbezier(200,140)(201,149)(204,151)
\put(180,135){\line(1,0){20}}
\put(180,105){\line(1,0){20}}
\put(200,95){\line(0,1){45}}
\put(180,95){\line(0,1){45}}

\qbezier(200,0)(200,-20)(230,-20)
\qbezier(230,-20)(260,-20)(260,0)
\qbezier(180,0)(180,-40)(230,-40)
\qbezier(230,-40)(280,-40)(280,0)

\put(280,0){\line(0,1){100}}
\put(260,0){\line(0,1){100}}
\put(200,180){\line(0,1){95}}
\put(180,180){\line(0,1){95}}

\put(-80,-180){
\qbezier(280,180)(275,200)(265,190)
\qbezier(260,190)(265,170)(275,180)
\put(0,30){\qbezier(280,180)(275,200)(265,190)
\qbezier(260,190)(265,170)(275,180)
}
\put(260,190){\line(0,1){20}}
\put(280,190){\line(0,1){20}}

\put(0,85){\qbezier(280,180)(275,200)(265,190)
\qbezier(260,190)(265,170)(275,180)
}
\put(260,255){\line(0,1){10}}
\put(280,255){\line(0,1){10}}
\put(260,220){\line(0,1){10}}
\put(280,220){\line(0,1){10}}

\put(264,240){\mbox{$\ldots$}}
}

\qbezier(170,152.5)(163,148.5)(141,150.5)
\qbezier(170,152.5)(180,157)(180,169)
\qbezier(172,157)(169,166)(174,170)
\qbezier(208,170)(190,185)(174,170)
\qbezier(210,157)(213,166)(208,170)
\qbezier(210,152.5)(200,157)(200,169)

\put(249,120){\line(-1,0.83){40}}
\qbezier(260,100)(260,110)(255,115)
\qbezier(256,100)(245,110)(256,124)
\qbezier(256,124)(259,126)(260,135)

\put(260,170){\line(1,0){20}}
\put(260,140){\line(1,0){20}}
\put(260,135){\line(0,1){45}}
\put(280,135){\line(0,1){45}}

\qbezier(280,180)(275,200)(265,190)
\qbezier(260,190)(265,170)(275,180)

\put(0,30){\qbezier(280,180)(275,200)(265,190)
\qbezier(260,190)(265,170)(275,180)
}
\put(260,190){\line(0,1){20}}
\put(280,190){\line(0,1){20}}

\put(0,85){\qbezier(280,180)(275,200)(265,190)
\qbezier(260,190)(265,170)(275,180)
}
\put(260,255){\line(0,1){10}}
\put(280,255){\line(0,1){10}}
\put(260,220){\line(0,1){10}}
\put(280,220){\line(0,1){10}}

\qbezier(260,275)(260,295)(230,295)
\qbezier(230,295)(200,295)(200,275)
\qbezier(280,275)(280,315)(230,315)
\qbezier(230,315)(180,315)(180,275)

\qbezier(280,135)(281,126)(284,124)
\qbezier(280,100)(280,110)(285,115)
\qbezier(284,100)(295,110)(284,124)
\qbezier(264,97)(270,95)(276,97)
\qbezier(291,120)(300,130)(320,135)
    
\end{picture}
\caption{\footnotesize $l$ and $k$ horizontal locks in 2-strand braids of $p$ and $q$ crossings correspondingly can be substituted to $l$ and $k$ vertical opposite locks. A precursor diagram remains the same (see Fig.\ref{fig:KanenobuJones}) and the $(p+q)$-property still holds. Note that in general such a generalization is a link.}
    \label{fig:Kan-gen}
\end{figure}
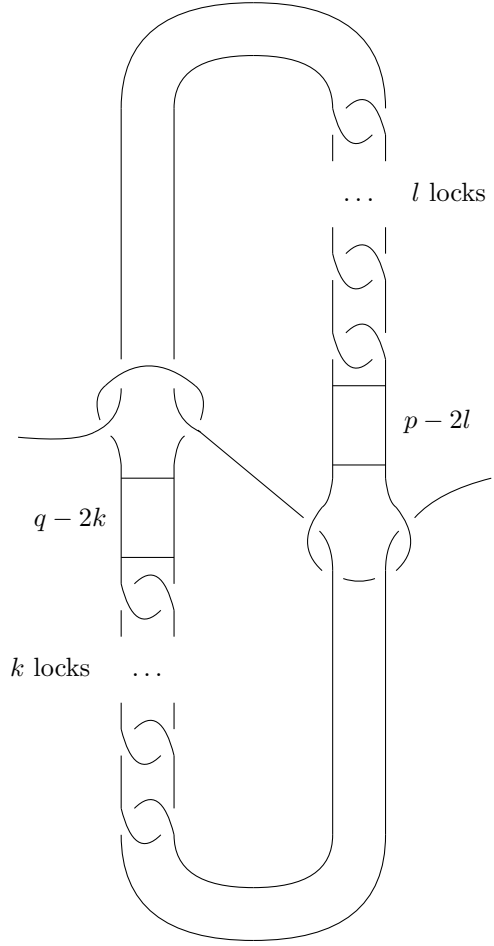

We can replace the chain parts with $m$ and $n$ AP locks of Kanenobu knot Kan$(2m+\varepsilon_1,2n+\varepsilon_2)$, $\varepsilon_{1,2}=0,1$, with the corresponding general vertices of the bipartite hypercube shown in Fig.\ref{fig:Gen-vert-hyp} (or their mirror). Namely, we replace the chain part with $m$ AP locks with a general vertex with $l$ vertical opposite edges and $m$ AP locks in total and the chain part with $n$ AP locks with a general vertex with $k$ opposite vertical edges and $n$ AP locks in total, see Fig.\ref{fig:Kan-gen}. Then, we get the same four resolutions as in Fig.\ref{fig:Kanenobu-resolved-1} but with factors coming from~\eqref{theta-bar}, and the HOMFLY polynomial is 

\begin{equation}\label{KanGenHOMFLY}
\begin{aligned}
H_\Box^{{\rm KanGen}(2m+\varepsilon_1,2n+\varepsilon_2,2k,2l)}&=A^{-2m-2n+4k+4l}\left( \bar{\phi}^{k+l} H_\Box^{{\rm Kan}(\varepsilon_1,\varepsilon_2)}+ \underline{D^2(\bar{\phi}^l\bar{\Theta}_{n,k}+\bar{\phi}^k\bar{\Theta}_{m,l}+ \bar{\Theta}_{n,k}\bar{\Theta}_{m,l}D)}\right)\overset{\eqref{rel-gen-vert}}{=}\\
&=A^{-2m-2n+4k+4l}\left(\bar{\phi}^{k+l} H_\Box^{{\rm Kan}(\varepsilon_1,\varepsilon_2)}+ D^2\bar{\Theta}_{n+m,k+l}\right) 
\end{aligned}
\end{equation}
where $\varepsilon_{1,2}=0,1$. Note that again, for fixed $k+l$, there are three families of Kanenobu knots dependent only on the sum of parameters -- KanGen$(2m,2n,2k,2l)$, KanGen$(2m+1,2n,2k,2l)$ = KanGen$(2m,2n+1,2k,2l)$ and KanGen$(2m+1,2n+1,2k,2l)$, and only the family KanGen$(2m,2n,2k,2l)$ is bipartite.

\begin{figure}
\subfigure[]{\includegraphics[width=0.42\linewidth]{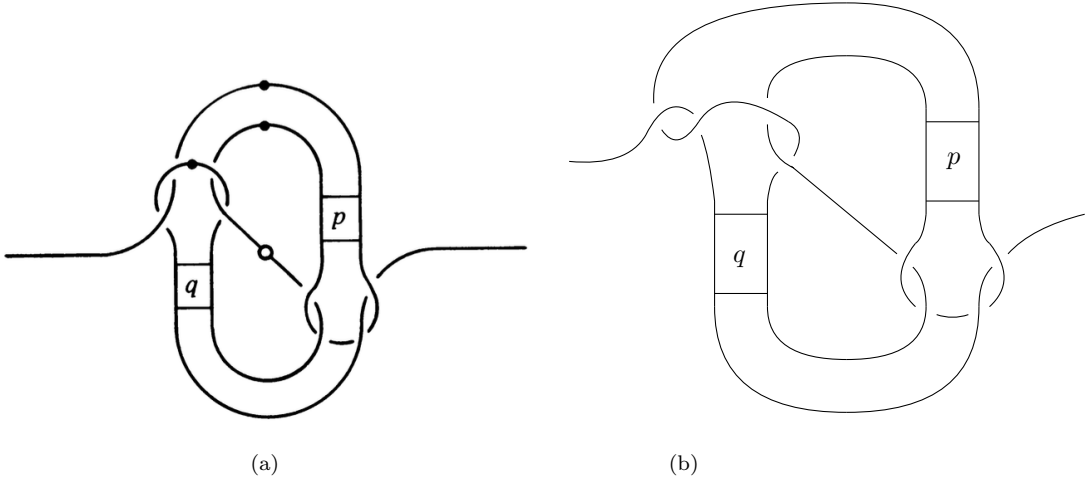}}
\subfigure[]{\begin{picture}(100,150)(120,50)

\put(187,117){\mbox{$q$}}
\put(268,154){\mbox{$p$}}

\qbezier(180,140)(178,157)(175,165)
\qbezier(200,140)(201,149)(204,151)
\put(180,135){\line(1,0){20}}
\put(180,105){\line(1,0){20}}
\put(200,100){\line(0,1){40}}
\put(180,100){\line(0,1){40}}
\qbezier(200,100)(200,80)(230,80)
\qbezier(230,80)(260,80)(260,100)
\qbezier(180,100)(180,60)(230,60)
\qbezier(230,60)(280,60)(280,100)

\qbezier(154,165)(147,153)(125,155)
\qbezier(172,173)(162,181)(154,165)
\qbezier(174,170)(166,158)(160,167)
\qbezier(208,170)(184,185)(174,170)
\qbezier(210,157)(215,166)(208,170)
\qbezier(210,152.5)(200,157)(200,169)
\put(249,120){\line(-1,0.83){40}}
\qbezier(260,100)(260,110)(255,115)
\qbezier(256,100)(245,110)(256,124)
\qbezier(256,124)(259,126)(260,135)

\put(260,170){\line(1,0){20}}
\put(260,140){\line(1,0){20}}
\put(260,135){\line(0,1){40}}
\put(280,135){\line(0,1){40}}
\qbezier(260,175)(260,195)(230,195)
\qbezier(230,195)(200,195)(200,179)
\qbezier(280,175)(280,215)(230,215)
\qbezier(230,215)(155,215)(157,177)

\qbezier(280,135)(281,126)(284,124)
\qbezier(280,100)(280,110)(285,115)
\qbezier(284,100)(295,110)(284,124)
\qbezier(264,97)(270,95)(276,97)
\qbezier(291,120)(300,130)(320,135)
    
\end{picture}}
\caption{\footnotesize (a) Kanenobu tangle from Fig.\ref{fig:Kanenobu} copied for convenience to compare with Fig.\ref{fig:Kan-def}(b). (b) The left horizontal opposite lock in Fig.\ref{fig:Kan-def}(a) is changed to a vertical lock. The chain parts in $p$- and $q$- boxes can be changed to a general vertex of bipartite hypercube in Fig.\ref{fig:Gen-vert-hyp}, as it is done in Fig.\ref{fig:Kan-gen}. In these cases, the precursor diagram stays the same but the $(p+q)$-property is spoiled. Emphasize that this family does not contained in one from Fig.\ref{fig:Kan-4}.}
    \label{fig:Kan-def}
\end{figure}

However, if one changes one of the four horizontal locks explicitly shown in Fig.\ref{fig:Kanenobu}, the $m+n$ symmetry breaks. For example, when the left horizontal opposite lock is changed to a vertical lock (see Fig.\ref{fig:Kan-def}), the framing factor still depends on $m+n$ while another $(m,\,n)$-dependent part underlined in~\eqref{KanGenHOMFLY} changes to
\begin{equation}
    \bar{\phi}^l\bar{\Theta}_{n,k}P_\Box^{\,\rm def}+\bar{\phi}^k\bar{\Theta}_{m,l}P_\Box^{DB(2,2)}+ \bar{\Theta}_{n,k}\bar{\Theta}_{m,l}P_\Box^{DB(2,2)}D\overset{\eqref{rel-gen-vert}}{=}\bar{\phi}^l\bar{\Theta}_{n,k}(P_\Box^{\,\rm def}-P_\Box^{DB(2,2)})+\bar{\Theta}_{n+m,k+l} P_\Box^{DB(2,2)}
\end{equation}
where $P^{DB(2,2)}$ is the double braid polynomial from~\eqref{mncyc}:
\begin{equation}
    P_\Box^{DB(2,2)}=D(1+\phi^2+2\phi D)\,,
\end{equation}
and
\begin{equation}
    P_\Box^{\,\rm def}=D(D+\phi)\left\{(D+\bar{\phi})\left(1+\phi(D+\phi)\right)+\phi\right\}+D\phi\bar{\phi}(1+\phi D)
\end{equation}
Obviously, $P_\Box^{\,\rm def}-P_\Box^{DB(2,2)}\neq 0$. Thus, for this deformed Kanenobu knot, the HOMFLY polynomial is separately dependent on $m$ and $n$. 

Therefore, the analysis shows that we cannot start from the precursor of a Kanenobu diagram shown on the left of Fig.\ref{fig:KanenobuJones} in order to prove dependence of the HOMFLY polynomial for a Kanenobu knot only on the sum of parameters. The reason is that this precursor diagram generates a whole hypercube of bipartite diagrams (see Section~\ref{sec:hypercube}, fourth option), and we have shown that for some of these bipartite diagrams the $(m+n)$-dependence of the HOMFLY polynomial breaks.

Note that all bipartite Kanenobu-like links considered in this section are from the same hypercube, so that their precursor Jones polynomials are given by~\eqref{Jones-Kan-precursor}.

\subsubsection{Replacing four locks explicitly shown in Figs.\ref{fig:Kanenobu},\ref{fig:Kanenobu-2}}\label{sec:Kan-factor}

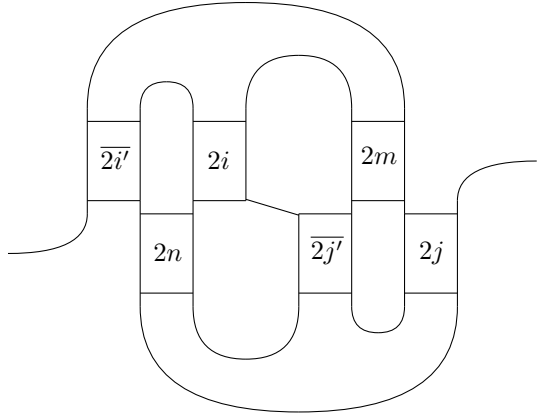
\begin{figure}[h!]
\begin{picture}(100,165)(-20,55)

\put(165,152){\mbox{$\overline{2i'}$}}
\put(205.5,152){\mbox{$2i$}}
\put(185,117){\mbox{$2n$}}
\put(263.5,154){\mbox{$2m$}}
\put(285.5,118){\mbox{$2j$}}
\put(244.5,118){\mbox{$\overline{2j'}$}}

\put(180,135){\line(1,0){20}}
\put(180,105){\line(1,0){20}}
\put(200,100){\line(0,1){40}}
\put(180,100){\line(0,1){40}}

\put(60,0){
\put(180,135){\line(1,0){20}}
\put(180,105){\line(1,0){20}}
\put(200,100){\line(0,1){35}}
\put(180,100){\line(0,1){35}}
}

\put(100,0){
\put(180,135){\line(1,0){20}}
\put(180,105){\line(1,0){20}}
\put(200,100){\line(0,1){40}}
\put(180,100){\line(0,1){35}}
}

\put(20,35){
\put(180,135){\line(1,0){20}}
\put(180,105){\line(1,0){20}}
\put(200,105){\line(0,1){35}}
\put(180,105){\line(0,1){35}}
}

\put(-20,35){
\put(180,135){\line(1,0){20}}
\put(180,105){\line(1,0){20}}
\put(200,105){\line(0,1){35}}
\put(180,100){\line(0,1){40}}
}

\put(260,170){\line(1,0){20}}
\put(260,140){\line(1,0){20}}
\put(260,135){\line(0,1){40}}
\put(280,135){\line(0,1){40}}

\qbezier(200,100)(200,80)(220,80)
\qbezier(220,80)(240,80)(240,100)
\qbezier(180,100)(180,60)(240,60)
\qbezier(240,60)(300,60)(300,100)

\qbezier(260,100)(260,90)(270,90)
\qbezier(270,90)(280,90)(280,100)

\qbezier(260,175)(260,195)(240,195)
\qbezier(240,195)(220,195)(220,175)
\qbezier(280,175)(280,215)(220,215)
\qbezier(220,215)(160,215)(160,175)

\qbezier(200,175)(200,185)(190,185)
\qbezier(190,185)(180,185)(180,175)

\put(240,134.5){\line(-1,0.3){20}}

\qbezier(300,140)(300,155)(330,155)
\qbezier(130,120)(160,120)(160,135)
    
\end{picture} 
    \caption{\footnotesize In the case $p=2m$, $q=2n$, one can replace explicitly shown AP locks in Fig.\ref{fig:Kanenobu} to AP chains of the same orientation. Overlining in $\overline{2i'}$, $\overline{2j'}$ means taking opposite locks. The HOMFLY polynomial depends on the sum $p+q$ but not on each parameter separately only in the case $i=i'$, $j=j'$.}
    \label{fig:Kan-4}
\end{figure}

The computations in Section~\ref{sec:another-Kan} are easily promoted to a 3-parametric family of knots.
Indeed, the Kanenobu diagrams in Figs.\ref{fig:Kanenobu},\ref{fig:Kanenobu-2} contains 4 AP explicitly shown locks. Each lock can be substituted by the chain of $j$ locks of the same orientations. The resulting tangle is shown in Fig.\ref{fig:Kan-4} where $i=i'=j'=j$. This correspond to changing elsewhere (but not inside $\Phi_n$, $\Phi_m$!)
\be\label{subs-Kan-3}
\phi \to \Phi_j \quad \text{and} \quad \bar{\phi} \to \overline{\Phi}_j\,.
\ee
This simple correspondence works because $\Phi_j$, $\overline{\Phi}_j$ obey the same relation as $\phi$, $\overline{\phi}\,$: 
\be
\phi+\bar{\phi}+D\phi\bar{\phi}=0\quad \longrightarrow \quad \Phi_j+\overline{\Phi}_j+D\Phi_j\overline{\Phi}_j=0\,.
\ee
Thus, making substitutions~\eqref{subs-Kan-3} in equation~\eqref{Kan-P}, we get
\be
\boxed{P_\Box^{\,{\rm Kan}(2m,2n,2j)}=
D^2\Phi_{m+n}+D\big(1+(D-D^{-1})(\Phi_j+\overline{\Phi}_j)\big)^2\,.}
\ee
For the colored HOMFLY polynomial, one has
\begin{equation}
\begin{aligned}
    H_\Box^{\,{\rm Kan}(2m,2n,2j)}&=A^{-2m-2n}\cdot D\left(D\Phi_{m+n}+\big(1+(D-D^{-1})(\Phi_j+\overline{\Phi}_j)\big)^2\right)\overset{\eqref{compolocks}}{=} \\
    &=D\left(1+\{Aq\}\{A/q\}\cdot A^{-2m-2n}\left(\frac{\{A^j\}}{\{A\}}\right)^2\left(2+\{Aq\}\{A/q\}\left(\frac{\{A^j\}}{\{A\}}\right)^2\right)\right),
\end{aligned}
\end{equation}
and the differential expansion~\eqref{DEdef} dictates that
\begin{equation}
   F_{[1]}^{\,{\rm Kan}(2m,2n,2j)}=A^{-2m-2n}\left(\frac{\{A^j\}}{\{A\}}\right)^2\left(2+\{Aq\}\{A/q\}\left(\frac{\{A^j\}}{\{A\}}\right)^2\right)
\end{equation}
so that the dependencies on $m+n$ and $j$ factorize like in the double-braid case~\eqref{DB-F}, see the discussion in Section~\ref{sec:diffexp}. The question for future research is to find out whether this factorization property holds for higher representations.

Using the method from Section~\ref{sec:p+q-dep}, it can be easily shown that the greatest generalization saving $(p+q)$-property in the direction of substitution of these 4 AP locks (as shown in Fig.\ref{fig:Kan-4}) is done by substituting vertical chain and opposite chain of $j$ locks instead of two left locks in Fig.\ref{fig:Kanenobu} and vertical chain and opposite chain of $i$ locks instead of two right locks in Fig.\ref{fig:Kanenobu}, see Fig.\ref{fig:Kan-4} in the case of $i=i'$, $j=j'$. Again, a trick is that after resolution of $(p,\,q)$-chain parts, the resulting 4 diagrams untie to a collection of unknots as in Fig.\ref{fig:Kanenobu-resolved-1}. As a result, we get
\begin{equation}
   \boxed{ P_\Box^{\,{\rm Kan}(2m,2n,2j,2i)}=D^2\Phi_{m+n}+D\big(1+(D-D^{-1})(\Phi_j+\overline{\Phi}_j)\big)\big(1+(D-D^{-1})(\Phi_i+\overline{\Phi}_i)\big)+D^2(D-D^{-1})\Phi_{i-j}\overline{\Phi}_{i-j}\,,}
\end{equation}
and restoring the framing factor for the HOMFLY polynomial:
{\small \begin{equation}\label{H-Kan-4}
\begin{aligned}
    H_\Box^{\,{\rm Kan}(2m,2n,2j,2i)}&=A^{-2m-2n}\Big(D^2\Phi_{m+n}+\overbrace{D\big(1+(D-D^{-1})(\Phi_j+\overline{\Phi}_j)\big)\big(1+(D-D^{-1})(\Phi_i+\overline{\Phi}_i)\big)+D^2(D-D^{-1})\Phi_{i-j}\overline{\Phi}_{i-j}}^{H^{\,{\rm Kan}(0,0,2j,2i)}}\Big)\overset{\eqref{compolocks}}{=} \\
    &=D\left(1+\{Aq\}\{A/q\}\cdot A^{-2m-2n}\left(\left(\frac{\{A^j\}}{\{A\}}\right)^2+\left(\frac{\{A^i\}}{\{A\}}\right)^2+\{Aq\}\{A/q\}\left(\frac{\{A^j\}}{\{A\}}\right)^2\left(\frac{\{A^i\}}{\{A\}}\right)^2-\left(\frac{\{A^{i-j}\}}{\{A\}}\right)^2\right)\right).
\end{aligned}
\end{equation}}
Again, the dependencies of the cyclotomic function on $m+n$ and $i,\,j$ factorize:
\begin{equation}\label{factor-Kan-cyc}
    F_{[1]}^{\,{\rm Kan}(2m,2n,2j,2i)}=A^{-2m-2n}\left(\left(\frac{\{A^j\}}{\{A\}}\right)^2+\left(\frac{\{A^i\}}{\{A\}}\right)^2+\{Aq\}\{A/q\}\left(\frac{\{A^j\}}{\{A\}}\right)^2\left(\frac{\{A^i\}}{\{A\}}\right)^2-\left(\frac{\{A^{i-j}\}}{\{A\}}\right)^2\right)\,.
\end{equation}
Actually, the same way the HOMFLY polynomial~\eqref{KanGenHOMFLY} are generalized:
\begin{equation}\label{KanGenHOMFLY-2}
\begin{aligned}
H_\Box^{{\rm KanGen}(2m+\varepsilon_1,2n+\varepsilon_2,2k,2l,2j,2i)}&=A^{-2m-2n+4k+4l}\left(\bar{\phi}^{k+l} H_\Box^{{\rm Kan}(\varepsilon_1,\varepsilon_2,2j,2i)}+ D^2\bar{\Theta}_{n+m,k+l}\right) 
\end{aligned}
\end{equation}
where $\varepsilon_{1,2}=0,1$. Note that only the family ${\rm KanGen}(2m,2n,2k,2l,2j,2i)$ is bipartite. It can be shown that for $k\neq 0$, $l\neq 0$, there is no factorization property of the cyclotomic function like in~\eqref{factor-Kan-cyc}.

The Jones precursor diagram in this case is not so topologically simple as in Fig.\ref{fig:KanenobuJones}, so that even in the Jones case it is not clear that the family depends only on $m+n$ but not on $m,\,n$ separately. The Jones polynomial of ${\rm KanGen}(2m,2n,2k,2l,2j,2i)$ precursor is:
\begin{equation}\label{J-Kan-4}
\begin{aligned}
    J_\Box^{\,{\rm Kan}(m,n,j,i)}&=(-1)^{\frac{1}{2}(w+n+m)}\cdot q^{-\frac{3}{2}w-\frac{1}{2}(n+m)}D\Big((-q^2)^{n+m}-1+\left(1+\frac{(-q^2)^j+(-q^2)^{-j}-2}{(q+q^{-1})^2}(q^2+q^{-2}+1)\right)\times \\
    &\times\left(1+\frac{(-q^2)^i+(-q^2)^{-i}-2}{(q+q^{-1})^2}(q^2+q^{-2}+1)\right)-\frac{(-q^2)^{i-j}+(-q^2)^{-i+j}-2}{(q+q^{-1})^2}(q^2+q^{-2}+1)\Big)\,.
\end{aligned}
\end{equation}

\subsubsection{Kanenobu family of an arbitrary number of parameters~\cite{Kanenobu-gen}}

\begin{figure}[h!]
\center{\includegraphics[width=0.5\linewidth]{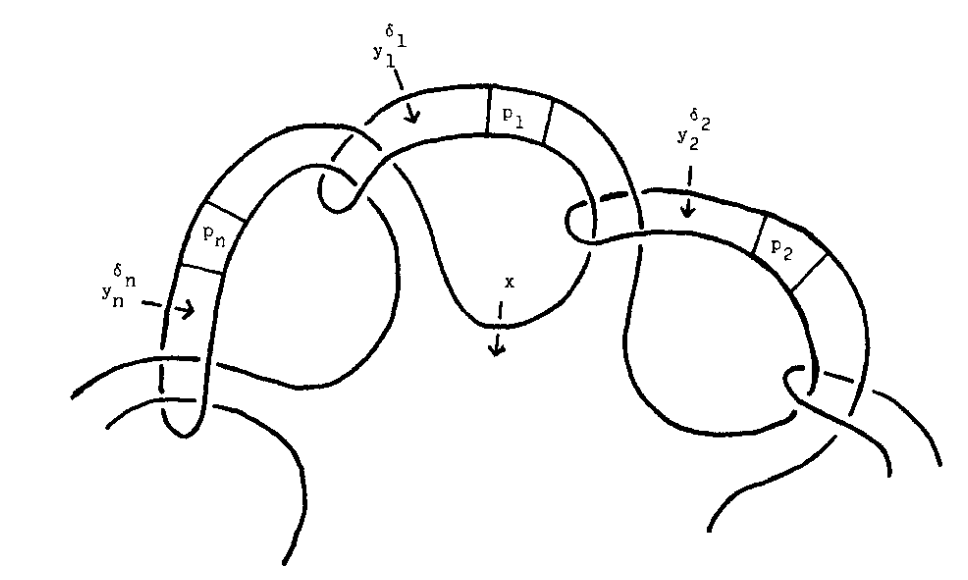}}
    \caption{\footnotesize 
    A part of a knot providing Kanenobu knots $K(p_1,\dots,p_n)$ of an arbitrary number of parameters considered in~\cite{Kanenobu-gen}.}
    \label{fig:Kanenobu-gen}
\end{figure}

In~\cite{Kanenobu-gen}, T. Kanenobu has suggested another family of knots $K(p_1,\dots,p_n)$ shown in Fig.\ref{fig:Kanenobu-gen}. The HOMFLY polynomial for these knots also have a property to depend only on the sum of the parameters. The interesting cases in this sense are ones of $n\geq 3$ as $n=1$ gives just the unknot, and for $n=2$, the knots turn out to be the same for any fixed $p_1+p_2$ what can be seen by easy topological manipulations. Note that the $n=2$ case does not give Kanenobu knots from Fig.\ref{fig:Kanenobu-2},
i.e. describes a different $2$-parametric family,
which has a far-going generalization to $n$ parameters in Fig.\ref{fig:Kanenobu-gen}.
Moreover, this general family in Fig.\ref{fig:Kanenobu-gen} is not realized in a bipartite way.
Still, it can be handled with the help of our planar
decomposition method, because all the constituent 2-braids
in boxes $p_1,\ldots, , p_n$
possess antiparallel orientation --
and explicit gluing of boxes is a simple procedure.

Consider the case of three-parametric family $K(p_1,p_2,p_3)$ with $p_i=2m_i+\varepsilon_i$ where $\varepsilon_i=0,1$ depending on the parity of the corresponding $p_i$. Resolutions of chain parts, i.e. locks from boxes $p_1,\,p_2,\,p_3$, gives (analogously to Section~\ref{sec:p+q-dep}) two unknots with the coefficients $\Phi_{m_i}$, three unknots with the coefficients $\Phi_{m_i}\Phi_{m_j}$, $i\neq j$, four unknots with the coefficient $\Phi_{m_1}\Phi_{m_2}\Phi_{m_3}$ and the Kanenobu knot $K(\varepsilon_1,\varepsilon_2,\varepsilon_3)$. As the result, we get
\begin{equation}
    P_\Box^{\,K(2m_1+\varepsilon_1,2m_2+\varepsilon_2,2m_3+\varepsilon_3)}=P_\Box^{\,K(\varepsilon_1,\varepsilon_2,\varepsilon_3)}+(\Phi_{m_1}+\Phi_{m_2}+\Phi_{m_3})D^2+(\Phi_{m_1}\Phi_{m_2}+\Phi_{m_1}\Phi_{m_3}+\Phi_{m_3}\Phi_{m_2})D^3+\Phi_{m_1}\Phi_{m_2}\Phi_{m_3}D^4\,,
\end{equation}
and property~\eqref{compolocks} leads to
\begin{equation}
    P_\Box^{\,K(2m_1+\varepsilon_1,2m_2+\varepsilon_2,2m_3+\varepsilon_3)}=P_\Box^{\,K(\varepsilon_1,\varepsilon_2,\varepsilon_3)}+\Phi_{m_1+m_2+m_3}D^2\,,
\end{equation}
so that the HOMFLY polynomial depend on the sum $m_1+m_2+m_3$. In the case of generic $n$, the calculations are the same, and the final answer is
\begin{equation}\label{Kan-mul-par}
    P_\Box^{\,K(2m_1+\varepsilon_1,\dots,2m_n+\varepsilon_n)}=P_\Box^{\,K(\varepsilon_1,\dots,\varepsilon_n)}+\Phi_{m_1+\dots+m_n}D^2\,,
\end{equation}
where the polynomial also depends on the sum of the parameters, but not on the parameters separately.

\subsection{Back to Jones polynomial}

In this section, we gather knowledge about the precursor diagrams and their Jones polynomial of all discussed Kanenobu-like links.

$\bullet$ First, the precursor of the simplest Kanenobu knot, Fig.\ref{fig:Kanenobu}, is topologically the same for all $m$, $n$ with fixed $m+n$, and obviously depends only on $n+m$, see Fig.\ref{fig:KanenobuJones}.

$\bullet$ Second, one can think whether there is a straightforward lift from the $(m+n)$ precursor property to the one for the bipartite HOMFLY. The initial precursor diagram on the left of Fig.\ref{fig:KanenobuJones} generates a whole hypercube of $2^{m+n+4}$ bipartite diagrams, and the question is whether all these diagrams respect the $(m+n)$ property. We have shown that any bipartite diagrams of type in Fig.\ref{fig:Kan-gen} possess the $(m+n)$ property while for the diagram in Fig.\ref{fig:Kan-def}(b) the $(m+n)$ property is violated. Thus, not all the diagrams from the bipartite hypercube have the $(m+n)$ property, and the Jones precursor argument cannot be lifted to the HOMFLY case by such a way.

$\bullet$ Third, we have considered the bipartite generalization of Kanenobu knots from Fig.\ref{fig:Kanenobu}, see Fig.\ref{fig:Kan-4}. In the case of $i=i'$, $j=j'$ the HOMFLY polynomial possesses the $(m+n)$-property. However, the precursor diagram is of the same form, and the dependence of its Jones polynomial on $(m+n)$ is not obvious, as it was in Fig.\ref{fig:KanenobuJones}.

\subsection{Outcome}

In this section we provide a resulting table summarizing main properties of the Kanenobu-like links under consideration. By ``factorisation'' we mean the factorization property of the cyclotomic function discussed in Section~\ref{sec:diffexp}, and by $(p+q)$ we mean the property of the HOMFLY polynomial to depend on the sum of parameters but not on the parameters separately. From equation~\eqref{Kan-mul-par}, one gets the corresponding HOMFLY and Jones polynomials by substitutions from Table~\ref{tab:subs}. Note that the Kanenobu-like links KanGen$(p,q,2k,2l,2j,2i)$ do not possess the factorization property of the cyclotomic function in general, but in the case $k=l=0$ and $p=2m$, $q=2n$ they reduce to Kan$(2m,2n,2j,2i)$ for which the factorization property holds.

\begin{table}[h!]
    \centering
    \begin{tabular}{|c|c|c|c|c|c|c|}
    \hline
         \raisebox{-0.1cm}{notation} & \raisebox{-0.1cm}{diagram} & \raisebox{-0.1cm}{$(p+q)$} & \raisebox{-0.1cm}{biparticity} & \raisebox{-0.1cm}{HOMFLY} & \raisebox{-0.1cm}{Jones} & \raisebox{-0.1cm}{factorisation}  \\ [1.5ex]
    \hline
    \hline
         \raisebox{-0.1cm}{Kan$(p,q)$} & \raisebox{-0.1cm}{Fig.\ref{fig:Kanenobu}} & \raisebox{-0.1cm}{yes} & \raisebox{-0.1cm}{for $p=2m$, $q=2n$} & \raisebox{-0.1cm}{eq.~\eqref{KanHOMFLY}} & \raisebox{-0.1cm}{eq.~\eqref{Jones-Kan-precursor}} & \raisebox{-0.1cm}{--} \\ [1.5ex]
    \hline 
        \raisebox{-0.1cm}{KanGen$(p,q,2k,2l)$} & \raisebox{-0.1cm}{Fig.\ref{fig:Kan-gen}} & \raisebox{-0.1cm}{yes} & \raisebox{-0.1cm}{for $p=2m$, $q=2n$} & \raisebox{-0.1cm}{eq.~\eqref{KanGenHOMFLY}} & \raisebox{-0.1cm}{eq.~\eqref{Jones-Kan-precursor}} & \raisebox{-0.1cm}{no} \\ [1.5ex]
    \hline 
        \raisebox{-0.1cm}{--} & \raisebox{-0.1cm}{Fig.\ref{fig:Kan-def}(b)} & \raisebox{-0.1cm}{no} & \raisebox{-0.1cm}{for $p=2m$, $q=2n$} & \raisebox{-0.1cm}{--} & \raisebox{-0.1cm}{eq.~\eqref{Jones-Kan-precursor}} & \raisebox{-0.1cm}{--} \\ [1.5ex]
    \hline
        \raisebox{-0.1cm}{Kan$(2m,2n,2j,2i)$} & \raisebox{-0.1cm}{Fig.\ref{fig:Kan-4}} & \raisebox{-0.1cm}{yes} & \raisebox{-0.1cm}{yes} & \raisebox{-0.1cm}{eq.~\eqref{H-Kan-4}} & \raisebox{-0.1cm}{eq.~\eqref{J-Kan-4}} & \raisebox{-0.1cm}{yes, eq.~\eqref{factor-Kan-cyc}} \\ [1.5ex]
    \hline 
        \raisebox{-0.1cm}{KanGen$(p,q,2k,2l,2j,2i)$} & \raisebox{-0.1cm}{--} & \raisebox{-0.1cm}{yes} & \raisebox{-0.1cm}{for $p=2m$, $q=2n$} & \raisebox{-0.1cm}{eq.~\eqref{KanGenHOMFLY-2}} & \raisebox{-0.1cm}{eq.~\eqref{J-Kan-4}} & \raisebox{-0.1cm}{in general, no} \\ [1.5ex]
    \hline
        \raisebox{-0.1cm}{$K(p_1,\dots,p_n)$} & \raisebox{-0.1cm}{Fig.\ref{fig:Kanenobu-gen}} & \raisebox{-0.1cm}{yes} & \raisebox{-0.1cm}{no} & \raisebox{-0.1cm}{eq.~\eqref{Kan-mul-par}} & \raisebox{-0.1cm}{eq.~\eqref{Kan-mul-par}} & \raisebox{-0.1cm}{--} \\  [1.5ex]
    \hline 
    \end{tabular}
    \caption{\footnotesize Summary of main properties of the Kanenobu-like links considered in this section.}
    \label{tab:summary}
\end{table}

\section{Planar technique for non-bipartite families}

The above examples illustrate that sometimes the described planar decomposition method can be applied to calculate the HOMFLY polynomial for non-fully bipartite links, or at least to simplify their calculations. A trick is applicable to links having just tangles constructed by AP locks but which may be not fully bipartite. In this case, the planar decomposition method should be applied to these bipartite tangles, and an initial link turns to a collection of simpler links.

Let us briefly repeat the Kanenobu examples from Figs.\ref{fig:Kanenobu},\ref{fig:Kan-gen},\ref{fig:Kanenobu-gen}. In all these links, for any choice of parameters (even in not fully bipartite cases), 2-strand braids are always antiparallel and thus, form bipartite tangles. Resolutions of these tangles lead to just a collection of unknots giving $D$ to the power of number of unknots contributions so that the HOMFLY polynomial is easily computed, see formulas~\eqref{KanHOMFLY},~\eqref{KanGenHOMFLY},~\eqref{KanGenHOMFLY-2},~\eqref{Kan-mul-par}. In general, resolution of bipartite tangles does not give just unknots but an initial link always simplifies. 

\setcounter{equation}{0}
\section{On III Reidemeister moves for lock diagrams}\label{sec:RIII}

Knot diagrams can be bipartite, and it is a question, if a particular knot or link can have such a realization.
Diagrams result from projections on a plane (naturally arising in the temporal gauge $A_0=0$ in the Chern--Simons theory \cite{MoSmi}),
and associated with knots/links diagrams are ones modulo Reidemeister moves.
Bipartite property is not Reidemeister invariant, and only appropriate $2d$ projection can be bipartite (if any).
For example, the trefoil $3_1$ is not bipartite if realized as a 2-strand torus knot $T[2,3]$, but becomes bipartite
if represented as a twist knot ${\rm Tw}_2$.
Planar decomposition of a diagram is also not Reidemeister invariant.
Thus, though the HOMFLY polynomials are topological invariant, our planar calculus for the HOMFLY polynomial is not.
It is also unclear what are the obstacles for existence of bipartite realization for a given knot or link --
and thus, how broad (generically applicable) is actually this approach, see Section~\ref{sec:bipknots} for the discussion.

An amusing exercise -- of yet unclear value -- is to check what happens if Reidemeister moves are extended to lock diagrams --
we call these moves the \lR\ moves.
The point is that there are now an additional freedom --
to switch between vertical and horizontal
vertices,
and what happens is that \lR\ moves mix them.

\begin{figure}[h!]
\begin{picture}(100,70)(-100,25)

    \put(40,75){\line(1,0){57}}
    \put(113,75){\line(1,0){15}}
    \put(60,30){\line(1,1){60}}
    \put(50,90){\line(1,-1){10}}
    \put(70,70){\line(1,-1){10}}
    \put(90,50){\line(1,-1){20}}

    \thicklines
    \put(155,65){\line(1,0){12}}
    \put(155,60){\line(1,0){12}}

    \thinlines
    \put(200,30){\line(1,1){60}}
    \put(225,45){\line(1,0){57}}
    \put(187,45){\line(1,0){18}}
    \put(215,90){\line(1,-1){18}}
    \put(242,63){\line(1,-1){13}}
    \put(265,40){\line(1,-1){10}}

\end{picture}
\caption{\footnotesize The III Reidemeister move
} \label{IIIReid}
\end{figure}
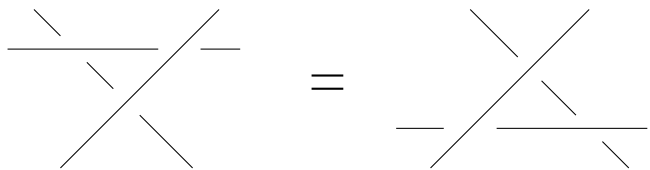

\begin{figure}[h!]
$$\begin{array}{c}
\includegraphics[width=6.5cm]{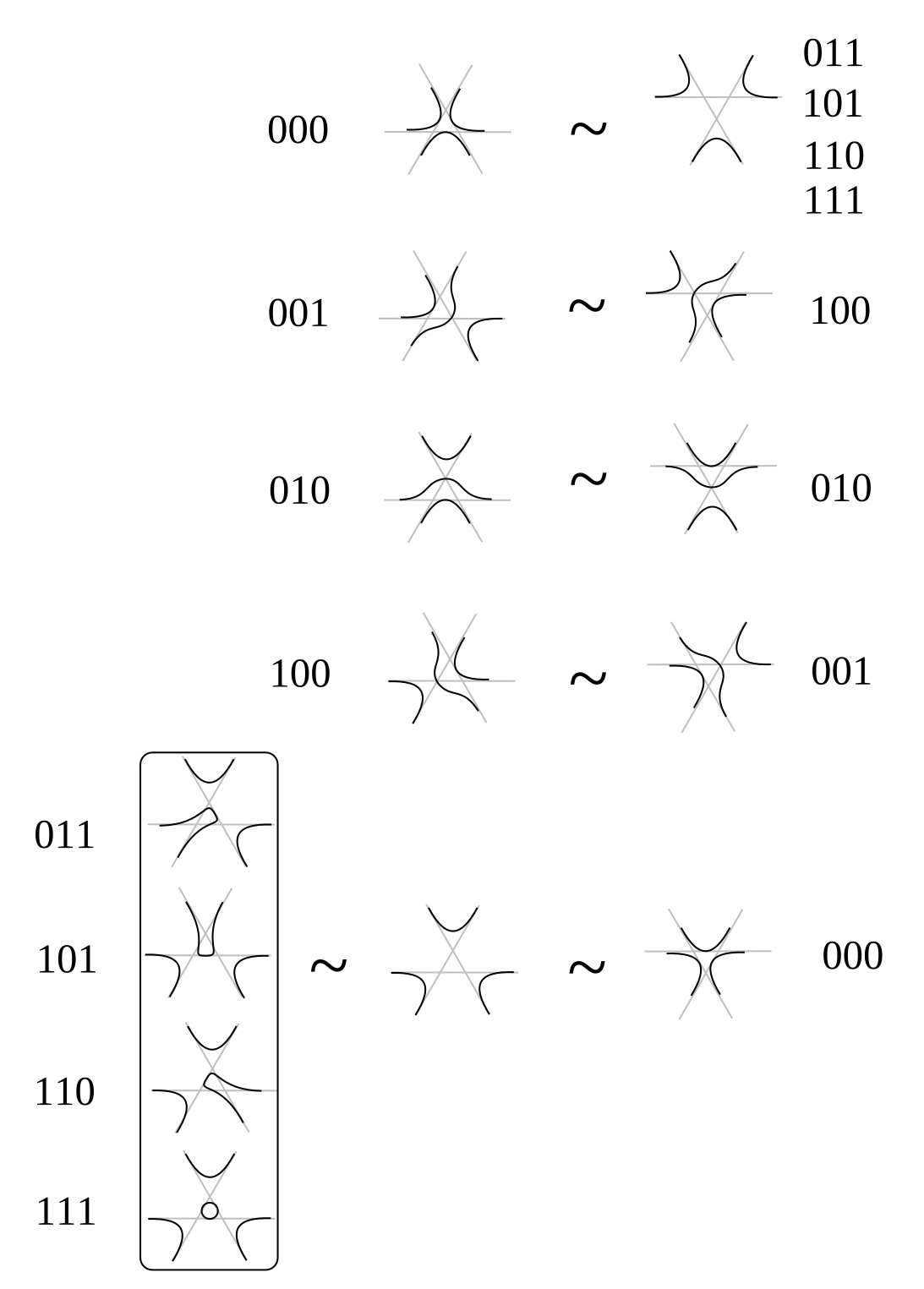}
\end{array}$$
\caption{\footnotesize Resolutions of the lock Reidemeister diagrams and their topological equivalences. On the left hand side, resolutions of the lock diagrams coming from the precursor diagram on the r.h.s. of Fig.\ref{IIIReid} are shown. On the right hand side, resolutions of the lock diagrams coming from the precursor diagram on the l.h.s. of Fig.\ref{IIIReid} are shown. The precursor diagrams are drawn as light gray avatars behind the corresponding resolutions. Numbers near diagrams enumerate resolutions. }\label{Fig:RIII}
\end{figure}

\begin{figure}[h!]
$$\begin{array}{c}
\includegraphics[width=11cm]{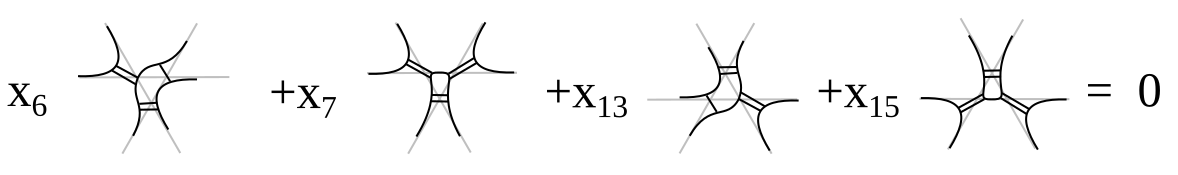}
\end{array}$$
\caption{\footnotesize The lock Reidemeister move. The coefficients are $x_6=-x_{13}=1$, $x_{15}=-x_7=-\cfrac{\bar{\phi}(1-\phi \bar{\phi})}{1+\bar{\phi}D+\bar{\phi}^2}\,$. The first two diagrams are shrinked to the left Jones precursor diagram in Fig.\ref{IIIReid}, and the other ones are shrinked to the right diagram in Fig.\ref{IIIReid}. In the Jones case $x_{15}\rightarrow q^{-1}$, and we return to the ordinary Reidemeister move in Fig.\ref{IIIReid}.}\label{Fig:RIII2}
\end{figure}

As an example, let us consider the third Reidemeister move shown in Fig.\ref{IIIReid}. Each diagram generates $2^3=8$ bipartite diagrams, and each bipartite diagram has the same resolutions shown in Fig.\ref{Fig:RIII}. Fig.\ref{Fig:RIII} also shows equivalences between various resolutions of the lock diagrams being the bipartite extensions of the l.h.s. and r.h.s. of the initial III Reidemeister move from Fig.\ref{IIIReid}. As the result, we get a system of 5 equations on 16 variables, which we denote $x_i$, $i=1,\dots,16$. Of course, the solution of this system is not unique, but among them there are 22 ``minimal" non-trivial solutions involving only four bipartite diagrams. One of these solutions is shown in Fig.\ref{Fig:RIII2}. The first two diagrams are shrinked to the left Jones precursor diagram in Fig.\ref{IIIReid}, and the other ones are shrinked to the right diagram in Fig.\ref{IIIReid}.
%

\begin{figure}[h!]
$$\begin{array}{c}
\includegraphics[width=10cm]{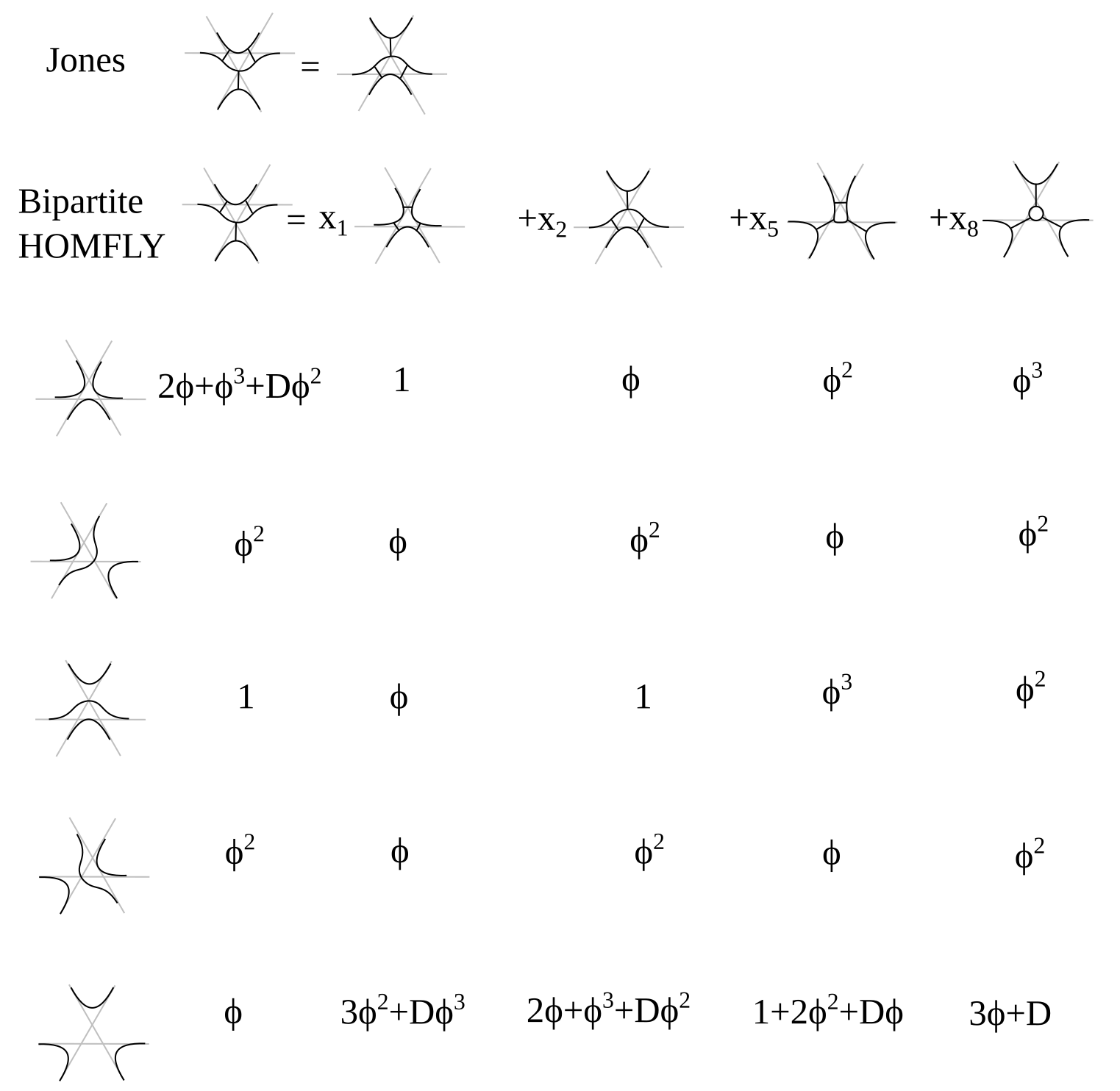}
\end{array}$$
\caption{\footnotesize In the second line, the lock Reidemeister move is shown. The coefficients $x_1$, $x_2$, $x_5$, $x_8$ are given by~\eqref{coef-2} and can be found by solving a system of equations~\eqref{x-system} formed from coefficients of resolutions below. In the top line, the bipartite diagrams contributing in the Jones case are shown. The corresponding Jones precursor diagrams give exactly the III Reidemeister move from Fig.\ref{IIIReid}.}\label{Fig:RIII1}
\end{figure}



An interesting peculiarity of lock Reidemeister moves is that there exist relations involving locks coming from opposite to ones shown in Fig.\ref{IIIReid} intersections in bipartite diagrams. If we include such locks, than each diagram in Fig.\ref{IIIReid} gives $4^3=64$ diagrams. Let us describe in detail the example in Fig.\ref{Fig:RIII1} where only locks but not opposite locks are included. We want to find a solution of the equation with 5 diagrams and 4 coefficients. The solution is unique because this equation leads to a system of 4 distinct equations on coefficients of 5 resolutions. In the Jones case, only two diagrams shown at the top of Fig.\ref{Fig:RIII1} survive. Shrinking the locks in these diagrams leads exactly to the III Reidemeister move shown in Fig.\ref{IIIReid}. From Fig.\ref{Fig:RIII1}, we write down 4 distinct linear equations on 4 variables:
\be\label{x-system}
2\phi+\phi^3+D\phi^2&=&x_1+x_2\phi+x_5\phi^2+x_8\phi^3,\nn\\
\phi^2&=&x_1\phi+x_2\phi^2+x_5\phi+x_8\phi^2,\nn\\
1&=&x_1\phi+x_2+x_5\phi^3+x_8\phi^2,\\
\phi&=&x_1(3\phi^2+D\phi^3)+x_2(2\phi+\phi^3+D\phi^2)+x_5(2\phi^2+1+D\phi)+x_8(3\phi+D).\nn
\ee
This system has the solution
\be\label{coef-2}
x_1=-x_5=x_8/\phi=\frac{\phi^2+D\phi+1}{(\phi^2-1)^2},\qquad x_2=\frac{-D\phi^3+3\phi^2-1}{(\phi^2-1)^2}.
\ee
In particular, this solution gives the regular third Reidemeister move in the Jones case because
\be
\phi=-q,\quad D=q+q^{-1}\qquad\Rightarrow\qquad x_1=x_5=x_8=0,\qquad x_2=1.
\ee


\setcounter{equation}{0}
\section{Towards planar calculus for other symmetric representations}\label{sec:symm}

%

Extension of our bipartite Kauffman calculus to the {\it colored} HOMFLY polynomial can seem problematic.
The problem is that cabling is inconsistent with planarity.
If we just substitute lines  by cables, then the cabled lock does not decompose into elementary locks --
non-bipartite diagrams arise and cabled \ld\ is not reduced to ordinary \ld s.
Thus cabling does not help -- the lock-intersection of multiple lines is not planarized.
However, projection to (at least) the symmetric representations is planar, see Fig.\ref{cable}.
The point is that the product of two symmetric representations contains $r+1$ irreducibles:
$[r]\otimes \overline{[r]} = \sum_{i=0}^{r} Adj_{(i)}$, where $Adj_{(i)}$ are representations from the adjoint sector -- and this is exactly the number of different
planar resolutions of the $r$-cable.
Thus, if we project out $[r]$ from the $r$-cables at each end, it will be possible to adjust the $r+1$
coefficients in the planar expansion of the colored lock.
For example, for $r=2$ we get
\be
\phi_{2,0}=1, \ \ \ \
\phi_{2,1} = A^2q^3\{q\}\{q^2\}, \ \ \ \ \phi_{2,2} = [2]Aq^2\{q^2\}\,.
\ee
Details involve somewhat different technique of a projector calculus and its quantization,
together with amusing parallels with \cite{DM3,AnoM}
-- and will be presented elsewhere.

\begin{figure}[h]
\begin{picture}(100,60)(-130,-75)

\put(0,-50){
\put(-105,18){\vector(1,-1){10}} \put(-85,6){\vector(1,1){10}}
\put(-105,15){\vector(1,-1){10}} \put(-85,3){\vector(1,1){10}}
\put(-105,12){\vector(1,-1){10}} \put(-85,0){\vector(1,1){10}}
\put(-95,0){\vector(-1,-1){10}} \put(-75,-12){\vector(-1,1){10}}
\put(-95,-3){\vector(-1,-1){10}} \put(-75,-15){\vector(-1,1){10}}
\put(-95,-6){\vector(-1,-1){10}} \put(-75,-18){\vector(-1,1){10}}
\put(-90,0){\circle{10}}

\put(-125,15){\mbox{$\Pi_{[3]}$}}  \put(-70,15){\mbox{$\Pi_{[3]}$}}
\put(-125,-21){\mbox{$\Pi_{[3]}$}}  \put(-70,-21){\mbox{$\Pi_{[3]}$}}

\put(-60,-2){\mbox{$= \ \ \sum_{i=0}^r \phi_{r,i} \cdot (i{\rm -th\ planar\ resolution})\ = $}}

\put(142,15){\vector(0,-1){30}} \put(145,15){\vector(0,-1){30}} \put(148,15){\vector(0,-1){30}}
\put(157,-15){\vector(0,1){30}} \put(160,-15){\vector(0,1){30}} \put(163,-15){\vector(0,1){30}}

\put(200,0){
\put(-30,-2){\mbox{$ +\  \phi_{3,1}   $}}
\put(2,15){\vector(0,-1){30}} \put(5,15){\vector(0,-1){30}}
\qbezier(8,15)(12.5,0)(17,15)\put(16,12){\vector(1,2){2}}
\qbezier(8,-15)(12.5,0)(17,-15)\put(9,-12){\vector(-1,-2){2}}
\put(20,-15){\vector(0,1){30}} \put(23,-15){\vector(0,1){30}}
}

\put(260,0){
\put(-30,-2){\mbox{$   +\  \phi_{3,2}    $}}
\put(2,15){\vector(0,-1){30}}
\qbezier(8,15)(12.5,0)(17,15)\put(16,12){\vector(1,2){2}}
\qbezier(8,-15)(12.5,0)(17,-15)\put(9,-12){\vector(-1,-2){2}}
\qbezier(5,15)(12.5,-10)(20,15)\put(19,12){\vector(1,2){2}}
\qbezier(5,-15)(12.5,10)(20,-15)\put(6,-12){\vector(-1,-2){2}}
\put(23,-15){\vector(0,1){30}}
}

\put(300,0){
\put(-10,-2){\mbox{$  + \ \phi_{3,3}    $}}
\put(20,15){\vector(1,0){20}} \put(20,12){\vector(1,0){20}} \put(20,9){\vector(1,0){20}}
\put(40,-15){\vector(-1,0){20}} \put(40,-12){\vector(-1,0){20}} \put(40,-9){\vector(-1,0){20}}
}
}

\end{picture}
\caption{\footnotesize If all the external legs are projected to the symmetric representation $[r]=[3]$,
than it is possible to adjust the coefficients at the r.h.s. to make this identity valid.
} \label{cable}
\end{figure}

\setcounter{equation}{0}
\section{Khovanov-like calculus}\label{sec:Kh-bip}

One more natural question to ask concerns Khovanov calculus \cite{Kho,BarN,DM12}.
Since we have a direct generalization of Kauffman decomposition and a hypercube formalism,
is it possible to raise it to homological level, i.e. perform a categorification?
The usual story is that at each vertex of hypercube we have a system of cycles,
which can be easily promoted to the product of linear spaces, each of quantum dimension $D$.
What remains to do is to promote the reshuffling of cycles into linear maps with
the nilpotent property -- what allows then to consider cohomologies of the emerging complex
and their generating function, known as Khovanov polynomial.
Usually things get much simplified by the fact that along every edge of the hypercube
the reshuffling reduces to a single operation of cutting or joining a single pair of cycles,
what allows to look for the differentials among the standard and easily constructed
cut-and-join operators, see \cite{DM12} for a presentation in these terms.

A straightforward lift of this construction from the Jones polynomial to the HOMFLY polynomial is not so easy \cite{DM3},
because one of the resolutions of ordinary $R$-matrix  vertices at $N>2$ involves a difference,
thus there are differences of cycles at hypercube vertices and their natural lifting
are cosets (factor spaces).
Generic theory of cut-and-join operators in this case remains obscure
(does not attract enough interest to be developed).
Just a guess-work instead of established theory is quite hard and not fully satisfactory \cite{AnoM}.

However, for bipartite diagrams things can be different: decompositions are in terms
of ordinary cycles and there is no need for factor spaces.
Thus, categorification should be much simpler, if not straightforward.
It will be in terms of {\it lock hypercube}, thus, its relation to full topological invariance
and to Khovanov-Rozansky calculus \cite{KhR,KhR-2,KhR-3} can be absent --
still it would be interesting to see what will emerge in this way and what can be its potential use.
These are also questions for the future. 

\setcounter{equation}{0}
\section{Conclusion}

In this paper we explained how Kauffman planar decomposition for $N=2$ can be lifted to an arbitrary $N$
at the expense of restriction to peculiar bipartite knot diagrams.
Many knots and links actually possess bipartite realizations --
and for them, the calculation of the HOMFLY polynomials and study of various evolutions become very simple.
Moreover, this calculus can be extended to the symmetrically colored HOMFLY polynomial. In another direction, Kauffman decomposition for the Jones polynomial is used in numerous attempts
to apply knot theory to anyon physics, to quantum computers and algorithms.
This means that our far-going generalization can probably be straightforwardly used
for applications, considered in \cite{MMMMMel}, \cite{AndMor}
and especially in their version in \cite{Mel}. There are numerous interesting questions, raised by the construction of planar calculus to the HOMFLY polynomial,
some of them are explained in the mini-sections of the above text, and we write down them briefly below.

Directions for future research include
\begin{itemize}
    \item lift from the Khovanov polynomials to the Khovanov--Rozansky polynomials for bipartite knots, as described in Section~\ref{sec:Kh-bip};
    \item explicit bipartite realization of torus knots and other knot families;
    \item developing non-biparticity criteria via Alexander ideals and Seifert matrices, see Section~\ref{sec:bipknots};
    \item study on factorization property of cyclotomic functions of higher representations for Kanenobu-like links, see Section~\ref{sec:Kan-factor};
    \item planar decomposition of the HOMFLY polynomial for virtual knots, see the last paragraph in Introduction;
    \item applications of lock Reidemeister moves, see Section~\ref{sec:RIII}.
\end{itemize}

\section*{Acknowledgments}

Our work is supported by the RSF grant No.24-12-00178.

\printbibliography

\newpage

\appendix
\def\theequation{A.\arabic{equation}}
\setcounter{equation}{0}
\section{Computation of second Alexander ideals}\label{sec:App}

In this section, we provide examples of computations of the second Alexander ideals needed to distinguish non-bipartite knots, see Section~\ref{sec:bipknots}. The Seifert matrices are taken from~\cite{knotinfo}.

\subsection*{Knot $3_1$}

For the trefoil knot, the Seifert matrix is
\begin{equation}
    S=\left(
\begin{array}{cc}
 -1 & 0 \\
 -1 & -1 \\
\end{array}
\right).
\end{equation}
Then, the Alexander matrix is
\begin{equation}
    \mathbb{A}=q^2 S-S^\vee=\left(
\begin{array}{cc}
 1-q^2 & 1 \\
 -q^2 & 1-q^2 \\
\end{array}
\right).
\end{equation}
Its minors of matrices of size 1 = generators of the 2-nd Alexander ideal in $\mathbb{Z}[q^2,q^{-2}]$ are
\begin{equation}
    \langle 1 - q^2,\, 1,\, -q^2,\, 1 - q^2 \rangle = \langle 1 \rangle\,. 
\end{equation}
The Alexander polynomial is
\begin{equation}
    {\rm Al}_\Box = \det \mathbb{A} = q^4-q^2+1\,.
\end{equation}

\subsection*{Knot $8_{18}$}

For the knot $8_{18}$, the Seifert matrix is
\begin{equation}
    S=\left(
\begin{array}{cccccc}
 1 & 0 & 0 & 0 & -1 & 0 \\
 0 & -1 & 0 & 0 & 0 & 0 \\
 0 & -1 & -1 & 0 & 0 & 0 \\
 1 & -1 & -1 & 1 & -1 & 1 \\
 0 & -1 & -1 & 0 & -1 & 0 \\
 1 & 0 & -1 & 0 & -1 & 1 \\
\end{array}
\right).
\end{equation}
Then, the Alexander matrix is
\begin{equation}
    \mathbb{A}=q^2 S-S^\vee=\left(
\begin{array}{cccccc}
 q^2-1 & 0 & 0 & -1 & -q^2 & -1 \\
 0 & 1-q^2 & 1 & 1 & 1 & 0 \\
 0 & -q^2 & 1-q^2 & 1 & 1 & 1 \\
 q^2 & -q^2 & -q^2 & q^2-1 & -q^2 & q^2 \\
 1 & -q^2 & -q^2 & 1 & 1-q^2 & 1 \\
 q^2 & 0 & -q^2 & -1 & -q^2 & q^2-1 \\
\end{array}
\right).
\end{equation}
Its minors of matrices of size 5 = generators of the 2-nd Alexander ideal in $\mathbb{Z}[q^2,q^{-2}]$ are
\begin{equation}
\begin{aligned}
    &\langle -(-1 + q^2)^3 (1 - q^2 + q^4),\, -q^2 (1 - q^2 + q^4),\, (-2 + q^2) q^4 (1 - q^2 + 
    q^4),\, (-1 + q^2)^2 (1 - q^2 + q^4),\, (-1 + q^2)^2 (1 - q^2 + 
    q^4),\, \\ &-(-1 + 2 q^2) (1 - q^2 + q^4), q^4 (1 - q^2 + q^4),\, (-1 + 
    q^2)^3 (1 - q^2 + q^4),\, -q^4 (1 - q^2 + q^4),\, -(-1 + 2 q^2) (1 - q^2 + 
    q^4),\, \\ &-(-1 + q^2) q^2 (1 - q^2 + q^4), (-1 + q^2)^2 q^2 (1 - q^2 + 
    q^4),\, (-1 + 2 q^2) (1 - q^2 + q^4),\, 
 q^2 (1 - q^2 + q^4),\, -(-1 + q^2)^3 (1 - q^2 + q^4),\, \\ &-q^2 (1 - q^2 + 
    q^4),\, -(-1 + q^2)^2 (1 - q^2 + q^4),\, (-1 + q^2) q^2 (1 - q^2 + 
    q^4),\, -(-1 + q^2)^2 q^2 (1 - q^2 + q^4),\, -(-2 + q^2) q^4 (1 - q^2 + q^4),\, 
 \\ & q^4 (1 - q^2 + q^4), (-1 + q^2)^3 (1 - q^2 + q^4),\, (-1 + 2 q^2) (1 - q^2 + 
    q^4),\, -(-1 + q^2)^2 q^2 (1 - q^2 + q^4),\, -(-1 + q^2)^2 q^2 (1 - q^2 + 
    q^4),\, \\ &-(-1 + q^2) q^2 (1 - q^2 + q^4),\, (-1 + q^2)^2 q^2 (1 - q^2 + q^4),\, (-2 +
     q^2) q^4 (1 - q^2 + q^4),\, (-1 + q^2)^3 (1 - q^2 + q^4),\, -q^4 (1 - q^2 + 
    q^4),\, \\ &-(-2 + q^2) q^4 (1 - q^2 + q^4),\, -(-1 + q^2)^2 (1 - q^2 + 
    q^4),\, (-1 + q^2) q^2 (1 - q^2 + q^4),\, (-1 + q^2)^2 (1 - q^2 + q^4),\, 
 q^2 (1 - q^2 + q^4),\, \\ & -(-1 + q^2)^3 (1 - q^2 + q^4) \rangle = \langle 1 - q^2 + q^4 \rangle \,.
\end{aligned}
\end{equation}
The Alexander polynomial is
\begin{equation}
    {\rm Al}_\Box = \det \mathbb{A} = -q^{12}+5 q^{10}-10 q^8+13 q^6-10 q^4+5 q^2-1\,.
\end{equation}

\subsection*{Knot $9_{35}$}

For the knot $9_{35}$, the Seifert matrix is
\begin{equation}
    S=\left(
\begin{array}{cc}
 -3 & -1 \\
 -2 & -3 \\
\end{array}
\right).
\end{equation}
Then, the Alexander matrix is
\begin{equation}
    \mathbb{A}=q^2 S-S^\vee=\left(
\begin{array}{cc}
 3-3 q^2 & 2-q^2 \\
 1-2 q^2 & 3-3 q^2 \\
\end{array}
\right).
\end{equation}
Its minors of matrices of size 1 = generators of the 2-nd Alexander ideal in $\mathbb{Z}[q^2,q^{-2}]$ are
\begin{equation}
    \langle -3 (-1 + q^2),\, 2 - q^2,\, 1 - 2 q^2,\, -3 (-1 + q^2) \rangle = \langle 3,\,1+q^2 \rangle 
\end{equation}
because
\begin{equation}
\begin{aligned}
    1-2q^2 &=3-2(q^2+1)\,, \\
    -3 (-1 + q^2)&=6-3(q^2+1)\,, \\
    2-q^2&=3-(q^2+1)\,.
\end{aligned}
\end{equation}
Thus, the second Alexander ideal of the knot $9_{35}$ contains the polynomial $1+q^2$, and thus, $9_{35}$ is not bipartite due to Theorem 1 of Section~\ref{sec:bipknots}.

The Alexander polynomial is
\begin{equation}
    {\rm Al}_\Box = \det \mathbb{A} = 7 q^4-13 q^2+7\,.
\end{equation}

\end{document}